%% file: main.tex
\documentclass[12pt,preprint]{aastex63}

\usepackage{amssymb}
\usepackage{amsmath}
\usepackage{epstopdf}
\usepackage{color}
\usepackage{amsmath}
\usepackage{graphicx}
\usepackage{hyperref}
\usepackage{graphics,graphicx}
\usepackage{natbib}
\usepackage{cleveref}
\usepackage{multirow}
\usepackage{longtable}
\usepackage{threeparttablex}
\usepackage{booktabs}
\usepackage{marginnote}
\usepackage{verbatim}
\usepackage{makecell}

\def\lapp{\ifmmode\stackrel{<}{_{\sim}}\else$\stackrel{<}{_{\sim}}$\fi}
\def\gapp{\ifmmode\stackrel{>}{_{\sim}}\else$\stackrel{>}{_{\sim}}$\fi}

\newcommand{\bonsai}{{\tt{bonsai}}}
\newcommand{\fitburst}{{\tt{fitburst}}}

\newcommand{\degrees}{\ensuremath{^{\circ}}}
\renewcommand{\vec}[1]{\boldsymbol{#1}}
\newcommand{\dm}{\mathrm{DM}}
\newcommand{\SNR}{\mathrm{SNR}}

\definecolor{changed}{rgb}{0, 0, 1.}

\newcommand{\arxivonly}[1]{#1}

\begin{document}
\title{The First CHIME/FRB Fast Radio Burst Catalog} 
\shorttitle{CHIME/FRB Catalog 1}
\shortauthors{CHIME/FRB Collaboration: M.~Amiri, \emph{et al.}}

\collaboration{1000}{The CHIME/FRB Collaboration}

\input{auth.tex}

\correspondingauthor{Kiyoshi W.\ Masui}
\email{kmasui@mit.edu}

\begin{abstract}

We present a catalog of 536 fast radio bursts (FRBs) detected by the Canadian
Hydrogen Intensity Mapping Experiment Fast Radio Burst (CHIME/FRB) Project
between 400 and 800\,MHz
from 2018 July 25 to 2019 July 1,
including 62 bursts from 18 previously reported repeating sources.
The catalog represents the first large sample, including bursts from
repeaters and non-repeaters, observed in a single survey
with uniform selection effects. This 
facilitates comparative and absolute studies of the FRB
population.
We show that repeaters and apparent non-repeaters have sky locations and
dispersion measures (DMs) that are consistent with being drawn from the same distribution.
However, bursts from repeating sources differ
from apparent non-repeaters in intrinsic temporal width and spectral
bandwidth. Through injection of simulated events into our detection
pipeline, we perform an absolute calibration of selection effects
to account for systematic biases.
We find evidence for a population of FRBs---comprising
a large fraction of the overall population---with a scattering
time at 600\,MHz in excess of 10\,ms, of which only a small fraction are
observed by CHIME/FRB. We infer a power-law index for the
cumulative fluence distribution of
$\alpha=-1.40\pm0.11(\textrm{stat.})^{+0.06}_{-0.09}(\textrm{sys.})$,
consistent with the $-3/2$ expectation for a non-evolving population in
Euclidean space. We find $\alpha$ is steeper for high-DM events and shallower for low-DM events, which is
what would be expected when DM is correlated with distance.
We infer a sky rate of
$[525\pm30(\textrm{stat.})^{+140}_{-130}({\textrm{sys.}})]/\textrm{sky}/\textrm{day}$
above a fluence of 5\,Jy\,ms at 600\,MHz, with
scattering
time at $600$\,MHz under 10\,ms, and DM above 100\,pc\,cm$^{-3}$.

\end{abstract}


\section{Introduction}
\label{sec:intro}

Although the first Fast Radio Burst (FRB) was discovered nearly a decade and a half ago \citep{lbm+07}, the nature
of these sources remains a mystery.  Now securely determined to originate from
external galaxies,
generally
from cosmological distances \citep[e.g.,][]{tbc+17,mpm+20}, FRBs inhabit a unique and extreme portion of radio luminosity/time scale
phase space \citep[e.g.,][]{cs19} compared to other radio
transients
and hence are of great interest.  Moreover, all-sky rates of $\sim 10^3$ per
day \citep{bkb+18} indicate that the phenomenon
is ubiquitous. The
mystery of FRBs therefore signals a common 
cosmic
phenomenon borne from extreme, unknown environments.

One major clue regarding the nature of FRBs is that some repeat
\citep{ssh+16a}, with periodic activity observed in two sources \citep{aab+20,
rms+20}.  Repetition rules out cataclysmic
models for at least the repeating FRB sources, though it remains
unclear whether all FRBs are repeating sources that come with vastly
different waiting times between repetitions \citep{rav19, jof+20b}.  Evidence for distinct 
emission phenomena
has come from repeat bursts being wider 
than
those from apparent non-repeaters \citep{ssh+16b,abb+19c,fab+20}.
Additionally, the two localized repeating FRBs whose hosts have measured
properties \citep{clw+17,mnh+20} are in
late-type galaxies that have star formation, whereas 
localizations of apparent non-repeaters indicate these
latter sources can sometimes reside in galaxies with modest or little star formation \citep{bsp+20}.
To date, one Galactic magnetar has shown both repeated X-ray bursts
and a radio burst of luminosity close to the FRB range
\citep{abb+20,bkr+20}. This suggests that repeaters may be young, active magnetars, a scenario consistent with localizations of repeating FRBs to star-forming locations \citep{clw+17,mnh+20}. Volumetric rate
comparisons between FRBs and giant magnetar flares have been used
to support the
magnetar scenario \citep[e.g.,][]{abb+20}; however, uncertainties in current rate estimates are large, being
dominated by either small number statistics or by systematics from including multiple different surveys having distinct
biases.  Detailed studies of a larger sample of FRBs from a single survey, repeating or not,
are clearly of great value.

A detailed study of large numbers of FRBs, in a single homogeneous survey
with a well-measured instrument selection function, is desirable for many additional reasons.
A wide-field survey of many FRBs could be used to probe large-scale
structure through spatial correlations \citep{2015PhRvL.115l1301M}, or combined with galaxy surveys to search for correlations and variations
with redshift \citep[e.g.,][]{rsm20}.  Furthermore, the FRB sky distribution can be correlated with Galactic direction to investigate claims of
Galactic Plane avoidance \citep{bb14,psj+14,kp15}.  Large FRB samples in different frequency ranges can help elucidate the population's
average spectrum \citep[e.g.,][]{kca+15,ckj+17} or effects of FRB radio wave propagation in local environments \citep{cwh+17}.  Moreover, a large
sample of FRBs can be used to determine the population's energy distribution function
\citep{vrhs16,lvl+17,jem+19,hgo+20,jpm+21a,jpm+21b}, which contains evidence of the redshift distribution of FRB sources, as well as their detectability as a function of survey sensitivity for a given telescope.  
Analyses of dispersion measure (DM) distributions,
especially comparing repeaters and apparent non-repeaters, can reveal different source class locations and
environments, as could searches for differences in scattering times or bandwidths.  Additionally, correlations among parameters can
be investigated with a large enough sample; for example, a DM--scattering
correlation could signal either that the local environment contributes to both
measures \citep{qsf+20}, or that galaxy halos along the line of sight cause radio wave scattering \citep{vp19}.
Alternatively, a width--DM correlation is expected due to Hubble expansion if DM is 
indeed a faithful proxy for cosmic distance as recent studies suggest
\citep{mpm+20}.  However, 
all
previous and current surveys have
limited fields-of-view, sensitivity, survey durations, and/or processing capabilities,
rendering them incapable of detecting sufficiently large numbers of FRBs to address many or all of the 
above possibilities.
Past efforts have required the combination of the results from
several
individual surveys to boost statistical power; but, these surveys have different and
largely undetermined instrumental transfer functions, which result in strong
biases in their data sets \citep{con19}.
The detection of a large number of events with a single instrument, for which a well-defined selection function can be robustly determined, can therefore enable significant progress in the field.

\begin{figure}
    \centering
    \includegraphics[width=0.95\textwidth]{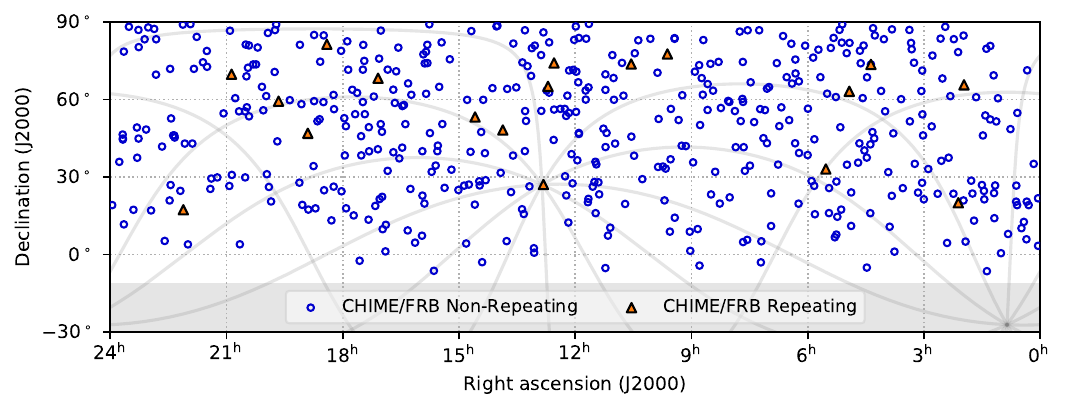}
    \caption{
    Sky distribution of 18 repeating sources and 474 sources that have not been
    observed to repeat. The gray shaded region at the bottom is outside the CHIME/FRB
    field of view, below declination $-11^{\circ}$. The gray lines
    in the background are the Galactic Plane and lines of constant Galactic
    longitude spaced by 30$^{\circ}$. CHIME, being a transit telescope, has
    relatively uniform exposure in this Mercator projection of equatorial
    coordinates.
    }
    \label{fig:catalog}
\end{figure}

The CHIME/FRB Project \citep{abb+18} uses the Canadian Hydrogen Intensity Mapping Experiment (CHIME) to detect FRBs 
in the 400--800-MHz band.  CHIME's large collecting area and wide field-of-view make it an excellent FRB detector.
Indeed, during a few initial weeks of pre-commissioning, CHIME/FRB detected over a dozen new FRBs,
demonstrating that the phenomenon exists down to 400 MHz, the lowest known frequency at that time \citep{abb+19a}. CHIME/FRB also detected the second
known source to emit repeat bursts \citep{abb+19b}.  Since then, CHIME/FRB has discovered an additional 17 repeaters \citep{abb+19c,fab+20},
as well as 
one that repeats regularly \citep{aab+20}.
These repeating sources, published rapidly to allow
the community to assist with localization efforts \citep[e.g.,][]{mnh+20}, 
are part of a larger number of FRBs detected by
CHIME/FRB during its first year of operation. 
Here we present the first FRB Catalog released by CHIME/FRB, hereafter
referred to as ``Catalog~1''.
We include 536 bursts
detected between 25 July 2018 and 1 July 2019 including all bursts from repeating sources previously published in other
works. The sky distribution of these bursts is shown in
Figure~\ref{fig:catalog}.

This paper is organized as follows.  In Section~\ref{sec:obs}, we describe the observing parameters, including a brief synopsis
of the instrument, pipeline and overall methodologies used, as well as our sky exposure, sensitivity thresholds,
and our determination of instrumental biases.
In Section~\ref{sec:catalog}, we describe the contents of our catalog and how they were determined, including burst localizations and properties, along with relevant tables and figures.
Section~\ref{sec:injection} describes our method for injecting synthetic
signals into the our detection pipeline for calibrating selection biases.
In Section~\ref{sec:repvsnonrep}, we compare parameter distributions for repeaters and apparent non-repeaters, in order to identify differences between these types of sources.
In Section~\ref{sec:abs_pop},
we show parameter distributions corrected for instrumental biases and deduce cosmic distributions for many key FRB parameters, including the FRB
fluence distribution.
In Section~\ref{sec:discussion} we discuss these results
in the context of other FRB findings in the literature, and briefly describe contemporaneous analyses of
these same data that are presented in accompanying papers.  We present our conclusions in Section~\ref{sec:conclusions}.

\section{Observations}
\label{sec:obs}

The CHIME telescope, its FRB detection instrument, and real-time pipeline
have been described in detail elsewhere \citep[see
Table~\ref{ta:chime};][]{abb+18}.  Briefly, the telescope is
located on the grounds of the Dominion Radio Astrophysical Observatory (DRAO)
near Penticton, British Columbia, Canada, and consists of four 20-m $\times$ 100-m
cylindrical paraboloid reflectors oriented N-S, with each cylinder axis
populated by 256 equispaced dual-linear-polarization antennas sensitive in the
frequency range 400--800~MHz.  CHIME has no moving parts.
The CHIME/FRB detector
views the entire
sky north of declination $-11^{\circ}$ (a configurable choice, see below) every day  as it transits overhead.
Sources northward of declination $+70^{\circ}$
are visible twice per day, on opposite sides of the North Celestial Pole.  The
2048 antenna signals are amplified, conditioned, digitized, and split into 1024 frequency channels at 2.56-$\mu$s time resolution
by the portion of CHIME's correlator called the ``F-Engine'', which uses 128
custom-built field programmable gate array (FPGA)-based ``ICE'' Motherboards
\citep{bbc+16} housed in two radio frequency interference (RFI)-shielded
shipping containers located under the reflectors.  The signals are then sent to the ``X-Engine'', which consists of 256 liquid-cooled GPU-based
compute nodes located in two custom RFI-shielded shipping containers located
adjacent to the reflectors.  Within the X-Engine, the spatial correlation is
performed and polarizations are summed, forming 1024 independent total
intensity sky beams along the N-S primary beam \citep[256 N-S $\times$ 4 E-W;][]{nvp+17}, as well as up-channelization to the 16k frequency channels at
0.983-ms time resolution used in the real-time FRB search.  Formed beams are
spaced evenly in $\sin\theta$ N-S from $\theta = -60^{\circ}$ to
$\theta = +60^{\circ}$, where $\theta$ is the angle from zenith along the meridian.
Formed beam  centers  are  separated  by  0.4$^{\circ}$; the FWHM of each beam is approximately 0.5$^{\circ}$ at 400 MHz and 0.25$^{\circ}$ at 800 MHz, though the aspect ratio E-W versus N-S changes with declination \citep[see][for details]{nvp+17}.

\begin{table}[t]
\begin{center}
\caption{Key Properties of the CHIME Telescope and CHIME/FRB Instrument}
\begin{tabular}{lc} \hline
    Parameter  &  Value \\\hline
    Collecting area & 8000 m$^2$ \\
    Longitude & $119^{\circ}37' $03\farcs00$ $ West \\
    Latitude & $49^{\circ}19' 13\farcs08$ North\\
    Altitude & 547.9\,m \\
    Frequency range & 400--800 MHz \\
    Polarization & orthogonal linear \\
    E-W FoV & 2.5$^{\circ}$--1.3$^{\circ}$ \\
    N-S FoV & $\sim$120$^{\circ}$ \\
    Focal ratio, $f/D$ & 0.25 \\ 
    Receiver noise temperature$^{\rm a}$ & $\sim$50 K \\
    Number of beams & 1024 \\
    Synthesized Beam width (FWHM) & 40$^\prime$--20$^\prime$ \\
    FRB search time resolution & 0.983 ms \\
    FRB search frequency resolution & 24.4 kHz \\
    Source transit duration & Equator: 10--5 min \\
                            & Dec = 45$^{\circ}$: 14--7 min \\
                            & North Celestial Pole:  24 hr \\\hline
\end{tabular}
\label{ta:chime}
\end{center}
    \textbf{Notes.} Reproduced from \citet{abb+18} for convenience with
    updates to reflect the current operating configuration. Where two numbers appear, they refer to the low and high frequency edges of the band, respectively.
    \\
    $^{\rm a}$ Including losses in the feeds, the full analog chain, and ground spill, although very
    approximate.
\end{table}

Each beam's data are searched in real time for FRBs using a custom-developed
triggering software pipeline consisting of four stages termed L0, L1, L2/L3 and
L4 \citep{abb+18}.  Briefly, L0 is effectively the spatial
correlation, beamforming, and up-channelization stage described above.  L1 is
the primary search workhorse, with initial cleaning of RFI (of either
anthropogenic or solar origin), followed by a highly
optimized tree-style dedispersion, spectral-weighting, and peak-search algorithm (called ``{\tt
bonsai}'').  L1 is executed on a dedicated cluster of 128 CPU-based nodes
located in a third, custom-built shipping container adjacent to the telescope.
L1 nodes are constantly buffering intensity data which can be saved upon
detection of a candidate FRB event.  L2/L3 combines results from all beams and
groups detections to identify likely unique events, as well as further reject RFI,
identify known sources, and verify a source's extragalactic nature via its DM
combined with the NE2001 \citep{cl02} and YMW16 \citep{ymw17} models of the maximum Milky Way DM.  Metadata
headers (containing detection beam locations, initial DM, pulse width and
signal strength estimates) for FRB candidates are stored in the L4 database, and raw intensity data buffered by L1 are saved to disk for offline analysis.

A significant change to the pipeline as described in \citet{abb+18}
was made to the event classification in L2/L3 in February 2019, where a high signal-to-noise ratio (SNR) classification bypass
was added in response to a number of very high SNR events being misclassified as
RFI. After the change, any event with $\SNR>100$ bypasses automatic
classification and proceeds directly to human classification.

Note that as of December 2018, the CHIME/FRB realtime pipeline also triggers the
recording of raw telescope baseband from CHIME's 1024 antennas
(which is buffered in memory $\sim$34\,s on the
X-engine) upon detection of a bright FRB
candidate.  A description of the baseband portion of the CHIME/FRB
instrument as well as its analysis pipeline is presented elsewhere
(\citealt{mmm+21}; Mckinven et al. 2021, submitted),
and
the analysis and results of the baseband data for the 153 events in Catalog~1 for which they were captured will be presented in a future work.

\subsection{Beam model}
\label{sec:beam_model}

Throughout our analysis, including when evaluating the exposure, sensitivity,
and when injecting synthetic events, we rely on a model of the CHIME/FRB's
beam, including the primary beam of an individual antenna element, and the
interferometric synthesized beams formed digitally in the X-engine.

The CHIME primary beam, owing to its N-S oriented cylindrical
reflectors, is narrow in the E-W direction and wide in the N-S direction.
The primary beam is the response of a single feed over the
cylinder, modulated by reflections off the ground plane and interactions
between neighbouring feeds. These interactions impart characteristic
spatial and spectral features to the primary beam.

\begin{figure}
    \centering
    \includegraphics[width=0.95\textwidth]{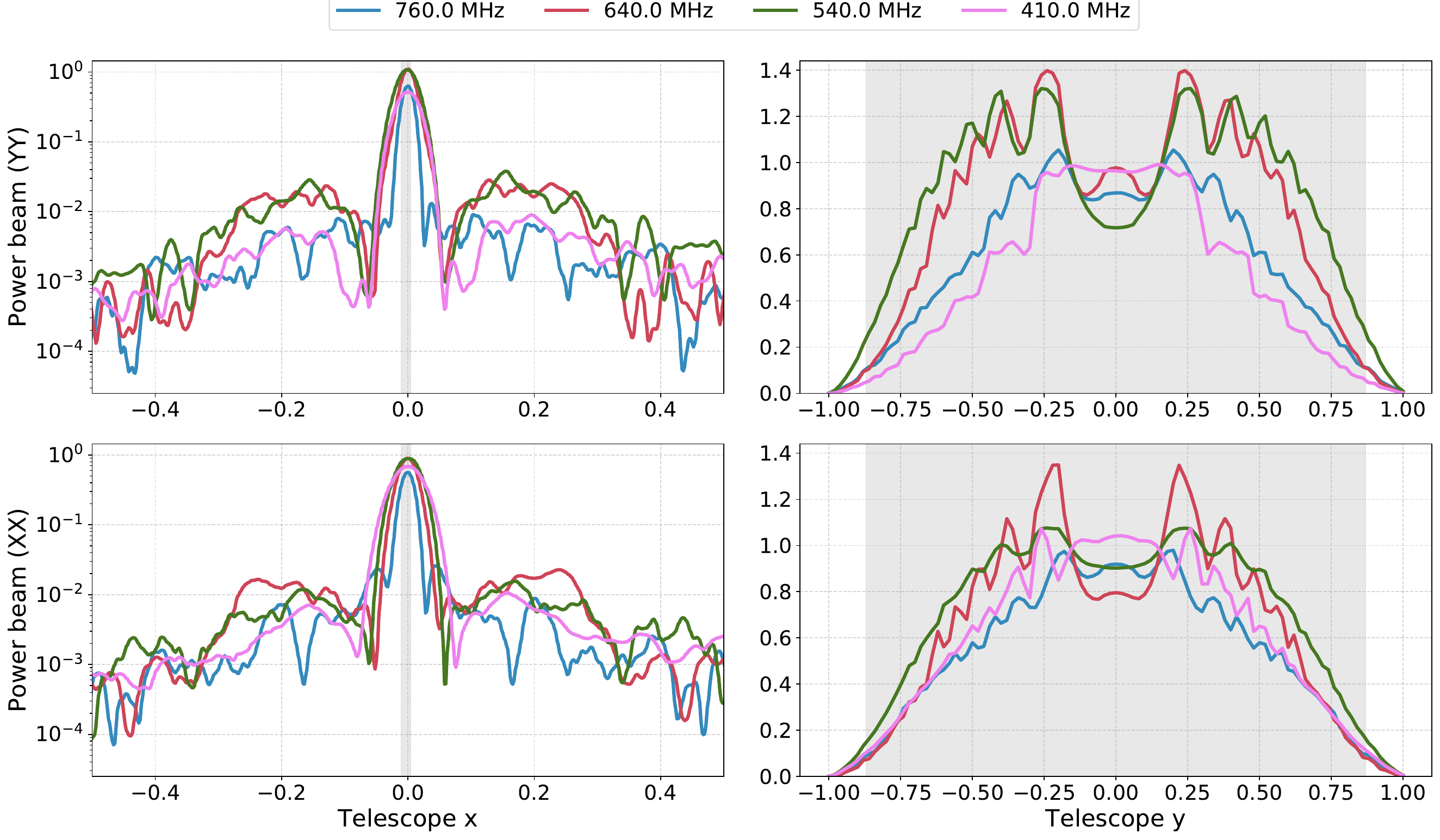}
    \caption{Slices through the primary beam model at four frequencies for each antenna
    polarization. Left panels show an East--West slice at the declination of
    Taurus A ($y=-0.46$). Right panels show a North--South slice along the
    meridian ($x=0$). Shaded regions represent the extent of the primary beam used to generate synthesized beams.    The Cartesian coordinates $x$ and
    $y$ are unitless, specifying sky locations on the unit sphere, as described
    in the text. The beam response is shown in beam-model units where the
    meridian response to Cygnus A is unity at all frequencies.
    }
    \label{fig:beam_slices}
\end{figure}

\begin{figure}
    \centering
    \includegraphics[width=0.95\textwidth]{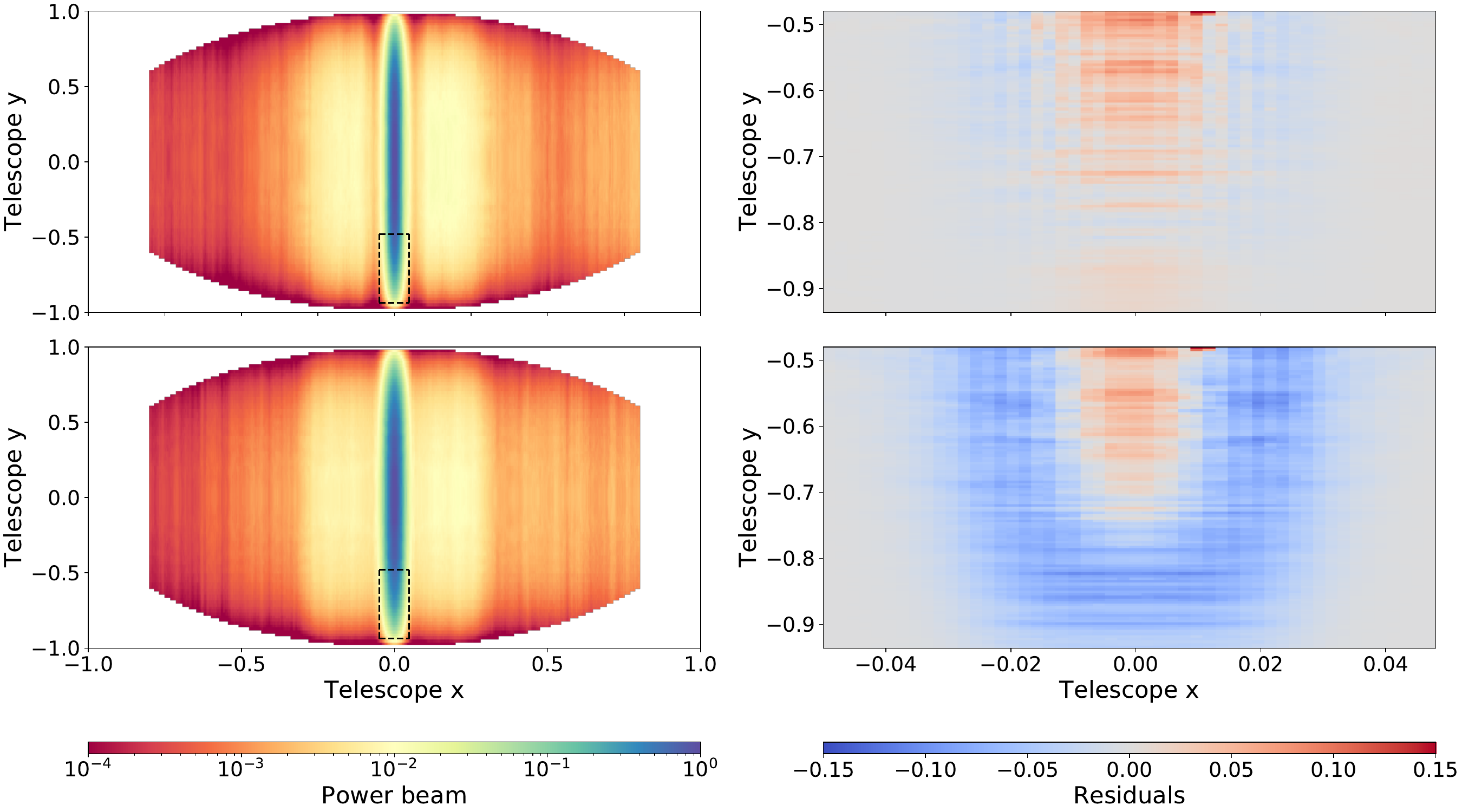}
    \caption{Frequency-averaged primary beam models and residuals from solar
    data comparison for each antenna polarization. The top and the bottom
    panels show the power response of Y and X polarizations respectively. Left
    panels shows the frequency median of the primary beam model for each antenna
    polarization. Right panel shows the frequency median of the difference
    between the model and measurements from transits of the Sun. The coordinate
    system and beam-model units are the same as in Figure~\ref{fig:beam_slices}. We take
    advantage of the Sun's seasonal declination change to measure the beam over
    a small area (region enclosed in the dashed line in the left panel) with a
    single source. In the region tested, the band-averaged primary beam model is
    accurate to better than 10\% of its peak response.
    }
    \label{fig:beam_freq_ave}
\end{figure}

A preliminary model is used for the primary beam in the analyses
presented here,
constructed from an outer product of independent estimates of the
E-W and N-S profiles of the beam. We term this our \texttt{v0} beam model,
which will be described in detail in future work. Reduced dimensionality
representations of the beam are shown in Figures~\ref{fig:beam_slices}
and~\ref{fig:beam_freq_ave}.
In brief, we define the beam model in telescope coordinates $x$ and $y$, which are two
Cartesian coordinate components of sky locations on the unit sphere, where the
$z$-axis lies at the telescope zenith and the $y$-axis points to the horizon
in the direction of local
celestial North.
The E-W ($x$) profile is
measured by tracking Taurus~A with the 26-m John A.\ Galt Telescope at DRAO, equipped with a CHIME feed,
as the source transits across the CHIME
primary beam, while correlating the 26-m signal with the signal from each of
the 1024 CHIME feeds \citep[see][]{bla+16}. This yields a high SNR measurement of the
E-W profile along the source track. Since the Galt telescope uses an equatorial
mount, the polarisation angle of the source can be kept fixed with respect to
its feed.  The profile used for the beam model in this work is an average
of multiple observations of Taurus~A, translated across all declinations, and
stacked over all CHIME feeds separately for each polarization and frequency.

The N-S ($y$) profile provides the normalization to the peak beam response at
each declination. This profile is estimated via a model that has been
developed to describe the cross-talk between feeds on the focal line,
referred to as the coupling response. Cross-talk can occur through several
paths: e.g., radiation broadcast by a feed being directly picked up
by nearby
feeds (direct path) or radiation being reflected by the
cylinder and reaching other feeds (1-bounce path). Each of these coupling paths
introduces a delayed copy of the broadcast radiation, and a superposition of
multiple such copies give rise to the coupling response. This model has a
number of free parameters associated with the coupling strength for each path
as a function of spectral frequency. We fit
these parameters
using observations of
37
bright radio point sources at
different declinations. This provides an estimate of the N-S profile
spanning all declinations and frequencies.
By internal convention, the primary beam model is scaled such that Cygnus A
(our most reliable calibrator)
has unit response (1 Jy/Jy) at each spectral frequency
when transiting the meridian. While we have not accounted for variability
in these point sources, we have checked that consistent results are obtained
using data collected six months apart.

The power responses for the two antenna polarization are averaged, meaning
our beam model applies best to unpolarized sources. While FRBs are typically
strongly polarized, they also typically have significant Faraday rotation which rotates
the polarization angle several times over the band. As such, we expect the
unpolarized beam response to be reasonably applicable when averaged over the
band.

To validate this primary beam model, we use
CHIME's intensity mapping data stream to observe the Sun, which allows a range
of declinations to be measured with a single source. We employ baselines
shorter than 10~m while beamforming to avoid resolving out the Sun. The primary
beam measurements were carried out over 2019-20, which is a known period of
solar minimum. Over this span, the daily flux variability of the Sun is
recorded to be $\leq$ 10\% in the CHIME frequency band (Wulf et al. in prep).
The comparisons are shown in
Figure~\ref{fig:beam_freq_ave}. In the region probed, we
find 10\% agreement between the measurements and our model for the primary
beam.
The low declination of the Sun compared to the CHIME
latitude means we probe relatively low elevations, where the beam model is
constrained by few point-source observations. As such, we consider this
comparison to be the most pessimistic case for the beam model performance.

The response of the FFT synthesized beams is precisely known as
they are synthesized digitally
\citep{nvp+17}.
We have measured the antenna-to-antenna phase variations from either
calibration errors or the primary beams to be at or below the 0.01 radian
level. Phase variations at this level have been shown to be negligible
\citep{msn+19}. Antenna-to-antenna amplitude variations are dominated by
variations in the primary beam and exist at the 10\% level. These are expected
to induce percent-level
perturbations to the synthesized beam. Observations of bright sources
such as Cygnus~A and Taurus~A
using the CHIME/FRB backend have been used to evaluate the composite beam
model, and in particular the FFT synthesized beams since the primary beam is
well characterized at these declinations. These observations match
expectations at the few percent level, implying our overall uncertainty is
dominated by the uncertainty in the antenna-mean primary beam.

\subsection{Sky exposure}
\label{sec:expo}

CHIME is a N-S-oriented transit telescope with cylindrical reflectors 
that yield a long $\sim$120$^{\circ}$  N-S primary beam on the sky. 
The telescope operates nominally 24 hours per day.
As such, CHIME/FRB's exposure to the sky is effectively uniform in right ascension, 
but not in declination. Additionally,
during the survey period,
CHIME was not fully operational 100\% of the time; there were occasional
shut-offs for maintenance or software upgrades, or for unexpected occurrences
like sudden power outages.  Even when operational, the nature of CHIME's
infrastructure means portions may be offline.  For example, a temporarily
non-functional GPU node in the X-Engine results a portion of the bandwidth (1 part
in 256, or 64 out of 16384 channels) being unavailable.
A temporarily non-functional CPU node in the FRB cluster results
in eight sky beams not being processed.  To quantify the exposure and
sensitivity of the telescope to FRBs, these effects must be accounted for.  Metrics of
all computing systems relevant to CHIME/FRB are
recorded for this purpose. Metrics for the L1 nodes are recorded whenever an event (astrophysical or RFI) is detected by the real-time pipeline. Maximum temporal separation between events, and thus L1 metrics, when the real-time pipeline is functioning nominally, is of the order of a few minutes. Monitoring of the L2/L3 and L4 stages was manual with the system being checked every few hours. The exposure on the sky for each detected FRB presented here can thus be determined, as can
the exposure for \emph{any} position on the sky.

For the purpose of computing exposure, we consider a sky location as being
detectable if it is within the FWHM region of a synthesized beam at 600 MHz and
the CPU node designated for processing data for that beam is operational. We
evaluate the exposure for daily transits of all sky locations with declination
$\delta > -11^{\circ}$ by querying the recorded system metrics. We exclude
transits observed in the pre-commissioning period (2018 July 25 to 2018 August
27) as the telescope was operating with a different beam configuration, resulting in the sensitivity to a given sky location being significantly different than that for the current configuration.
Additionally, we excise transits during which the system was not operating at
nominal sensitivity. System sensitivity variations arise due to changes in gain
calibration and RFI environment and are characterized by analyzing
distributions of SNR values for CHIME/FRB detections of Galactic radio pulsars,
as described in detail by \citet{jcf+19} and \citet{fab+20}. A transit is
excised if it occurs on a sidereal day for which the RMS noise derived from
pulsar detections exceeds the mean RMS noise in the period used for the
exposure calculation by more than 1$\sigma$.  On average, 7\% of all
transits were excised for each sky location. This is lower than the expected
excision fraction for a one-sided 1$\sigma$ cut (16\%) as the distribution
of daily RMS noise values is not perfectly Gaussian.
23 FRB events from excised periods are not included in population distributions and analyses, as their selection function and rate statistics cannot be well-characterized.

\begin{figure}
    \centering
    \includegraphics[width=0.8\textwidth]{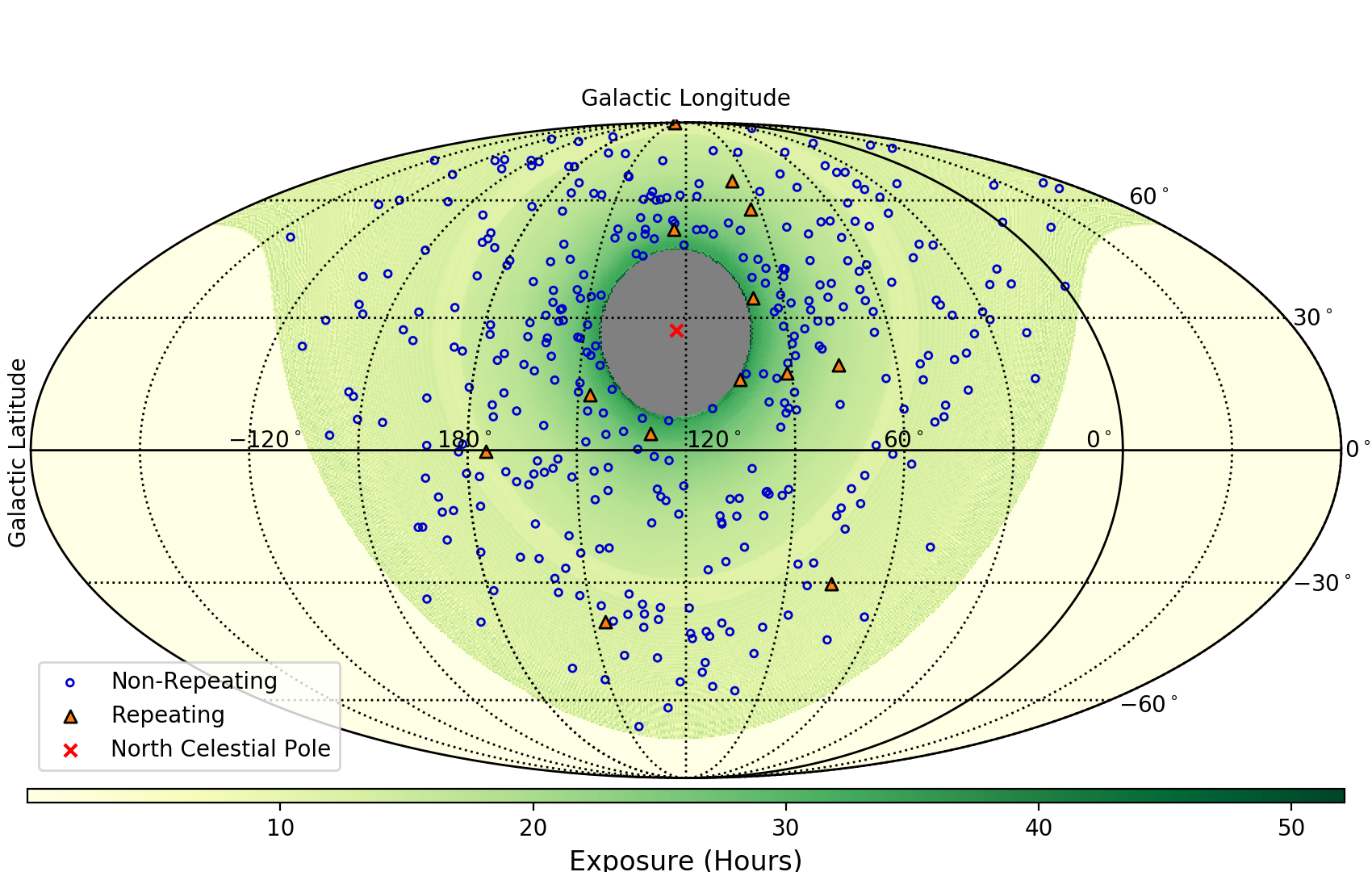}
    \includegraphics[width=0.95\textwidth]{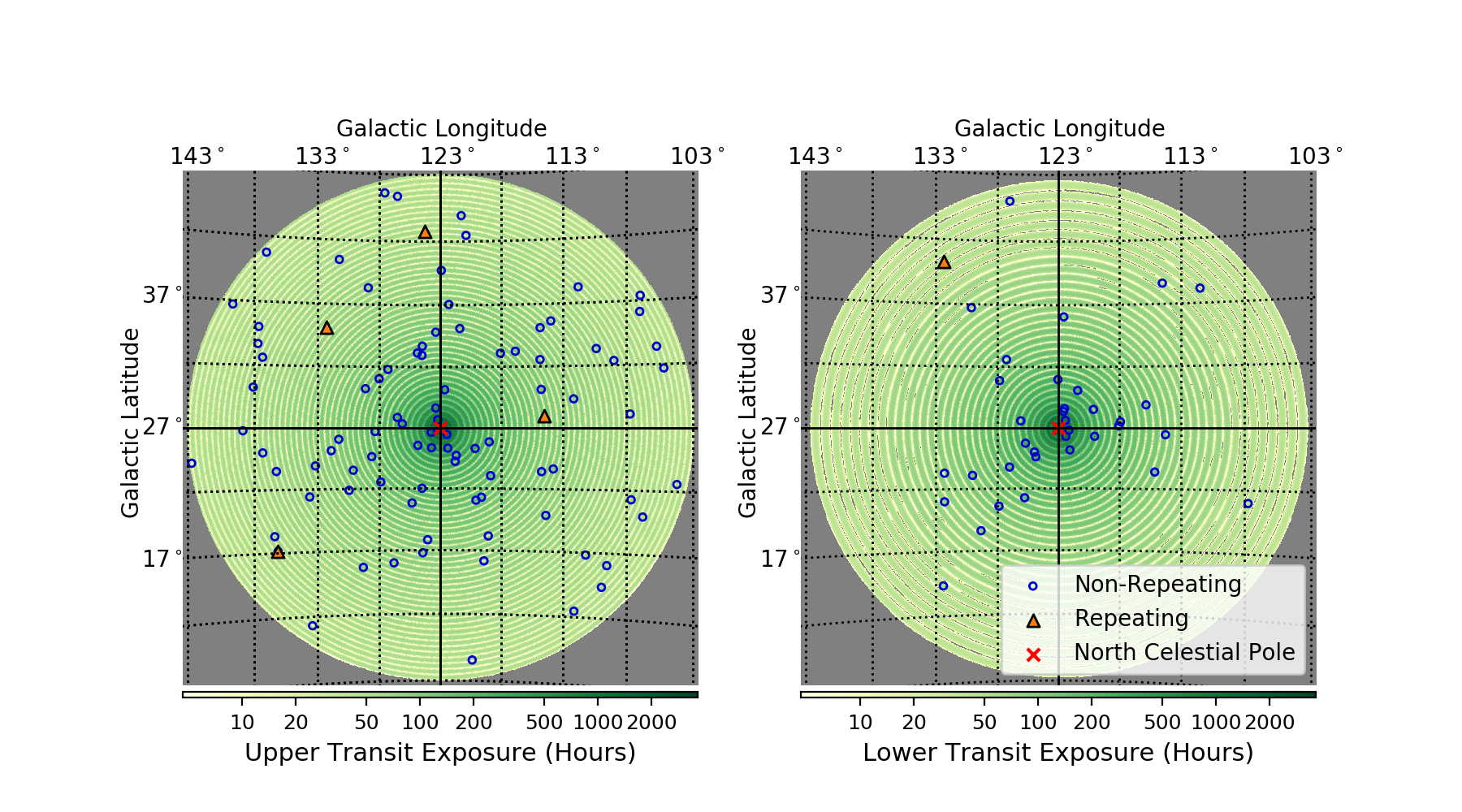}
    \caption{Sky maps in Galactic coordinates with locations of all repeating
    and apparent non-repeating FRB sources presented in this work
    overlaid on the total exposure of the CHIME/FRB system in the period from
    2018 August 28 to 2019 July 1. The top panel shows sky locations that
    transit across the primary beam of the telescope once per day ($\delta <
    70^\circ$) while the bottom panels show upper and lower transit exposures
    for locations which transit across the primary beam of the telescope twice
    per day ($\delta > 70^\circ$). Maps in the bottom panel are centered on the
    North Celestial Pole and have a logarithmic color scale. Despite comparable
    exposure for the two transits, there are fewer FRB detections in the lower transit
    due to reduced sensitivity of the primary beam as compared to the upper transit. The concentric circular patterns arise due to regions between synthesized beams having zero exposure.
    }
    \label{fig:exposure_map}
\end{figure}

\begin{figure}
    \centering
    \includegraphics[width=\textwidth]{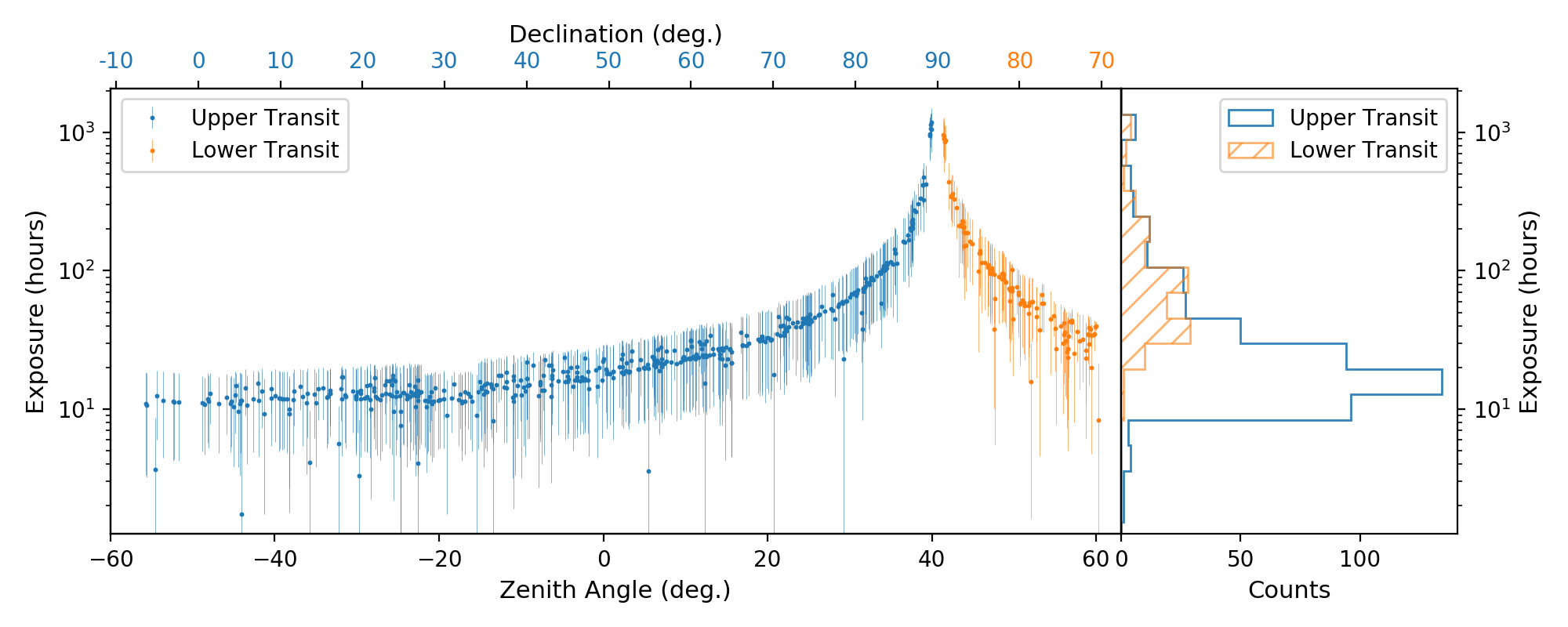}
    \caption{CHIME/FRB's exposure for each of the sources presented in this work for upper and lower transits (if observable) plotted as a function of declination ($\delta$) and zenith angle. Errors on the exposure are due to uncertainties in source declinations (see Section~\ref{sec:expo}). The reduced mean exposure for sources with declinations between $27^\circ$ and  $34^\circ$ is due to a time-limited failure of one of the four CPU nodes (see Section~\ref{sec:obs}) designated to process data for this declination range. A histogram of estimated exposure times for FRBs detected in the upper and lower transits are plotted in the right panel.}
    \label{fig:exposure_hist}
\end{figure}

An all-sky map of the total exposure is shown in Figure~\ref{fig:exposure_map}
with the circumpolar sky locations ($\delta > 70^\circ$) having the two
transits, upper and lower, plotted separately. We do not combine the exposure
for both transits as the primary and synthesized beam response varies
significantly between the two. The aforementioned sky map is then used to
compute the exposure for all detected FRBs. For each source, we calculate the
weighted average and standard deviation of the exposure over a uniform grid of
positions within its 90\% confidence localization region with the weights equal
to the sky-position probability maps (see Section~\ref{sec:loc}). The exposures for
all sources with the corresponding uncertainties are provided in Catalog~1
and shown in Figure~\ref{fig:exposure_hist}. 

The uncertainties in the exposure calculation are due to corresponding source
declination uncertainties as synthesized beam widths vary significantly with
declination. Therefore, we do not report any uncertainties on the exposures for
FRBs that have been localized with sub-arcsecond precision, FRB 20121102A and FRB 20180916B\footnote{Formerly known as FRB 121102 and
180916.J0158+65, respectively, prior to the establishment of the TNS naming convention
(see Section~\ref{sec:tns})}. \citep{clw+17,mnh+20}. We note that some sources have
exposures lower than the average value for their declination range (see
Figure~\ref{fig:exposure_hist}). This is due to a significant fraction of their
positional uncertainty region being located between the FWHM regions of two
synthesized beams (see details of localization in Section~\ref{sec:loc}).

\subsection{Sensitivity threshold}
\label{sec:sens}

Exposure on the sky is distinct from sensitivity---two beams on the sky that have equal exposure may not be equally sensitive.  Using recorded system metrics along with knowledge of the shapes of the primary beam and the formed beams, we have determined for each detected FRB in our catalog a sensitivity threshold for detection of FRBs.

We follow the fluence threshold methods detailed by \citet{jcf+19} and
\citet{abb+19c}. To estimate a sensitivity threshold across the quoted
exposure, we account for three sources of sensitivity variation by generating a
large number of detection scenarios in a Monte Carlo simulation. Day-to-day
variation is captured with detections of known pulsars; variation as the source
transits through the formed beams is computed using the beam model; and spectral sensitivity variation is estimated by combining simulated spectral profiles with the bandpass, which is obtained for each burst during the fluence measurement process, where steady-source transits provide a mapping between beamformer units and Janskys.
\citet{jcf+19} and \citet{abb+19c} used Gaussian profiles for the simulated spectra and drew the defining parameters uniformly around the fitted parameters of the reference burst.
In this work, we sample spectral parameters according to a Gaussian kernel
density estimation of the fitted parameters from all catalog bursts. After
assigning a date, position along transit, and spectrum, each simulated
detection scenario leads to a sensitivity scale factor, relative to the
observing conditions of the reference burst. The scale factors are then applied
to the fluence threshold inferred from the measured fluence and detection SNR, resulting in a distribution of fluence thresholds. We then associate a completeness confidence interval to the corresponding percentile of the distribution.
Completeness at the 95\% confidence interval is reported in Catalog~1 for each source.  For sources with $\delta > 70^\circ$, we simulate fluence threshold distributions for the upper and lower transit separately.  The median 95\% completeness across all bursts is approximately 5 Jy ms.

\section{CHIME/FRB Catalog 1}
\label{sec:catalog}

In this section, we present Catalog~1, including
for each event, the event name, arrival time, sky location, DM, pulse width,
scattering time, spectral parameters, and various measures of signal strength.
In Table~\ref{ta:catalog}, we provide a description of each field from the
Catalog. The Catalog itself is available in machine readable format
accompanying the online version of this article. It contains entries for each event or, in the case
of complex-morphology
bursts, each subcomponent of the event. A short excerpt from the Catalog can be
found in Appendix~\ref{app:cat_excerpt}.

During the period considered for Catalog~1, there were 28 occurrences where
a trigger from the real-time system fit all criteria for an FRB but, due to
a malfunction of the system, intensity data were not saved. There is no way to
determine whether these events would have been classified as true FRBs upon
human inspection.

Next, we discuss how we determine the values for each catalog field.

\begin{ThreePartTable}
\begin{TableNotes}
  \item{\textbf{Notes.} The data for Catalog~1\ in machine-readable format can
      be found accompanying the online version of this article as well as via
      the CHIME/FRB Public Webpage at \url{https://www.chime-frb.ca/catalog}. A small excerpt
    can be found in Appendix~\ref{app:cat_excerpt}.}
  \item[a]{All statistically significant \fitburst{} parameters (i.e., with parameter value $v$ and uncertainty $\sigma$ such that $v/\sigma > 3$) have their best-fit value and $1\sigma$ uncertainty reported; for marginal estimates, we report the $2\sigma$ upper limit obtained from \fitburst.}
\end{TableNotes} 
\begin{longtable}[l]{c l l l}
\bottomrule
\insertTableNotes
\endlastfoot
\caption{Description of CHIME/FRB Catalog~1 Data Fields} \endfirsthead
\caption{\textit{continued}} \endhead
\hline
Column Number & Unit & Column Name & Description
\\
\hline
    0 & ... & \texttt{tns\_name} & TNS name \\
    1 & ... & \texttt{previous\_name} & Previous name (if applicable) \\
    2 &... & \texttt{repeater\_name} & Associated repeater name (if applicable) \\
    3 & degrees & \texttt{ra} & Right ascension (J2000) \\
    4 & degrees & \texttt{ra\_err} & Right ascension error (see Section~\ref{sec:loc}) \\
    5 & ...& \texttt{ra\_notes} & Notes on right ascension \\
    6 & degrees & \texttt{dec} & Declination (J2000) \\
    7 & degrees & \texttt{dec\_err} & Declination error (see Section~\ref{sec:loc}) \\
    8 & ... & \texttt{dec\_notes} & Notes on declination \\
    9 & degrees & \texttt{gl} & Galactic longitude  \\
    10 & degrees & \texttt{gb} &  Galactic latitude \\ 
    11 & hour& \texttt{exp\_up} & Exposure for upper transit of the source  \\
    12 & hour& \texttt{exp\_up\_err} & Exposure error for upper transit of the source  \\
    13 & ... & \texttt{exp\_up\_notes} & Notes on exposure for upper transit of the source  \\
    14 & hour& \texttt{exp\_low} & Exposure for lower transit of the source \\
    15 & hour& \texttt{exp\_low\_err} & Exposure error for lower transit of the source \\
    16 & ... & \texttt{exp\_low\_notes} & Notes on exposure for lower transit of the source  \\
    17 & ...& \texttt{bonsai\_snr} & Detection SNR  \\
    18 & pc cm$^{-3}$ & \texttt{bonsai\_dm} & Detection DM \\
    19 & Jy ms  & \texttt{low\_ft\_68} & Lower limit fluence threshold (68$\%$ confidence) \\
    20 & Jy ms  & \texttt{up\_ft\_68} & Upper limit fluence threshold (68$\%$ confidence) \\
    21 & Jy ms  & \texttt{low\_ft\_95} & Lower limit fluence threshold (95$\%$ confidence)\\
    22 & Jy ms  & \texttt{up\_ft\_95} & Upper limit fluence threshold (95$\%$ confidence)\\
    23 & ... & \texttt{snr\_fitb} & SNR determined using the fitting algorithm \fitburst \\
    24 & pc cm$^{-3}$ & \texttt{dm\_fitb} & DM determined using the fitting algorithm \fitburst\tnote{a} \\
    25 & pc cm$^{-3}$ & \texttt{dm\_fitb\_err} & DM error determined using the fitting algorithm \fitburst\tnote{a} \\
    26 & pc cm$^{-3}$ & \texttt{dm\_exc\_ne2001} & \multirow{2}{*}{\parbox{3.8in}{DM excess between DM determined by \fitburst \ and NE2001 assuming the best-fit sky position of the source}} \\
     & & &\\
    27 & pc cm$^{-3}$ & \texttt{dm\_exc\_ymw16} & \multirow{2}{*}{\parbox{3.8in}{DM excess between DM determined by \fitburst \ and YMW16 assuming the best-fit sky position of the source}} \\
     & & &\\
    28 & s & \texttt{bc\_width} & Boxcar width of the pulse \\
    29 & s & \texttt{scat\_time} & Scattering time at 600\,MHz\tnote{a} \\
    30 & s & \texttt{scat\_time\_err} & Scattering time error\tnote{a} \\
    31 & Jy & \texttt{flux} & Peak flux of the band-average profile (lower limit)  \\
    32 & Jy & \texttt{flux\_err} & Flux error \\
    33 & ... & \texttt{flux\_notes} & Notes on the burst flux \\
    34 & Jy ms & \texttt{fluence} & Fluence (lower limit) \\
    35 & Jy ms & \texttt{fluence\_err} & Fluence error \\
    36 & ... & \texttt{fluence\_notes} & Notes on the burst fluence \\
    37 & ... & \texttt{sub\_num} & \multirow{3}{*}{\parbox{3.8in}{Sub-burst number (if applicable). If the FRB has only one burst, then the sub-burst number is 0. Sub-bursts listed in chronological order. }}  \\
     & & &\\
     & & &\\
    38 & MJD & \texttt{mjd\_400} & \multirow{2}{*}{\parbox{3.8in}{Time of arrival with reference to 400.1953125 MHz for the specific sub-burst. }} \\
    & & &\\
    39 & MJD & \texttt{mjd\_400\_err} & \multirow{2}{*}{\parbox{3.8in}{Time of arrival error with reference to 400.1953125 MHz for the specific sub-burst. }} \\
    & & &\\
    40 & MJD & \texttt{mjd\_inf} & \multirow{2}{*}{\parbox{3.8in}{Time of arrival with reference to infinite frequency for the specific sub-burst. }} \\
    & & &\\
    41 & MJD & \texttt{mjd\_inf\_err} & \multirow{2}{*}{\parbox{3.8in}{Time of arrival error with reference to infinite frequency for the specific sub-burst. }} \\
    & & &\\
    42 & s & \texttt{width\_fitb} & Width of sub-burst using \fitburst \\
    43 & s & \texttt{width\_fitb\_err} & Width error of sub-burst using \fitburst \\
    44 &... & \texttt{sp\_idx} & Spectral index for the sub-burst \\
    45 & ...& \texttt{sp\_idx\_err} & Spectral index error for the sub-burst \\
    46 &... & \texttt{sp\_run} & Spectral running for the sub-burst \\
    47 &...& \texttt{sp\_run\_err} & Spectral running error for the sub-burst \\
    48 & MHz & \texttt{high\_freq} &\multirow{2}{*}{\parbox{3.8in}{Highest frequency band of detection for the sub-burst at FWTM}} \\
    & & &\\
    49 & MHz & \texttt{low\_freq} & \multirow{2}{*}{\parbox{3.8in}{Lowest frequency band of detection for the sub-burst at FWTM}} \\
    & & &\\
    50 & MHz & \texttt{peak\_freq} & Peak frequency for the sub-burst \\
    51 & ... & \texttt{chi\_sq} & $\chi^2$ from \fitburst \\
    52 & ... & \texttt{dof} & Number of degrees of freedom in \fitburst  \\
    53 & ... & \texttt{flag\_frac} & Fraction of spectral channels flagged in \fitburst \\
    54 & ... & \texttt{excluded\_flag} & \multirow{3}{*}{\parbox{3.8in}{Flag for events excluded from parameter inference due to non-nominal telescope operation (1 = excluded, 0 = included).}} \\
     & & &\\
\label{ta:catalog}
\end{longtable}
\end{ThreePartTable}

\subsection{Event naming: transient name server}
\label{sec:tns}

Each of our detected FRBs has been assigned a name provided by
the Transient Name Server\footnote{\url{https://www.wis-tns.org.}} (TNS),
the official International Astronomical Union (IAU) mechanism for reporting new astronomical transients.
TNS names have format FRB~YYYYMMDDx  where YYYY is a 4-digit year,
MM is a 2-digit month code, DD is a 2-digit day (all in UTC), and x is a
string of 1 to 3 Latin letters, beginning with ``A'' for the first
source reported to the TNS for the relevant UTC day, ``Z'' for the 26th, and in lowercase letters after this i.e.
``aa'' for the 27th, and so forth, up to and including ``zzz'', for a total of 18278 possible unique FRBs reported on a given UTC day. The TNS functions as more than a name server, and in fact hosts basic data for all submitted FRBs. For Catalog~1 the hosted data are derived from the real-time CHIME/FRB detection pipeline. Chief among the hosted data are the DM, SNR, dispersion-corrected arrival time at $400~\text{MHz}$, and sky-position estimates with localization contours that can be downloaded in a machine-readable format.  For previously published CHIME/FRB-detected events that
are also in Catalog~1, we provide the previously published name (that followed
an \textit{ad hoc} and now outdated naming scheme) in the Catalog as well for reference, but we recommend henceforth referring exclusively to the TNS name.  To aid the community in acquiring TNS names for their FRBs (both those already and yet-to-be detected), we provide in Appendix~\ref{app:tns} instructions for doing so.

 \subsection{Event localization}
\label{sec:loc}

\begin{figure}[t]
	\centering
	\includegraphics[width=0.95\textwidth]{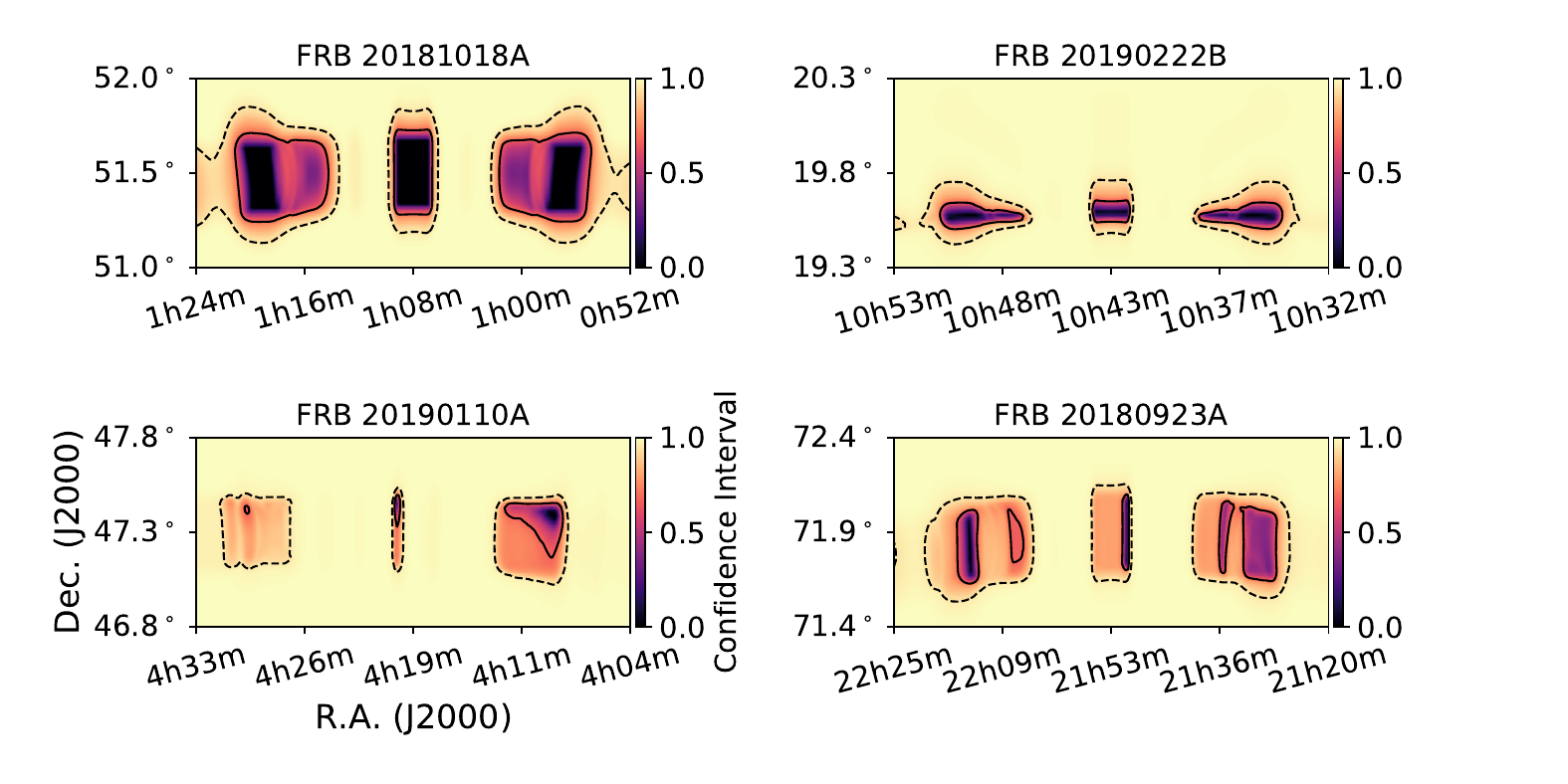}
    \caption{Example localization confidence interval
    plots for four
    different detection patterns. Clockwise from top left are: single beam,
    two beams N-S, two beams E-W, and four beams in a square. In each example, the frame spans 5$^{\circ}$
    in Right Ascension (scaled by cos(Dec.)), 1$^{\circ}$ in Declination, and
    is centered at the beam with the strongest detection. Localization is
    performed as a grid-search $\chi^2$ minimization, where confidence
    intervals are obtained from contours of constant $\Delta\chi^2$. The color
    scale encodes these intervals, such that the area enclosed by a given color
    defines the corresponding confidence interval. 
    The 68\% and 95\% intervals are shown with solid and dashed contours, respectively.
    Note that the common three-region pattern reflects the chromatically
    smeared side lobes of the formed beams.
    \arxivonly{Panels for all catalog bursts can be found at
    \url{https://www.canfar.net/storage/list/AstroDataCitationDOI/CISTI.CANFAR/21.0007/data/localizations/plots}.}
    }
    \label{fig:loc}
\end{figure}

We provide sky localizations for each of our events, determined via the header
metadata determined in real-time by L1 and stored in L4.  These
localizations are presented in Catalog~1 as their central coordinates and approximate uncertainties, with actual localization error regions presented as plots
in Figure~\ref{fig:loc}.

We follow the same localization method detailed in \citet{abb+19c}. Ratios
among per-beam SNR values are fit using least squares to 
beam model predictions for a grid of
model sky locations and model intrinsic spectra.
The mapping between $\Delta \chi^2$
and confidence interval is constructed from an ensemble of pulsar events
identified by the real-time system, such that true positions fall within
contours of a given confidence interval the appropriate fraction of the time.
While this uncertainty treatment is most appropriate for pulsar-like spectra,
we note that the true positions of the two localized repeaters (including
19 bursts from FRB20180916B observed over a range of hour angles), both emitting
band-limited and morphologically complex bursts, are contained in the
uncertainty regions of their respective CHIME/FRB SNR-based localizations. In
the E-W direction, the grid of model locations is chosen to contain the
main lobe of the primary beam. This span includes the first-order side lobes of
the formed beams, which leads to the disjointed uncertainty regions seen in
Figure~\ref{fig:loc}. Where tabulated, we report the extent of the 68\%
confidence interval closest to the beam with the strongest detection. The
disjointed contours, which include the near side lobes, can be found on the TNS
for a variety of common confidence intervals.

\subsection{Event morphologies}
\label{sec:properties}

The initial determination of DM provided by \bonsai{} in the L1 real-time
detection pipeline is only approximate due to the limited resolution with which
it is reported \citep[see][]{abb+18}.  For this reason, we
used the called-back, total-intensity data saved from our L1 buffers to
determine an improved DM via maximization of the SNR of
the burst using offline algorithms that also provide a determination of burst
time of arrival ($t_{\rm arr}$) prior to downstream model fitting. However,
even the SNR optimizing DM can be significantly biased due to chromatic
pulse broadening (DM-smearing and scattering) or chromatic complex burst
morphology.

The SNR-optimized DM and $t_{\rm arr}$ estimates were then provided as initial
guesses to a least-squares fitting routine, \fitburst{}\footnote{The
\fitburst{} code has not yet been made public, but the underlying model and
likelihood are the same as that used by \citet{mls+15a}, whose code is public.}, that directly models the two-dimensional dynamic spectra in terms of fundamental burst parameters. For a single burst, the parameters modeled by \fitburst{} are: DM; $t_{\rm arr}$, signal amplitude ($A$), temporal width ($w$), power-law spectral index ($\gamma$) and ``running" of the spectral index ($r$), and a timescale for multi-path scattering of the FRB signal \citep[$\tau$; e.g.,][]{mck14}. The composite model for a scattered, single-component dynamic spectrum with label $i$ ($S_i$) is defined as $S_i = A_i \times F_i \times T_i$, where: $A_i$ is the overall amplitude of the $i^{\rm th}$ burst component; $F_i \equiv F_i(\gamma_i, r_i)$ is a term that defines the time-independent spectral energy distribution as a function of frequency ($f$), relative to an arbitrary reference value ($f_0$):

\begin{equation}
     F_i(\gamma_i, r_i) = (f/f_0)^{-\gamma_i + r_i\ln{(f/f_0)}};
\end{equation}

and $T_i \equiv T_i({\rm DM}, t_{{\rm arr},i}, w_i, \tau)$ is a term that
models the temporal shape of the burst:

\begin{equation}
     T_i({\rm DM}, t_{{\rm arr},i}, w_i, \tau) = \frac{1}{2\tau}\exp{\bigg(\frac{w_i^2}{2\tau^2}\bigg)}\exp{\bigg[-\frac{(t({\rm DM}) - t_{{\rm arr}, i})}{\tau}\bigg]}\bigg\{1 + {\rm erf}\bigg[\frac{t({\rm DM}) - (t_{{\rm arr}, i} + w_i^2/\tau)}{w_i\sqrt{2}}\bigg]\bigg\}.
     \label{eq:fitburstT}
\end{equation}

The form of $T_i$ shown in Equation \ref{eq:fitburstT} is taken directly from \citet{mck14}, which represents the convolution between a Gaussian profile and a time-dependent exponential function, the latter function with characteristic decay timescale $\tau$ and truncated at $t = t_{{\rm arr},i}$ by a Heaviside function.

Using the above definitions, we modeled a multi-component burst as $S=
\sum_i^n S_i = \sum_i^n A_i \times F_i \times T_i$, where $n$ is the number of
distinct sub-bursts in the observed dynamic spectrum. We set $f_0 =$
400.1953125\,MHz which is the center of our lowest frequency channel, in order
to be consistent with L1 configuration settings. Both DM and $\tau$ are
considered to be ``global" parameters, such that all sub-burst components are
assumed to possess the same dispersion and scattering properties, while all
parameters with subscript $i$ indicate component-specific parameters. Moreover,
we assumed that $t({\rm DM}) = k{\rm DM}(f^{-2} - f_0^{-2})$, where $k =
(2.41\times10^{-4})^{-1}$\,s\,pc$^{-1}$\,cm$^3$\,MHz$^2$ (consistent with
physical expectations for dispersion in a cold plasma), and that $\tau \propto
f^{-4}$ \citep{lan71, lk05}, where we use 600\,MHz as the scattering reference
frequency.

For a given CHIME/FRB event with $n$ sub-bursts, we fitted for (2 + 5$n$)
parameters with {\tt fitburst} through $\chi^2$ minimization between the
$n$-component model and full-resolution L1 data. We accounted for intra-channel
dispersion smearing during each fit iteration by evaluating the model spectrum
$S$ at 8 and 4 times the data resolution in time and frequency, respectively,
and subsequently downsampling to the data resolution. Moreover, all CHIME/FRB
raw data were processed for automatic excision of narrowband RFI and noise-baseline subtraction prior to model fitting, though we did not explicitly calibrate the CHIME bandpass.

We generated two models with {\tt fitburst} for each CHIME/FRB event and
compared best-fit statistics in order to determine the significance of
multi-path scattering in spectra. One model was generated while simultaneously
fitting for all parameters discussed above, including $\tau$; for these models,
$w$ is interpreted as the width of the intrinsic, pre-scattered burst
component. A second model was generated assuming zero scattering, in which case
the function $T({\rm DM}, t_{{\rm arr},i}, w_i, \tau=0)$ in Equation
\ref{eq:fitburstT} is replaced with a Gaussian function of standard deviation
$w_i$ that reflects the full temporal width of profile component $i$. The
$\chi^2$ values for both models were then compared through an F-test for model
selection, and a $p$-value threshold of 0.1\% was used to declare the
significance of $\tau$. In cases where scattering is not significant, we quote
an upper limit on $\tau$ of $2 \times w$. In cases where the fit of the
width-scattering model is highly degenerate (i.e., when the covariance
matrix after least-squares optimization is singular), we default to the
no-scattering model as the superior description. Simulations have shown that
CHIME/FRB total intensity data can be used to robustly measure values of $w$
and $\tau$ larger than 100~$\mu$s only; for cases where the fitted value is
smaller than this we quote 100~$\mu$s as an upper limit.

The above procedure was performed automatically on each burst. However, manual
intervention was frequently required to adjust the parameter initial guesses
when the least-squares optimizer failed to converge on a satifactory result. In
addition, for bursts visually determined to have a complex morphology, the
value of $n$ was chosen manually.

The fitting procedure described here has a number of limitations, including
that the model may be an imperfect description of the intrinsic burst
morphologies; inhomogeneities in the spectral frequency response from the beam and
non-uniform noise; limited ability of the least-squares optimizer to converge on a
global best fit and represent uncertainties; and the reliance on
human judgement to assess adequate convergence and determine component count
for complex bursts. These limitations are discussed in detail in
Appendix~\ref{app:fitburst}, where we also describe metrics that can be used to
assess the quality of the fits on a burst-by-burst basis. Improvements to this
procedure, including the use of Markov Chain Monte Carlo (MCMC) techniques and
an automatic determination of $n$, are ongoing and will be the
subject of future CHIME/FRB catalogs.

Best-fit parameters from the above modeling procedure are provided in Catalog~1.  Tabulated uncertainties denote the 68\% confidence level unless otherwise specified. Upper limits are denoted with a ``$<$'' symbol and represent 95\% confidence upper limits unless otherwise specified.

We also derive a full-width-tenth-maximum (FWTM) emission bandwidth from
the model fits, capped at the top and bottom of the CHIME band. We measure a
total burst duration in the dedispersed and frequency-averaged time series.
Each time series is convolved with boxcar kernels with durations equal to
integer multiples of the sampling time up to 128 samples (although the search
range was manually tweaked in a few cases) and
normalized by the square-root of their respective widths. The burst duration is
defined as the width of the boxcar that results in the highest peak SNR after convolution. FRBs 20181019A, 20181104C, 20181222E, 20181224E, 20181226B, 20190131D, 20190213B and 20190411C have two distinct peaks in their time series (without a ``bridge'' in emission) and for those FRBs we report two burst durations.

\begin{figure}
    \centering
    \includegraphics[width=0.90\textwidth]{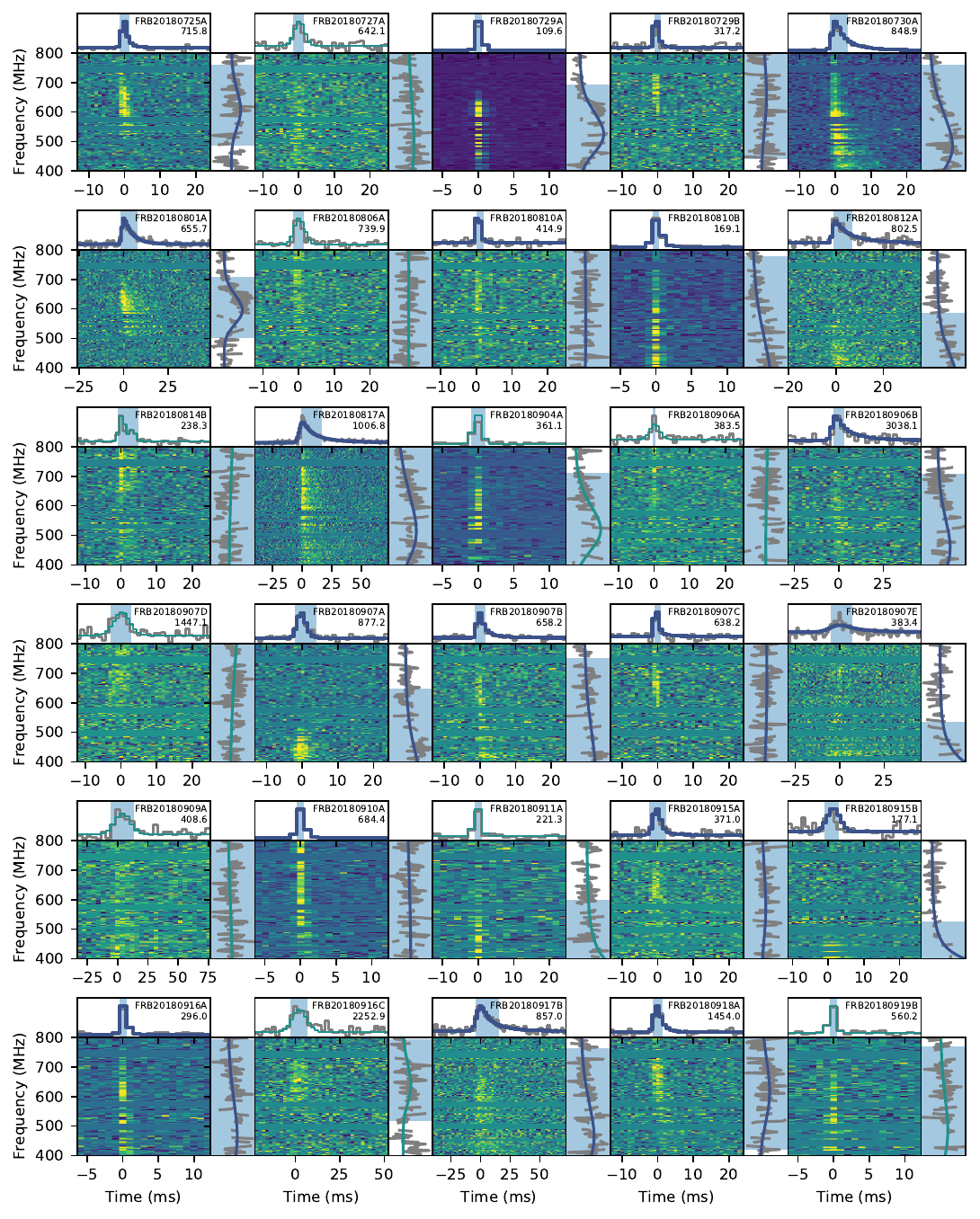}
    \caption{Dynamic spectra (``waterfall plots''), frequency-averaged time
    series and time-averaged spectra for all one-off FRBs in the Catalog,
    ordered by time. The TNS name and best-fit DM in units of pc\,cm$^{-3}$ are in the top right corner
    of each panel. Model fits are overlaid on the time series and spectra, in
    green (thin lines) if scattering was not significant and in blue (thick
    lines) if scattering was
    significant. The blue shaded regions in the time series and spectra indicate the burst durations and emission bandwidths FWTM, respectively. Bursts indicated with an asterisk are published repeaters, for which only one burst was detected before the Catalog cut-off date.
    \arxivonly{Panels for all catalog one-off bursts can be found at
    \url{https://www.canfar.net/storage/list/AstroDataCitationDOI/CISTI.CANFAR/21.0007/data/additional_figures/waterfalls}.}
    }
    \label{fig:waterfall_oneoffs}
\end{figure}

\begin{figure}
    \centering
    \includegraphics[width=0.90\textwidth]{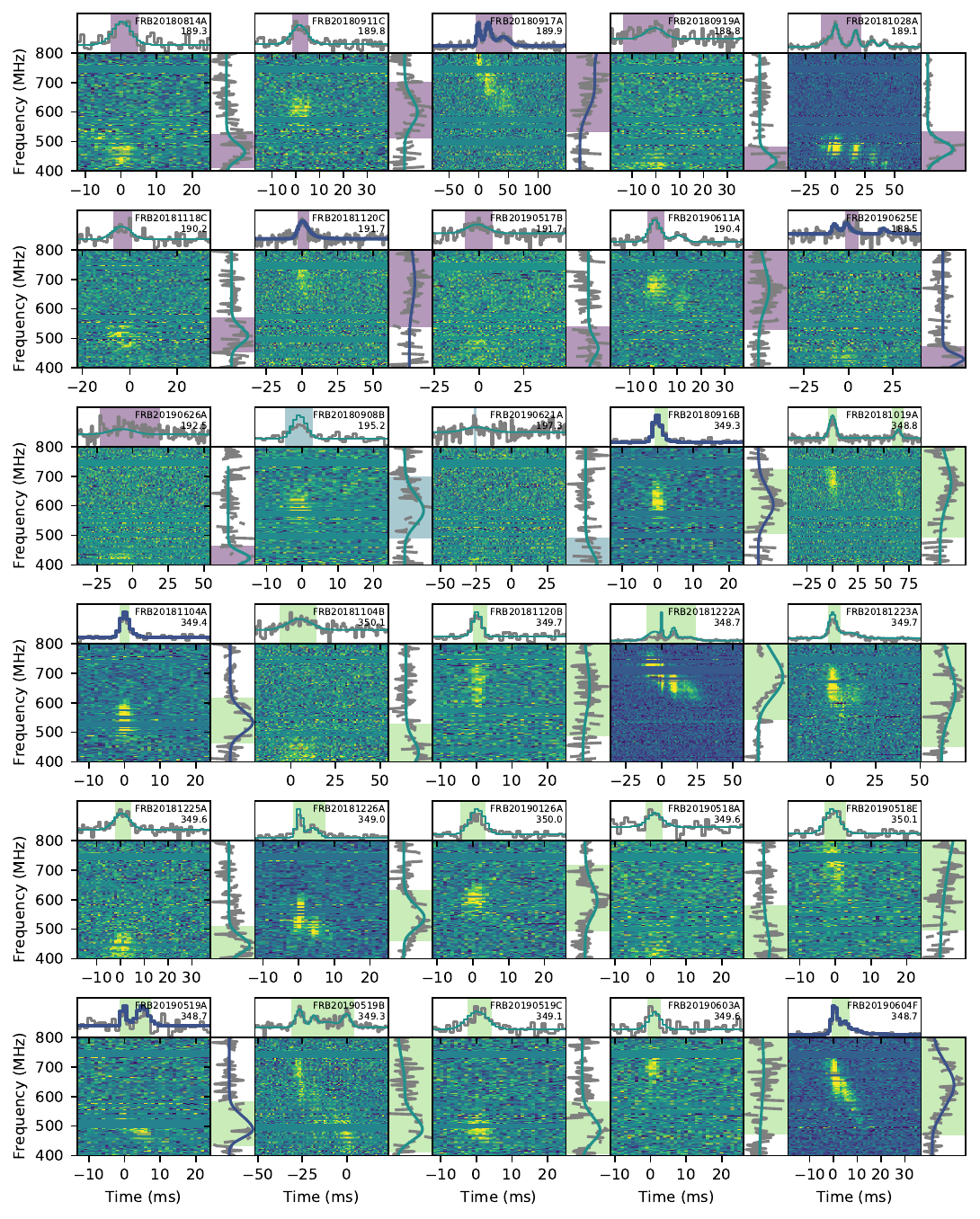}
    \caption{Same as Figure~\ref{fig:waterfall_oneoffs}, but for all 
    sources exhibiting more than one burst in this
    Catalog. Sources are ordered by their first detections, and bursts from any one source are ordered by time of detection. Differently colored shaded regions are used for different repeater sources.
    \arxivonly{Panels for all catalog repeating bursts can be found at
    \url{https://www.canfar.net/storage/list/AstroDataCitationDOI/CISTI.CANFAR/21.0007/data/additional_figures/waterfalls}.}
    }
    \label{fig:waterfall_repeaters}
\end{figure}

Time series depicting each burst, along with its dynamic spectrum (or ``waterfall plot'') and spectrum, with all three dedispersed to the optimal {\tt fitburst}-determined DM, are provided in Figures~\ref{fig:waterfall_oneoffs} and~\ref{fig:waterfall_repeaters}.
In these plots, we have overlaid the frequency-averaged and time-averaged
fitted models on the time series and spectra, respectively. We also show the
burst duration and emission bandwidth FWTM\@. For all FRBs, we show 128 frequency subbands. Time windows are multiples of 12.5 ms, based on the FRBs' width and scattering time scale.

For better visualization, we mask subbands with variance $> 3 \times$ the mean variance, and subbands with time-averaged values $<$ Q1$ - 1.5\times$IQR or $>$ Q3$ + 1.5\times$IQR, where Q1 and Q3 are the first and third quartiles, respectively, and IQR is the interquartile range. The color scales are capped to the 1$^\mathrm{st}$ and 99$^\mathrm{th}$ percentiles.

\subsection{Event signal strength}
\label{sec:fluences}

To characterize signal strength for each event,
we provide the SNR of the initial real-time pipeline detection, along with a fluence and flux determined in offline analyses.  

In  Catalog  1,  our  ability  to  determine  burst  fluences  is  limited  by
the  uncertainty  of  our  burst localization combined with CHIME’s complex and
rapidly varying beam pattern. In particular, the spectral structure of the beam
pattern and overall beam response can change significantly over the extent of the header localization
region obtained for each burst, making it difficult to reliably correct fluence
measurements for beam attenuation. Localization uncertainty, and to
a lesser extent beam model uncertainty, introduces an unknown primary beam
response that is a strong function of a bursts uncertain hour angle (see
Figure~\ref{fig:loc}). As such, we assume that
each burst was detected along the meridian of the primary beam (at the peak
sensitivity of the burst declination arc). Thus, our fluence
measurements are biased low, as bursts off-meridian will experience beam
attenuation that we are not accounting for. Note that the errors on the
fluences, discussed below, do not quantify this bias---the measurements
we provide are most appropriately interpreted as lower limits, with an
uncertainty on the limiting value.
A detailed description of the automated fluence calibration pipeline, including an explanation of current limitations, will be provided elsewhere. 
Here, we summarize the procedure, which is similar to that used in previous CHIME/FRB papers \citep{abb+19a, abb+19b, jcf+19, abb+19c, fab+20}.

Transit observations of steady sources with known spectral properties are used to sample the conversion from CHIME/FRB beamformer units to Janskys as a function of frequency across the primary beam. We pair each burst with the calibration spectrum of the nearest steady source transit, closest first in declination, then in time. We assume N-S beam symmetry, so that sources on both sides of zenith can be used for each event. By applying the calibration spectrum to the total-intensity data for each burst, we derive a dynamic spectrum in physical units roughly corrected for N-S primary beam variations. The fluence is then derived by integrating the burst extent in the band-averaged time series, while the peak flux is the maximum value within the burst extent (at 0.98304 ms resolution). 

The error due to differences in the primary beam between the calibrator and the
assumed FRB location along the meridian is estimated by using steady sources
from a single day to calibrate each other and measuring the average fractional
error compared to known flux values. This contributes a relative error on the
order of 20\% to the flux measurements. The error due to temporal variation in
the calculated beamformer unit to Jansky conversion spectra is determined by
measuring the RMS variation over a period of roughly two weeks surrounding the
burst arrival. This also captures uncertainty due to calibrator source
variability on that time scale, and contributes a relative error on the order of 13\% to the flux measurements, depending on the calibrator used. These two errors are also combined with the RMS of the off-pulse in the band-averaged time series to form the overall errors presented in Catalog~1. We note again that the errors estimated here do not encapsulate the bias due to our assumption that each burst is detected along the meridian of the primary beam, which causes our fluence measurements to be biased low.

During the period from the beginning of the Catalog to February 2019, the flux calibration pipeline was still being commissioned and steady source observations were sparse. We conservatively estimate the time error for bursts detected during this time by taking the fractional RMS variation in the calibration spectrum over the entire period, yielding errors typically on the order of 26\%. An additional error is included in the fluence estimates of the first 13 CHIME/FRB bursts to account for the phase-only complex gain calibration used during the pre-commissioning period when they were detected, as described in \cite{abb+19a}.

A total of 6 bursts were detected directly after a system restart, when we were not able to obtain steady source transits before upstream complex gain calibration was applied. Since we could not measure proper beamformer unit to Jansky scalings during these times, we do not provide fluences or fluxes for these bursts. We also note that early detected bursts previously presented in \cite{abb+19a}, \cite{abb+19b}, and \cite{jcf+19} have been re-analyzed using the automated Catalog~1 pipeline, and their reported fluences have changed significantly due to updates in our RFI mitigation methods.

\section{Synthetic signal injection}
\label{sec:injection}

As for any astrophysical instrument, CHIME/FRB has a transfer function,
introduces selection biases, and adds noise due both to the nature of the
telescope and the software detection pipeline.  These instrument
characteristics need to be carefully characterized so that they can be
accounted for in any population analysis of FRB events and their distributions, as is performed in Section~\ref{sec:abs_pop} below.
We account for these biases through careful measurements of the telescope beam,
calibration, and noise properties, and by probing the selection function
using Monte Carlo techniques with synthetic events injected into the
CHIME/FRB software system.  This strategy mimics the Monte Carlo event
generator techniques used in particle physics, with the exception that real-time telescope noise and the RFI environment are incorporated by injecting the
events \emph{in situ} while the telescope is operating.  The Monte Carlo
injection system was designed to allow synthetic FRBs to be
injected into the real-time pipeline with user-defined properties. Injected
pulses (hereafter ``injections'') are suitably flagged to ensure none are
mistaken as genuine astrophysical signals.
In this way, we measure instrumental biases, and using this knowledge, in Section~\ref{sec:abs_pop} determine actual cosmic FRB property distributions.

\begin{figure}
    \centering
    \includegraphics[width=\textwidth]{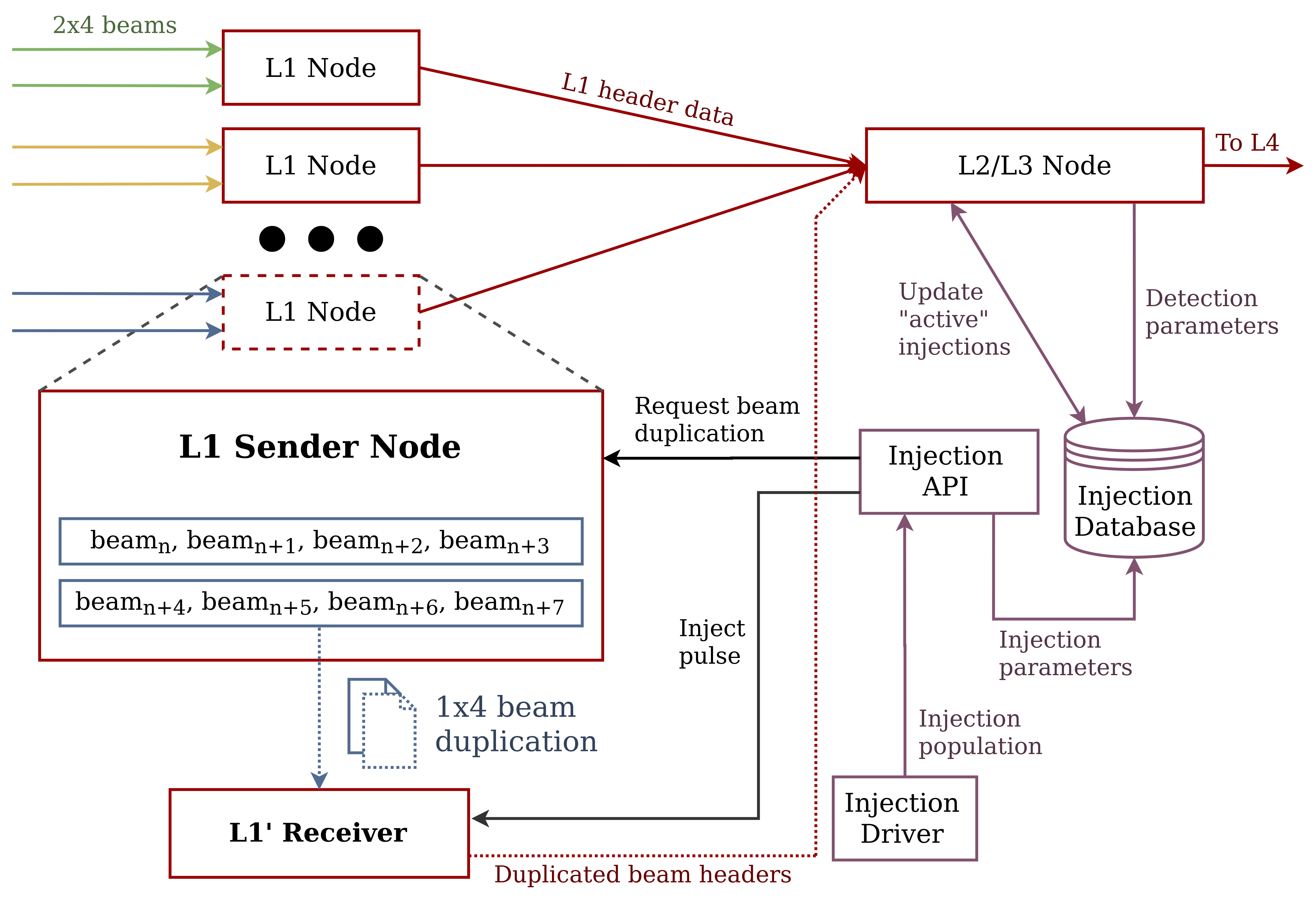}
    \caption{{Schematic of the CHIME/FRB Injection System. Each L1 node handles
    two data streams containing full-resolution intensity data for four beams.
    The injection system interacts with the L1 nodes through
    remote procedure calls. Its capabilities are outlined in
    Section~\ref{sec:signalgen}. The system injects a population of FRBs into
    the on-sky data streams of four duplicated beams at a time, and measures
    the events' detection properties, storing the injection and detection
    parameters in a database. The green, yellow, and blue arrows indicate
    intensity data streamed from sets of four beams. Red boxes and arrows
    indicate components of the real-time pipeline and header data,
    respectively. Purple boxes and arrows indicate components and data flow
    related to the injection system. Black boxes and arrows indicate HTTP request handling interfaces and requests.
    }}
    \label{fig:injection_system}
\end{figure}

The details of the injection system will be described elsewhere.
Here we provide a brief description of the use of the injection system
to quantify our instrumental and software detection pipeline biases.
Figure~\ref{fig:injection_system} shows a schematic drawing of the injection system as it is currently set up in the full CHIME/FRB system (see also Figures 4 and 6 in \citealt{abb+18}).

\subsection{Signal generation}
\label{sec:signalgen}

FRB signals are generated using the internally-developed
\texttt{simpulse}\footnote{\url{https://github.com/kmsmith137/simpulse}}
library. \texttt{simpulse} generates FRBs at the CHIME/FRB frequency channel
width and sampling time with an intrinsic running power-law spectrum,
DM, pulse width, and scattering time. The \texttt{simpulse}
library accounts for intrachannel dispersion smearing and other sampling
effects that would occur at the correlator stage for an astrophysical signal.
The injected FRB signal is multiplied with the complex spectral signature of the CHIME
telescope's primary beam and FFT synthesized beams evaluated at the chosen
position in the sky. 

Signals are scaled to the same absolute flux (Jy) units as the live telescope data stream.
Prior to beamforming, an absolute calibration (derived from daily observations
of bright continuum point-sources through CHIME's visibility data stream) is applied 
to the baseband data in the X-engine. Thus, we generate our simulated signals in
flux units, taking care to also apply factors introduced in the beamforming and
upchannelization process.

\subsubsection{Injections population}

Here we describe how we generate a population of FRBs for injecting into the CHIME/FRB real-time detection pipeline using the system described in 
Section~\ref{sec:injection}.
We start by sampling locations in the sky where we will evaluate our beam model
and place simulated FRBs.
We randomly sample $10^6$ locations uniformly distributed
on the celestial sphere
in telescope coordinates. Of these, we discard all locations: that are below the
horizon; for which the band-averaged primary beam response is less than $10^{-2}$ in beam model units
(see Section~\ref{sec:beam_model}); and for which the
band-averaged response does
not reach $10^{-3}$ in any of the 1024 synthesized beams. As such, we are
not injecting bursts into the far side lobes; however, this does not incur a
bias since such events are cut from the catalog for population inferences.
The fraction of sky locations surviving these cuts is
$f_{\rm sky} = 0.0277$.
Note this ``forward-modeling'' method of accounting for the telescope's beam
response is distinct from the simpler analyses done in other rate estimations;
it is important in our case because of the complex CHIME beam.

We then draw $5\times10^6$ FRBs and randomly assign them to the surviving sky
locations. The properties of these FRBs are drawn from initial probability
density functions $P_{\rm init}(F)$, $P_{\rm init}(\dm)$, $P_{\rm init}(\tau)$, $P_{\rm init}(w)$,
and $P_{\rm init}(\gamma, r)$ designed to both fully sample the range of
observed properties and more densely sample parts of phase space populated by
the catalog. The FRB properties in these distributions are uncorrelated
except for $\gamma$ and $r$. These distributions are described in more detail
in Section~\ref{sec:abs_pop} and Appendix~\ref{app:fid}.
After drawing from the initial distributions, we perform a cut of events that have little chance of being detected based
on the FRB properties, our beam model, and a conservative noise model. This
left 96\,942 events scheduled for injection. Due to overlapping
sensitivity, true FRBs can be detected in multiple beams simultaneously.
The injections system does not currently support multi-beam injections; instead, we inject a given event into the beam with the highest predicted SNR based on the
noise and beam model.

\subsubsection{Injection and detection}
\label{sec:inj_pop}

One of the 128 L1 nodes has been outfitted as a ``receiver node'' (L1$^\prime$) for the purposes of injections.  L$1^{\prime}$ receives a stream of duplicated data for four North-South adjacent intensity beams. These data are processed using the same software as the rest of the L1 nodes. Synthetic pulses are injected into the duplicated data through an interfacing server. This server manages beam duplication, and is capable of selecting which set of four beams are being streamed to L1$^\prime$ in the live system. Careful flagging of injected events in the duplicated data streams ensures that none of the injected signals are misclassified as true astrophysical events.

The injection system injects FRB signals using user-defined parameters. 
The FRBs to be injected are grouped by the beams in which they are expected to
have the highest SNR based on the beam model. A module known
as the injection driver chooses a set of four consecutive beams at random from
the 1024 CHIME/FRB intensity beams and requests the injection server to start
the duplication of these four beams to the L1$^{\prime}$ receiver node. The
injection system then waits for 300 seconds for the running estimates of
the noise properties using in L1 to achieve steady-state. The injection signals prescribed for these beams are generated and injected with a minimum interval of 1 second. However, the typical interval is 2--3 seconds, 
the actual time required to generate and inject an FRB.

Every injection successfully injected into the data stream without software failure is noted in a database and a unique ID is generated. An ``injection snatching'' module in the L2/L3 pipeline is provided with a list of ``active'' injections that are expected in the near future along with their unique IDs, expected DM, expected arrival time, and beam number. An FRB trigger that is detected at the same time and DM  (within a threshold based on the size of the \texttt{bonsai} DM bins) and from the same beam number is marked as an identified injection and the detection parameters are reported to the injection database tagged with the unique ID.

Of the 96\,942 events scheduled for injection, we were able to inject 84\,697 for an
injection efficiency of $\epsilon_{\rm inj} = 0.874$. Failures to inject events
were due to system errors and affect an essentially random subset of
injections. We injected into the predicted maximum-SNR beam
during a campaign in August 2020. Of these, 39\,638 events were detected and
assigned a \texttt{bonsai} SNR.

The sensitivity of the telescope during the injections campaign in August 2020
is not perfectly representative of the sensitivity during the catalog period one
to two years earlier. Based on the detection SNRs of pulsars (tabulated daily),
we estimate our noise levels have improved
by 6\% since the beginning of the survey and 3\% since midpoint of the survey
period. These changes are accounted for in our population analysis and
systematic error budget as described in Appendix~\ref{app:rate_alpha_sys}.
Furthermore, several periods of low sensitivity or differing instrument
configurations (including the pre-commissioning period over which our first 13
bursts were discovered) are not well represented by the injections campaign. These
periods, and the bursts discovered therein, are thus excised from further
analyses that rely on injections.
Finally, numerous tweaks to the operations of the instrument have occurred
over and since the observation period. These tweaks mostly served to
streamline observations and to increase the instrument uptime (for which we
have a separate accounting) and have caused only small changes in our
completeness. However, since our observations occurred prior to the
availability of the injections system, changes in our completeness over
time are difficult to quantify. Such effects should be better quantified in
future data releases where injections can be performed throughout the
observations.

\section{Comparison of Repeaters versus Apparent Non-Repeaters}

\label{sec:repvsnonrep}

This catalog represents by far the largest number of FRBs collected in a uniform manner
using a single telescope and detection pipeline.  This uniformity is helpful for studying FRB property distributions, as past analyses have been complicated by using FRBs from multiple surveys having very different 
survey parameters \citep[e.g.][]{lvl+17}.

The central challenge in studying the FRB population from our data set is compensating for selection effects (e.g. it is more difficult to measure a narrow intrinsic burst width in the presence of strong scattering)
and instrument-induced biases (e.g. it is more difficult to measure narrow intrinsic burst widths due to our finite time resolution)
in event reconstruction.
For some FRB properties (e.g.,~fluence, scattering), selection effects are strong and our fractional completeness varies by
orders of magnitude across the range of detected values for the property. For other properties (e.g.,~DM),
selection effects are at the 
factor-of-two
level. 

We use two strategies for dealing with these selection
effects. In this section, we compare repeater burst properties to those of
apparent non-repeaters, under the reasonable assumption both suffer the same
selection biases, subject to minor caveats discussed below.  In this way we can
deduce in a direct way differences in properties between the two observational
classes. However, this comparative method does not permit an absolute measurement
of the characteristics of either population, for example the fluence
distribution or overall sky rate.  In contrast,
in Section~\ref{sec:abs_pop}, we explicitly measure and compensate for selection
effects using injections, but only for the total population for which we
have the best statistics.

For both analyses, we perform a set of cuts on the catalog to remove events which 
are especially susceptible to selection effects that are challenging 
to quantify. These include the following:
\begin{enumerate}
    \item Events with {\tt bonsai} $\SNR<12$ are rejected, since 
    below this threshold there could have been
        real events detected by our pipeline but subsequently classified as noise upon
        human inspection. During human classification, events with $\SNR\geq12$ are
        visually unambiguous as either FRBs or RFI.
    \item Events having $\dm < 1.5 \max(\dm_{\rm NE2001}, \dm_{\rm YMW16})$ are rejected. This cut
        is more stringent than that used for classifying events as
        extragalactic FRBs. The purpose is to reduce unquantified
        incompleteness coming from misidentifying FRBs when localization errors
        induce an error in the estimated Galactic DM. It also reduces any dependence our results may have on the poorly understood systematic errors associated with the Galactic DM models.
    \item Events detected in far side-lobes are rejected,
    as our primary beam
    is poorly understood in this regime.
    These far side-lobe events have visually-identified ``spiky'' signatures in the burst spectrum
        \citep[e.g.,][]{abb+20}.
\end{enumerate}
These cuts eliminate 205 Catalog~1 FRBs (dominated by the SNR cut) from the following analysis.

The assumption of identical biases for non-repeaters and repeaters is certainly
untrue since we reduce our trigger threshold for the directions and DMs of
previously detected FRBs, to be additionally sensitive to repeat bursts.
For this reason, unless specified otherwise, we compare only the first-detected repeater events
for each repeating source, since that event's trigger threshold was at the nominal value, thereby
eliminating any possible disparity, and avoiding statistical complications of having multiple events per source.
More subtly, the assumption of identical biases for repeaters and apparent
non-repeaters, even with identical thresholds, is also likely untrue given the
differences in burst widths and bandwidths shown below and described in
detail by Pleunis et al. (2021, submitted) and previously reported
\citep{abb+19c,fab+20}, coupled with the fact that selection effects are
correlated (as discussed in detail in Section~\ref{sec:abs_pop}).
Nevertheless, in this analysis we are only sensitive to differences in the
selection-induced correlations between the two sub-populations, which, while we
have not explored this effect in detail, we expect it to be small and unlikely to
affect the conclusions of our comparison.
Note that although we consider only Catalog~1 events with $\SNR\geq12$,  we have verified that all conclusions below hold when all catalog events, regardless of SNR, are included.

For all distribution comparisons, we report probabilities from both
Anderson--Darling ($AD$) and Kolmogorov--Smirnov ($KS$) tests, where a $p$\,-value
$<0.01$ implies $>99$\% confidence that the two samples are drawn from different
underlying distributions.

\subsection{Sky distribution comparisons}
\label{sec:skycomp}

\begin{figure}
    \centering
    \includegraphics[width=1.3\textwidth]{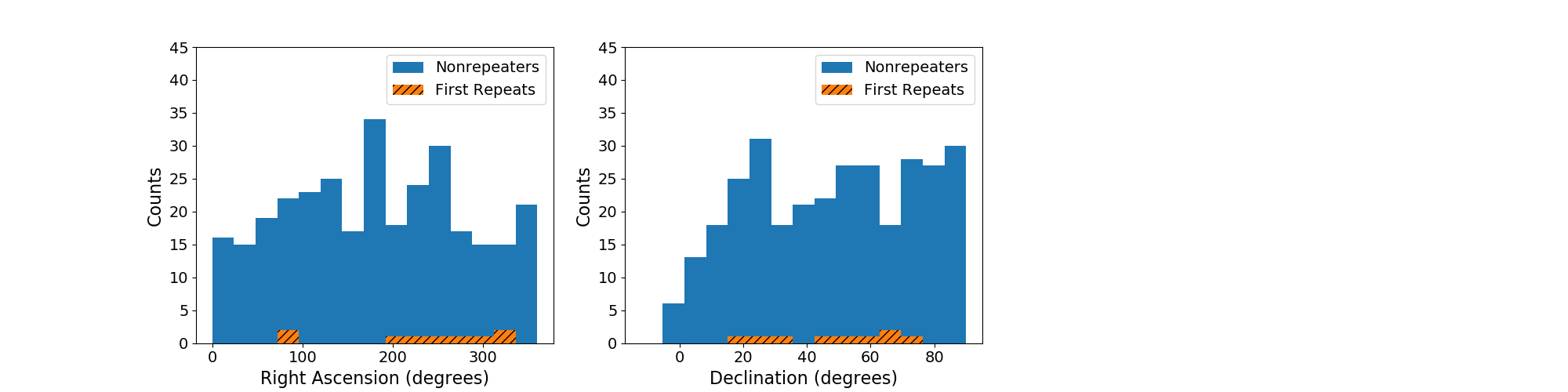}
    \caption{Observed distributions in right ascension (left) and declination (right) of apparent non-repeaters and the first-detected repeater events from Catalog~1.  Note neither is corrected for exposure.  In both cases, the first-detected repeat bursts and apparent non-repeater bursts are statistically consistent with having come from the same underlying distribution.
    }
    \label{fig:repcomp_radec}
\end{figure}

First we compare the sky distributions of repeaters and non-repeaters, specifically
their right ascension and declination distributions (see
Figure~\ref{fig:repcomp_radec}).  For right ascension, we find no difference in
the distributions ($p_{AD} = 0.22$, $p_{KS} = 0.24$), with both consistent with
a uniform distribution when including bursts at all declinations.  Similarly for declination, 
the two distributions are statistically consistent ($p_{AD} = 0.55$, $p_{KS} = 0.49$).
Note that the declination distributions in Figure~\ref{fig:repcomp_radec} are not corrected for exposure and sensitivity, but such corrections affect both repeaters
and non-repeaters similarly.  One caveat is that near the North Celestial Pole, our source density is high due to the long exposure (see Figure~\ref{fig:exposure_map}) which results in confusion that makes repeater identification more difficult than at lower declinations.  Ignoring the Polar region only strengthens the conclusions that the declination distributions of repeaters and apparent non-repeaters are statistically consistent with arising from the same sky distribution.  
We note that the apparent peak in the declination distribution of non-repeating FRBs at $\sim$ 28$\degrees$ is consistent within 2$\sigma$ with the remainder of the distribution.
Separately, we have performed detailed analyses 
of the sky distribution of our Catalog~1 sources. 
Specifically, Josephy et al. (2021, submitted)
search for evidence of correlation with Galactic latitude as has been previously claimed \citep{bb14,psj+14,mj15, bkb+18}.
We also report on a search for
correlation with large scale structure in future work.

\subsection{DM comparisons}
\label{sec:dmcomp}

\begin{figure}
    \centering
    \includegraphics[width=1.3\textwidth]{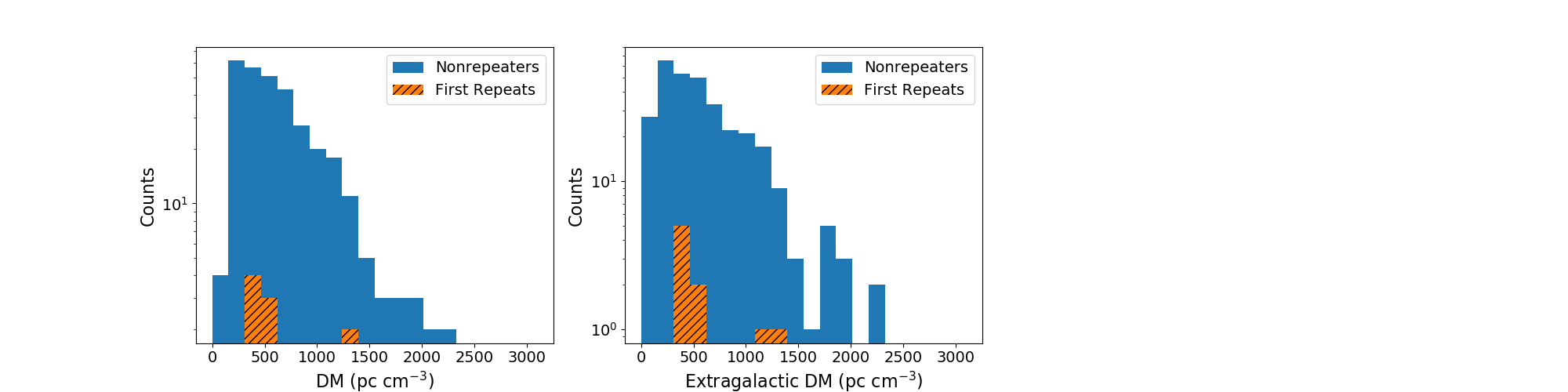}
    \caption{Observed distribution of DMs (left) and extragalactic DM (subtracting the maximal NE2001 component) of apparent non-repeaters and the first-detected repeater events from Catalog~1.  In both cases, the first-detected repeat bursts and apparent non-repeater bursts are statistically consistent with having come from the same underlying distribution.}
    \label{fig:repcomp_dm}
\end{figure}

Next, we consider the observed and extragalactic DMs of apparent non-repeaters and first-detected repeater events from Catalog~1, where extragalactic DM is defined as the observed DM minus the maximal line-of-sight component predicted by NE2001; see Figure~\ref{fig:repcomp_dm}.  We find that 
the distributions are consistent with being
drawn from the same underlying distribution
for DM ($p_{AD} = 0.35$, $p_{KS} = 0.33$)
and for extragalactic DM ($p_{AD} = 0.34$, $p_{KS} = 0.24$).  

\subsection{Signal strength comparisons}
\label{sec:strengthcomp}

\begin{figure}
    \centering
    \includegraphics[width=1.3\textwidth]{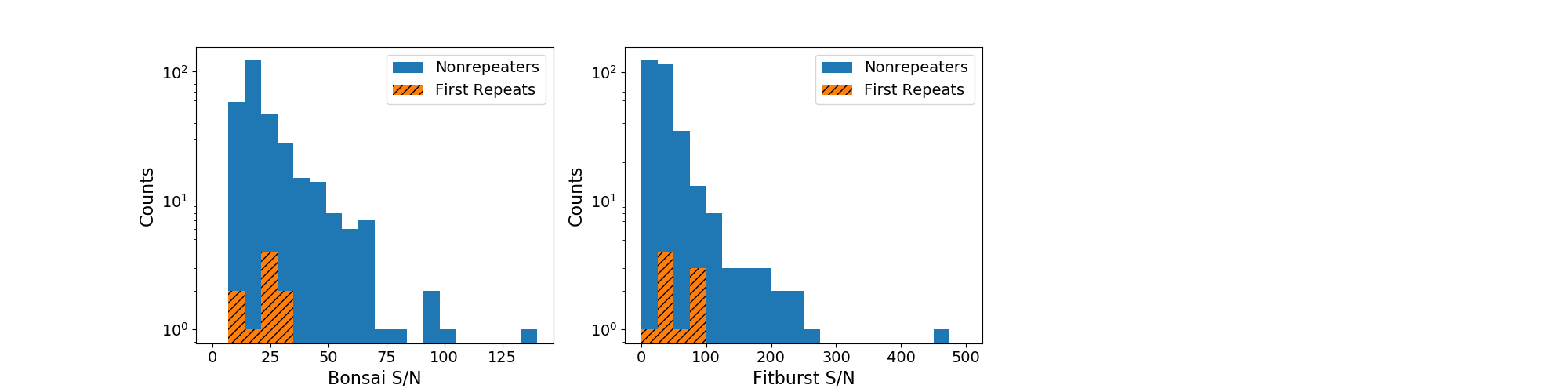}
    \caption{Distributions of {\tt bonsai} (left) and {\tt fitburst} (right) SNRs for apparent non-repeaters and first-detected repeater events.
    The repeater and apparent non-repeater distributions are consistent with being drawn from the same population for both SNR measures, although  the p-values (see text) are somewhat low for the {\tt fitburst} SNR  comparison,  with  a  possible  relative  paucity  of  repeater  bursts  at  the  very  lowest SNRs.  }
    \label{fig:repcomp_snr}
\end{figure}

Next we compare direct measures of signal strength, SNR, as measured by the initial trigger SNR from our real-time FRB search code {\tt
bonsai} \citep{abb+18} and also by
our intensity data burst code, {\tt fitburst} (see Section~\ref{sec:properties}).  Note that neither of these two SNR measurements is a
faithful representation of the true signal strength at the telescope aperture,
because of the complex, frequency-dependent CHIME beam response.
Moreover, {\tt bonsai} SNR is corrupted by RFI mitigation (clipping) for very
bright bursts, an effect with complex behavior in time and spectral
frequency.
The repeater and non-repeater samples could be differentially affected by
the beam, clipping, or other effects, since the two populations have intrinsically different
spectro-temporal properties (studied in detail below). Even so, the comparison is interesting since an
observed difference is indicative of an intrinsic difference in the
populations, even if it might be indirect through correlated observational
effects.
The distributions are shown in Figure~\ref{fig:repcomp_snr}.  The repeater and apparent non-repeater distributions are consistent with being drawn from the same population for both SNR measures ($p_{AD} = 0.65$, $p_{KS} = 0.44$ for {\tt bonsai} SNR and $p_{AD} = 0.08$, $p_{KS} = 0.26$ for {\tt fitburst} SNR).

\begin{figure}
    \centering
    \includegraphics[width=1.3\textwidth]{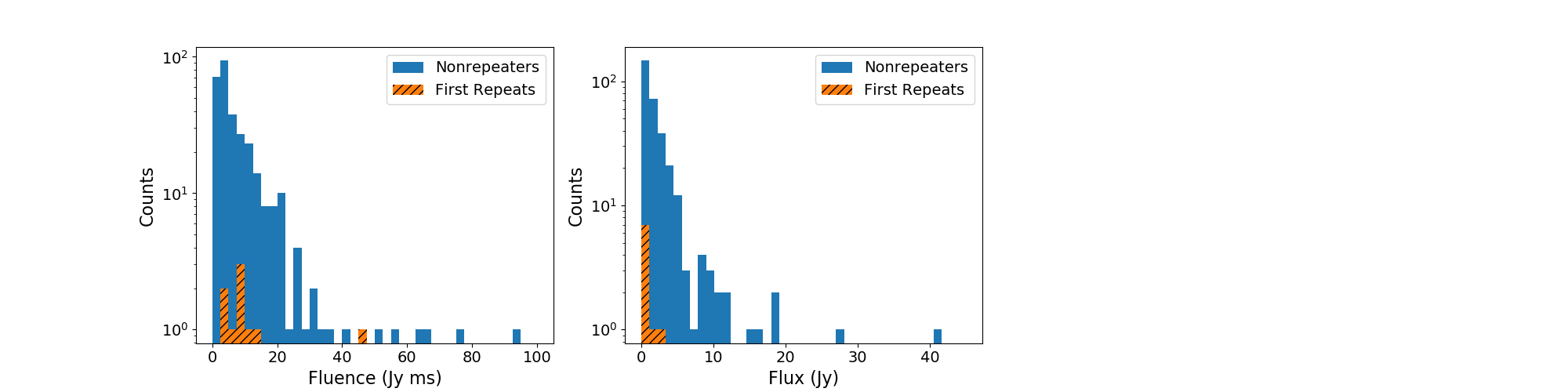}
    \caption{Fluence (left) and flux (right) distributions for apparent non-repeaters and first-detected repeater events.  The fluence distributions are statistically consistent with originating from the same underlying distribution.  For flux, there is marginal ($p_{AD}=0.028$, $p_{KS}=0.068)$ evidence for a difference, possibly related to the broader widths of the first-detected repeater events (see Fig.~\ref{fig:repcomp_width}).
    }
    \label{fig:repcomp_flu}
\end{figure}

We can also compare signal strength distributions using calibrated 
fluence and flux, noting, however, that our values have substantial
uncertainties and are biased low, mainly due to the unknown location of each event within the detection beam
(see Section~\ref{sec:fluences}).  The distributions are shown in
Figure~\ref{fig:repcomp_flu}. The fluence distributions are 
consistent with being drawn from the same underlying sample,
($p_{AD} = 0.070$, $p_{KS} = 0.066$), 
as are the flux distributions, though with lower $p$-values ($p_{AD} = 0.028$, $p_{KS} = 0.068$).
A possible origin for this putative, slight difference is the broader widths for repeaters (see below).
We note that also including Catalog~1 events with $\SNR\geq12$ results in similarly low but still inconclusive $p$\,-values
($p_{AD} = 0.028$, $p_{KS} = 0.021$ for fluence and
$p_{AD} = 0.040$, $p_{KS} = 0.044$ for flux), so does not lend additional support to the distributions being different.

A possible fluence or flux anti-correlation with extragalactic DM is expected since more distant sources should, on average, be fainter.  A simple observed
anti-correlation (as we are aware is present in our data for flux versus extragalactic DM) is insufficient to address this question due to the significant
instrumental biases (see Section~\ref{sec:cordists}).  However, one
can ask whether any naive correlation seen among apparent non-repeaters is seen
for repeaters, since both would suffer similar biases.  To do this, we compare the 2D fluence versus extragalactic DM distributions of apparent non-repeaters and first-detected repeaters using the 2D KS test\footnote{https://github.com/syrte/ndtest/blob/master/ndtest.py}
described by \citet{pea83} and
refined by \citet{ff87}.  We do the same for the 2D flux versus extragalactic DM distributions.  In both cases, the 2D distributions for apparent non-repeaters and for repeaters are consistent with originating from the same underlying distribution ($p_{2DKS} = 0.099$ for fluence and $p_{2DKS} = 0.43$ for flux). However, the sample size for first-detected repeaters is small and relatively minor differences in either fluence or flux distributions may not be detectable.  
Inclusion of
$\SNR<12$ events yields lower p-values: 
$p_{2DKS} = 0.015$ for fluence and $p_{2DKS} = 0.051$ for flux, still not significant at the $>99$\% level, but possibly noteworthy.
Whether the population as a whole exhibits such an anti-correlation, once selection biases are accounted for, is discussed in detail in Section~\ref{sec:abs_pop}.

\subsection{Burst temporal width and bandwidth distribution comparisons}
\label{sec:widthcomp}

\begin{figure}
    \centering
    \includegraphics[width=1.3\textwidth]{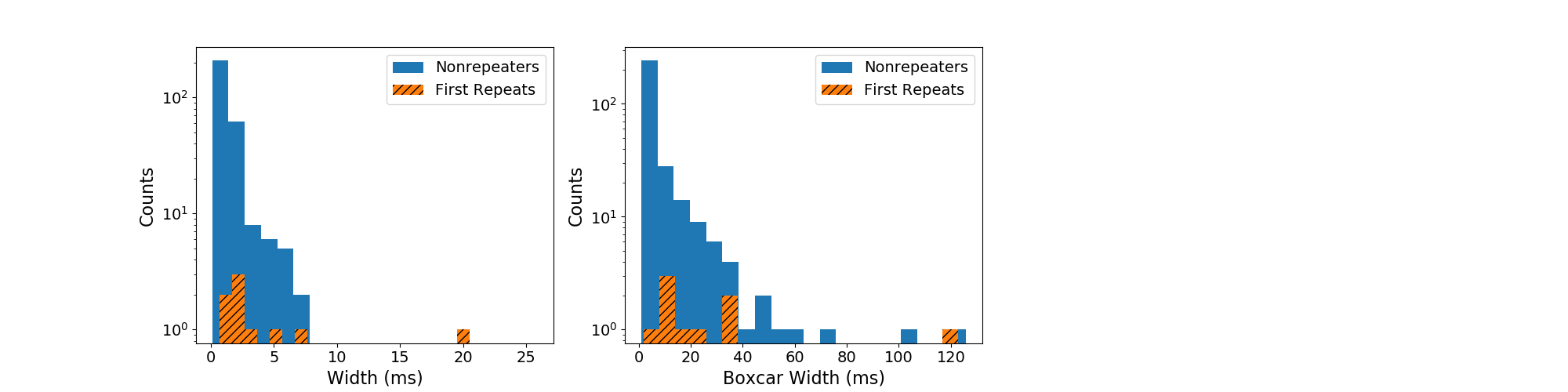}
    \caption{Distributions of \fitburst-measured intrinsic pulse width (left) for apparent non-repeaters and 
    first-detected repeater events in Catalog~1. Burst upper limits are not plotted.  For multi-component bursts, we plot the mean of individual component intrinsic widths, unless one is an upper limit (2 cases) in which case we plot the width of the first sub-component for which it is measurable.  (Right) Distributions of boxcar widths.  All of our events have a measured boxcar width (i.e., there are no upper limits), though they include intra-channel dispersion broadening and scattering, so do not necessarily reflect the intrinsic burst width.  For both panels there is strong evidence for different underlying distributions for first-detected repeater bursts and apparent non-repeater bursts (for intrinsic width 
    $p_{AD} = 7.3 \times 10^{-5}$ and $p_{KS} = 5.6 \times 10^{-5}$, and for
    boxcar width $p_{AD} = 1.5 \times 10^{-4}$, $p_{KS} = 2.2 \times 10^{-4}$.)
    }
    \label{fig:repcomp_width}
\end{figure}

Next, we look at distributions of burst intrinsic widths.  Figure~\ref{fig:repcomp_width} shows the distributions of measured widths (i.e., no upper limits, with scattering and DM smearing from the finite frequency channel size omitted) for first-detected repeater events and apparent non-repeaters.  For multi-component bursts, we have plotted the mean of each component width, unless one sub-component width is an upper limit (2 cases), in which case we plot the width of the first sub-component for which it is measurable.  The distributions are statistically extremely unlikely to have arisen from identical underlying distributions, with
$p_{AD} = 7.3 \times 10^{-5}$ and $p_{KS} = 5.6 \times 10^{-5}$, with repeaters on average broader.  This difference in repeater and apparent non-repeater burst widths was previously reported
based onlimited data
\citep{abb+19c,fab+20}
and is 
strongly supported by the Catalog~1 data.
The result strongly persists when including $\SNR<12$ bursts, and also when including upper limits on burst widths.
Because omission of upper limits represents a loss of information,
we also applied
two statistical tests that can incorporate upper limits  (i.e. ``left-censored'' data in survival analysis parlance): the log-rank test \citep{HarringtonFleming1982} and
Peto \& Peto's modification \citep{peto1972asymptotically} of the
Gehan-Wilcoxon test \citep{gehan1965generalized}, both implemented using the R NADA package's {\tt cendiff} routine\footnote{https://rdrr.io/cran/NADA/man/cendiff.html} \citep{helsel2005nondetects,Rsoftware,NADApackage}. Both tests yield
$p$-values that strongly support different underlying populations, with $p=5\times 10^{-10}$ and $7\times 10^{-10}$ , respectively.

Because the mean of the widths of individual sub-components in multi-component bursts does not necessarily reflect the overall burst length in those cases, 
we also show distributions of boxcar widths in Figure~\ref{fig:repcomp_width}.
Every Catalog~1 event has a measured boxcar width, i.e., there are no upper
limits.  These widths include intra-channel dispersion smearing and scattering,
so are not robust proxies for burst intrinsic width, but are equally non-robust for both repeaters and non-repeaters.
Again, the difference in distributions is highly significant ($p_{AD} = 1.5 \times 10^{-4}$, $p_{KS} = 2.2 \times 10^{-4}$).
Pleunis et al.\ (2021, submitted) present a more detailed analysis of the morphological properties of our Catalog~1 bursts.

\begin{figure}
    \centering
    \includegraphics[width=1.3\textwidth]{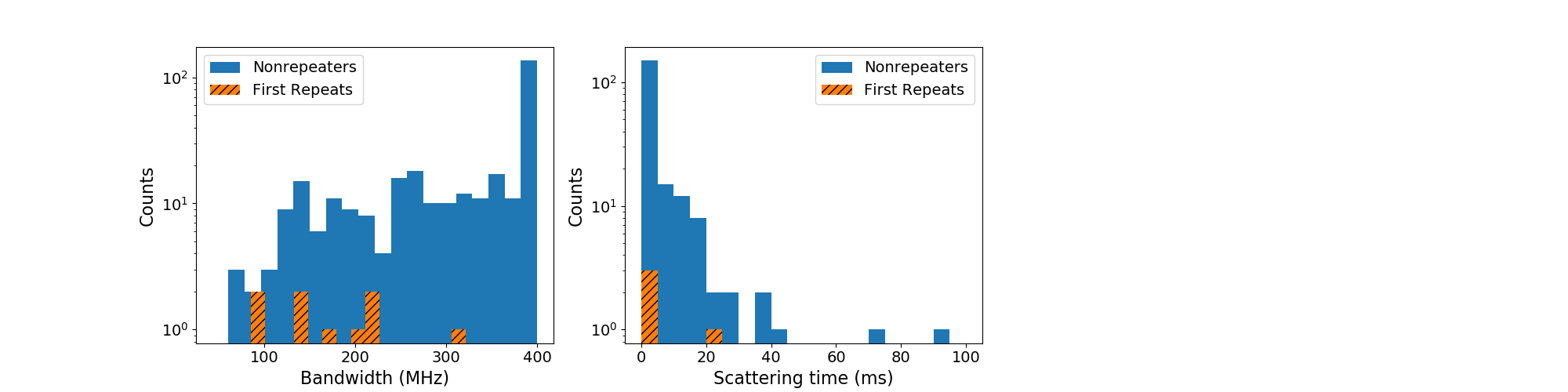}
    \caption{(Left) Distribution of bandwidth for first-detected repeater events and for apparent non-repeaters. Note that bandwidths have not been corrected for instrumental response, which is complex and strongly dependent on declination and location in the detection beam, but on average, is identical for non-repeaters and repeaters.  The distributions are inconsistent with arising from the same underlying sample ($p_{AD} = 1.3 \times 10^{-4}$, $p_{KS} = 2.3 \times 10^{-4}$).
    (Right) Distributions of scattering times for first-detected repeater
    events and apparent non-repeaters for which the time scale was measurable.
    Upper limits are not included in the plot.  For repeaters, the scattering
    time plotted is the most constraining of those measured of all repeat
    bursts (but still not plotting upper limits).  The samples are
    statistically consistent with arising from the same underlying
    distributions. }
    \label{fig:repcomp_bwscatt}
\end{figure}

For each burst, Catalog~1 contains both the lowest and highest frequencies at
which the burst was detected, and hence the difference, which is approximately the event
bandwidth.  The Catalog~1 values are uncorrected for instrumental bandpass response, however.  Under the reasonable assumption that on average, the correction is the same for non-repeaters and repeater bursts, we can compare the bandwidth distributions for the two groups.  This is shown in Figure~\ref{fig:repcomp_bwscatt} (left).  A substantial difference in distributions is apparent by eye, and confirmed statistically
($p_{AD} = 1.3 \times 10^{-4}$, $p_{KS} = 2.3 \times 10^{-4}$).  
The bandwidth properties of repeaters versus apparent non-repeaters are
discussed in more detail by \citet{pgk+21}.

We can also compare distributions of scattering times for apparent non-repeaters and repeaters in Catalog~1.  A difference might be expected if the local source environment between repeaters and non-repeaters differed, and if scattering in the local environment dominated over other sources of scattering.
Figure~\ref{fig:repcomp_bwscatt} (right) shows measured scattering times
(ignoring upper limits) for the
repeaters and apparent non-repeaters.  For repeaters, the scattering time plotted is the most constraining from all of the sources' repeat bursts.  Our statistical tests
indicate no evidence for the distributions being from different underlying populations
($p_{AD} = 0.42$, $p_{KS} = 0.32$).
We also verified this result using the log-rank test and
the modified
Gehan-Wilcoxon test. Both tests yielded
$p$-values indicating consistent underlying populations, with $p=0.1$ and 0.3, respectively.

A correlation between scattering times and extragalactic DMs might be expected if scattering is dominated by a component in the intergalactic medium and extragalactic DMs are not dominated by host contributions, or conversely if both scattering and extragalactic DM are dominated locally at the source.  Any correlation detected in Catalog~1 requires correction due to instrumental biases as discussed below
(Section~\ref{sec:cordists}).  However, such biases should
be the same for non-repeaters and repeaters so it is fair to ask here whether similar correlations exist for both groups.  To investigate,  we compared the 2D scattering time versus extragalactic DM distributions for apparent non-repeaters and for first-detected repeaters using the 2DKS test, and found that they are consistent with the distributions originating from the same underlying population ($p_{2DKS} = 0.10$).  However, the sample size for first-detected repeaters is small and minor distribution differences might be yet undetectable.  Inclusion of
$\SNR<12$ events yields no interesting difference.
We will report on a more detailed analysis of this possible correlation in future work,
but discuss it briefly
in Section~\ref{sec:disc_intrin_dist}.

\subsection{Summary of repeater vs apparent non-repeater comparisons}

In summary, we find strong evidence for significant differences in the intrinsic burst widths and bandwidths of repeating FRBs compared to a population yet to be seen to repeat.  In contrast, we do not find significant differences when comparing the two populations for sky distribution, DM and scattering distributions, or signal strengths.  A summary of the results of our comparison are provided in Table~\ref{ta:comparisons}.

\begin{table}[t]
\begin{center}
\caption{Summary of Repeater vs Apparent Non-Repeater Distribution Comparisons}
\begin{tabular}{lccccc} \hline
    Property  &  \S & Figure No. & $p_{AD}^{a}$  & $p_{KS}^{b}$ & $p_{2DKS}^{c}$  \\\hline
    Right Ascension & \ref{sec:skycomp} & \ref{fig:repcomp_radec} & 0.22 & 0.24 & ...  \\
    Declination & \ref{sec:skycomp} & \ref{fig:repcomp_radec} & 0.55 & 0.49 & ... \\
    DM & \ref{sec:dmcomp} & \ref{fig:repcomp_dm} & 0.35 & 0.33 & ... \\
    eDM$^{d}$ & \ref{sec:dmcomp} & \ref{fig:repcomp_dm} & 0.34 & 0.24 & ... \\
    {\tt bonsai} SNR & \ref{sec:strengthcomp} & \ref{fig:repcomp_snr} & 0.65 & 0.44 & ... \\
    {\tt fitburst} SNR & \ref{sec:strengthcomp} & \ref{fig:repcomp_snr} & 0.08 & 0.26 & ... \\
    Fluence & \ref{sec:strengthcomp} & \ref{fig:repcomp_flu} & 0.070 & 0.066 & ... \\
    Flux & \ref{sec:strengthcomp} & \ref{fig:repcomp_flu} & 0.028 & 0.068 & ... \\
    2D fluence vs eDM & \ref{sec:strengthcomp} & ... & ... & ... & 0.099 \\
    2D flux vs eDM & \ref{sec:strengthcomp} & ... & ... & ... & 0.43 \\
    Width$^e$ & \ref{sec:widthcomp} & \ref{fig:repcomp_width} &  $7.3 \times 10^{-5}$ & $5.6 \times 10^{-5}$ & ... \\
    Boxcar width & \ref{sec:widthcomp} & \ref{fig:repcomp_width} & $1.5 \times 10^{-4}$ & $2.2 \times 10^{-4}$ & ... \\
    Bandwidth & \ref{sec:widthcomp} & \ref{fig:repcomp_bwscatt} & $1.3 \times 10^{-4}$ & $2.3 \times 10^{-4}$ & ... \\
    Scattering time$^e$ & \ref{sec:widthcomp} & \ref{fig:repcomp_bwscatt} & 0.42 & 0.32 & ... \\
    2D scattering time$^e$ vs eDM & \ref{sec:widthcomp} & ... & ... & ... & 0.10
    \\\hline
\end{tabular}
\label{ta:comparisons}
\end{center}
    \textbf{Notes.} \\
    $^a$Anderson-Darling probability of originating from same underlying population\\
    $^b$Kolmogorov-Smirnov probability of originating from same underlying population\\
    $^c$2D Kolmogorov-Smirnov probability of originating from same underlying population\\
    $^d$Extragalactic DM \\
    $^e$excludes upper limits; results qualitatively the same when including them -- see text.\\
\end{table}

\section{Intrinsic characteristics of the FRB population}
\label{sec:abs_pop}

Here we infer the properties of the intrinsic FRB population from
the observed CHIME/FRB Catalog~1 data. The central challenge is to account for
selection biases, i.e.,\ the fact that the probability to detect an FRB depends
on its properties in a complicated way, and to account for instrument-induced errors in the measured quantities. To correct the observed property
distributions for these effects, we use the injection system described in Section~\ref{sec:injection}. Here we give
a
brief overview of the methods used to account for selection effects. A more
detailed description of our FRB inference pipeline, including additional
methods used for cross-checks, will be described elsewhere.
We present distributions of FRB properties corrected for selection biases and
perform a more detailed examination of the data in the property space of
fluence and DM: inferring the overall FRB sky rate, the fluence distribution,
and examining how the fluence distribution depends on DM.

For the present population analysis, we will consider six FRB properties:
fluence ($F$), DM, scattering timescale ($\tau$),
intrinsic width ($w$), spectral index ($\gamma$), and spectral running ($r$).
Using injections to compensate for selection effects is complicated by the fact
that both the selection effects,
and the intrinsic FRB population, may be correlated in the high-dimensional phase space of FRB properties.
For instance, we expect a selection bias against high-DM bursts because DM-smearing dilutes the burst signal
in time. However, this bias is weaker if FRBs have a wider intrinsic pulse profile since the
smearing would then have a smaller relative effect.  It is also weaker if FRBs have 
flatter spectra,
since a larger fraction of the signal would come from higher frequencies where
the effect of smearing is reduced.
There is also an interplay with signal loss from our data filtering and
flagging, which more adversely affects low-DM events.
Thus, in principle, the distributions of all three of these properties (and in fact all FRB properties) should
be modeled and fit simultaneously to be fully consistent.

We instead make a number of simplifying assumptions, and defer
a full multi-dimensional intrinsic correlations analysis to future work.
To simplify the analysis, we study FRB properties one or two at a time, holding the distributions
for the rest of the properties fixed
at a fiducial population model that provides a reasonable overall description of
the data. As we show in the next section, it is possible to robustly compensate
for correlations in the selection effects so long as correlations in the
intrinsic population are small.  

\subsection{Selection bias-corrected FRB property distributions}
\label{sec:cordists}

First, we set up our formalism and outline our procedure. We
wish to make inferences about the \emph{intrinsic} \emph{property rate function} of these FRBs:
$R(F, \dm, \tau, w, \gamma, r)$. However, observational effects
mean that not all regions of property space are observed with the same
efficiency. We define the \emph{observation function} to be
$P(\SNR | F, \dm, \tau, w, \gamma, r)$, which describes the stochastic
mapping
from event properties to SNR (the stochastic mapping is
because of a burst's random location in the beam and occurrence relative to
time-variable effects such as RFI).
In our usage, the observation function is
averaged over time and sky location, and is affected by the beam,
system sensitivity, detection pipeline efficiency, RFI, and other effects.

Our main simplifying assumption is that the FRB properties are intrinsically
uncorrelated (other than
$\gamma$ and $r$ for which we observe strong correlations) such that their
distributions factorize:
\begin{equation}
\label{eq:propertyratefunc}
    R(F, \dm, \tau, w, \gamma, r) = R_0 P(F)P(\dm)P(\tau)P(w)P(\gamma, r),
\end{equation}
where $R_0$ is the overall sky rate (with units events per sky per day)
and the other factors are the individual probability density functions for each
property.\footnote{For brevity of notation, we denote probability density
functions (PDFs)
by $P$ and distinguish them by their arguments, such that e.g.,\ $P(F)$ and
$P(\dm)$ are different functions. Note that as PDFs, $P(F)$ has units $1/F$,
$P(\dm)$ has units $1/\dm$, 
$P(\SNR | F, \dm, \tau, w, \gamma,
r)$ is unitless because $\SNR$ is unitless, and  $P(F, \dm)$ appearing in
Equation~\ref{eq:p_obs_snr_dm} has units $1/(F\,\dm)$.}
Rigorously testing
this assumption is beyond the scope of this paper and will be deferred 
to future work, except for $F$ and $\dm$, which we study briefly in
Section~\ref{sec:flux_rate}.  Given the limited statistical power of our sample of
$\sim250$ events,
we expect such correlations to have a small impact on our
results. We do, however, check for intrinsic correlations
through a series of jackknife tests described in
Section~\ref{sec:jackknife}, finding some evidence that such
correlations may exist.

One complication is that the fluence we measure for a given event is a
highly uncertain estimate of the true fluence (due primarily to the uncertain
localization and beam sensitivity, see Section~\ref{sec:fluences}). In addition, we do not currently have the ability to
robustly forward model the fluence measurement processes using the injections
system. For these reasons, we do not use fluence measurements for our inferences. We can
nonetheless make inferences about the intrinsic fluence distribution since the
detection SNR (which is robustly modeled using injections) strongly correlates
with fluence. Although it has been shown that under certain
assumptions SNR and fluence are distributed with the same power-law index
\citep{ocp16, con19, jem+19},
our analysis does not rely on this result, only
requiring that fluence and SNR
correlate. Integrating out the fluence, the observed rate of FRBs is:
\begin{equation}
\label{eqn:factorize}
    \mathcal{R}_{\rm obs}(\SNR, \dm, \tau, w, \gamma, r) = \int_0^\infty dF\, P(\SNR | F, \dm, \tau, w, \gamma, r)
    R(F, \dm, \tau, w, \gamma, r),
\end{equation}
where $\mathcal{R}_{\rm obs}$ denotes the observed, as opposed to the intrinsic,
rate function.

The observed distribution of any single
property
is the integral over the
other properties including property ranges within any cuts made on the data.
For example, the observed distribution of DM would be:
\begin{equation}
    P_{\rm obs} (\dm) \propto \int_{\SNR_{\rm thres}}^{\infty} d\SNR\,\int
    d\tau\, dw\,\,d\gamma\,dr\,\mathcal{R}_{\rm obs}(\SNR, \dm, \tau, w, \gamma, r),
\end{equation}
which---given our assumption of uncorrelated intrinsic distributions---can be
corrected to the intrinsic
distribution:
\begin{align}
    P(\dm) &= P_{\rm obs} (\dm) s(\dm)^{-1},
    \label{e:selfuncdef}
    \\
    s(\dm) &\propto \int_{\SNR_{\rm thres}}^{\infty} d\SNR 
    \int d\tau\,dw\,d\gamma\,dr\,dF\,P(\SNR | F, \dm, \tau, w, \gamma, r)
    P(F)P(\tau)P(w)P(\gamma, r),
    \label{e:selfunc}
\end{align}
where the constant of proportionality in the second line is set by the requirement that
the intrinsic distribution, $P(\dm)$,
and the observed distribution, $P_{\rm obs}(\dm)$, be normalized. 
Above, $s(\dm)$ is the DM selection function,
i.e.,\ the probability for an FRB with a given DM to be detected, marginalized
over all other FRB properties.

Note that
under our assumption of uncorrelated intrinsic distributions,
$s(\dm)$ does not in itself depend on the intrinsic distribution of FRB DMs
(see Eq.~\ref{e:selfunc}), but does
depend on the intrinsic distributions of the other properties. However, if
$P(\SNR | F, \dm, \tau, w, \gamma, r)$ factorizes into separate functions of
the FRB properties, this dependence vanishes. In practice we expect that
most selection effects induce correlations in the observed sample
at the $\lesssim$10\% level,
although we defer a detailed study of these correlations to future work.
This dependence generalizes to other
properties: for property $\xi$, $s(\xi)$ is independent of $P(\xi)$ and
depends on $P$ of the other properties only through correlations in the
observation function, which we expect to be modest.

Having argued that the selection
function should be weakly dependent on the underlying intrinsic property
distributions,
we can calculate an accurate
selection function given a \emph{fiducial} model (hereafter denoted with the
subscript ``fid'', e.g\@. $R_{\textrm{fid}}$, $P_\textrm{fid}$) for the FRB property
distributions that is a reasonable, albeit imperfect, match to the data. Likewise, below we study
the FRB population in the property subspace of DM--$F$, fixing the other
properties at the fiducial model. Our results for a given property distribution
(e.g.,\ the DM distribution)
should thus be interpreted in the context of a weak dependence on our
model for other properties (e.g.,\ the scattering distribution) as well as our overall assumption
of uncorrelated intrinsic properties.

In Appendix~\ref{app:fid} we detail our assumed functional forms for the
property distribution functions appearing in Equation~\ref{eq:propertyratefunc}
and our procedure for using injections to find fiducial model parameters that match our
observations.
Here we describe only
the details that are
critical to understanding further results.
Rather than iteratively injecting a new population for every
candidate model we wish to test, we use property dependent weights $W(F, \dm,
\tau, w, \gamma, r)$ to convert a single injected population to any other
population model.

We find that, due to a strong selection bias against highly scattered
events, the population with scattering time above 10\,ms is very poorly
constrained. In order for this part of the population to not dominate uncertainties,
we cut them from further analysis except for the measurement
of $s(\tau)$, which is independent of $P(\tau)$.
In addition, at low $\tau$ and $w$,
there is significant measurement uncertainty in these properties, with many
measurements in the $\sim1$\,ms range being upper limits. We deal with this uncertainty by
using wider $0.5$\,ms bins below 2\,ms, and assigning a value
of half the 2$\sigma$ upper limit where these occur
($24$ events for $w$ and $257$ events for $\tau$, before cuts).
This treatment is far
from ideal, but is likely sufficient since the purpose of the fiducial model is
only to roughly describe the property distributions for dealing with correlations in
the selection effects.

We exclude 39 events that are detected either during pre-commissioning, epochs of
low-sensitivity, or on days with software upgrades. In the catalog data these events are noted in the
\texttt{excluded\_flag} field.

After applying the cuts discussed in Section~\ref{sec:repvsnonrep},
the cuts on $\tau > 10$\,ms, and the cuts on days with system concerns,
the remaining sample still contains repeaters, with multiple bursts
making these cuts.
For these sources, only the first burst is kept; all subsequent events are
excluded from the analysis.
We do this because for some properties (DM, scattering), repeat bursts from the
same source should have the same value. Including repeat bursts would thus skew the statistics of
our distributions. As such, we are effectively studying the distributions of
\emph{sources} rather than bursts, a small but conceptually important distinction.

The post-cut sample includes $22$ events with complex morphology and thus no unique
value of $w$.
For these events, we estimate an ``effective'' pulse width by
using a value proportional to the boxcar width, with the constant of
proportionality (equal to 0.17) calibrated by comparing the pulse widths and the boxcar widths
of single-component events.

We consider only the DM range in excess of 100\,pc\,cm$^{-3}$, since even
after classifying events using the Galactic DM models, the extragalactic
nature of sources below 100\,pc\,cm$^{-3}$ is somewhat ambiguous. This
restriction
excludes no FRBs from the Catalog~1 sample but does exclude a number of injected events.
In total, 270 Catalog~1 events were excluded (265 events remained) after applying the above cuts.

\begin{figure}
    \centering
    \includegraphics[width=0.9\textwidth]{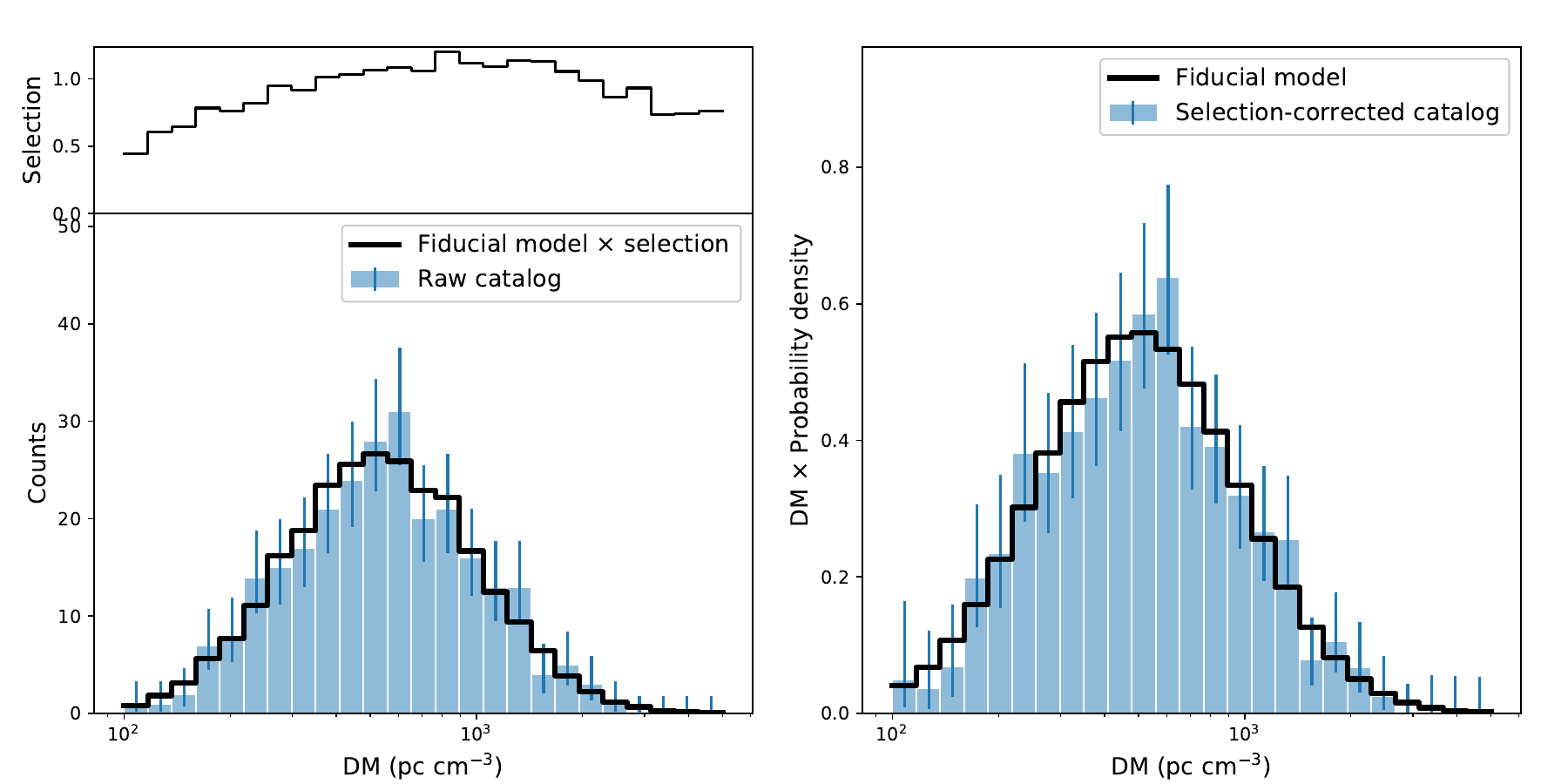}
    \caption{
        {DM distribution before (left) and after (right)
        compensating for selection effects.} Blue histograms are the catalog
        data with error bars representing the Poissonian 68\% confidence
        interval on the underlying bin mean. When plotted as a probability
        density in the right panel, the quantity $\dm \times P(\dm)$ is unitless and is
        equivalent to the probability density function re-parameterized in terms
        of $\ln \dm$. This scaling aids in visual interpretations of the
        area under the probability density when using a logarithmic horizontal axis.
        The selection function plotted in the top left is normalized such that
        Equation~\ref{e:selfuncdef} holds for the fiducial model.
        The fiducial model is the best-fit log-normal distribution
        resulting from the iterative fitting procedure described in
        Section~\ref{app:fid},
        with the appropriate selection function applied as per
        Equation~\ref{e:selfunc}. The selection function varies by
        more than a factor of two over the range of observed DMs,
        with biases against low-DM events from detrending and flagging
        being a larger effect than that of DM smearing affecting high DMs.
        A log-normal function provides a good description of the data once accounting for
        selection effects.
    }
    \label{fig:dm_dist}
\end{figure}

\begin{figure}
    \centering
    \includegraphics[width=0.9\textwidth]{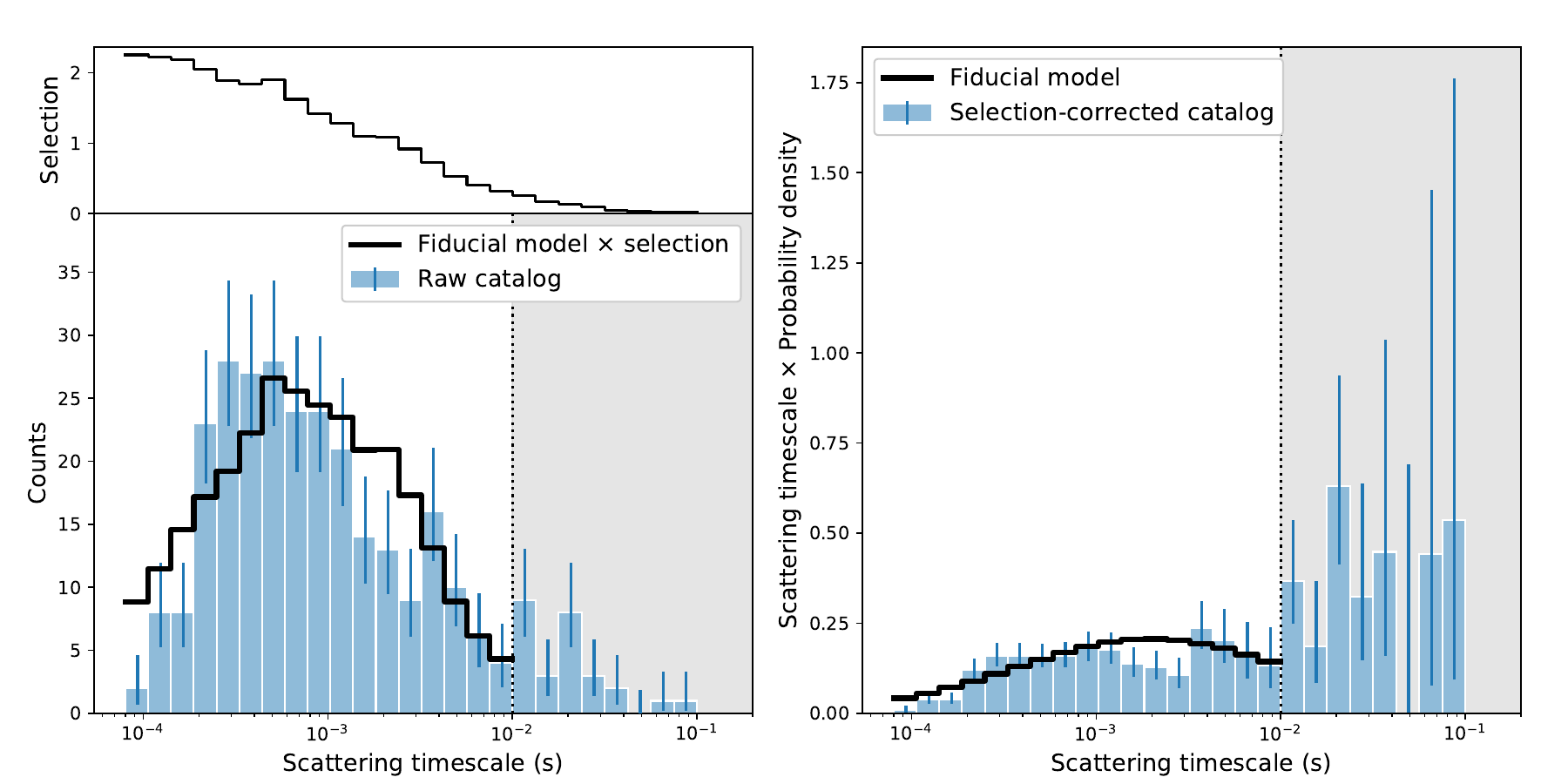}
    \caption{
        {Scattering time ($\tau$, scaled to 600\,MHz) distribution before (left) and after (right)
        compensating for selection effects.} Histogram, error bar, and line
        meanings are analogous to Figure~\ref{fig:dm_dist}. Because of the
        poorly constrained, apparently rising distribution, the gray region
        with scattering above 10\,ms is not included in fiducial model fit,
        subsequent analysis, or histograms of other properties.
        Nonetheless, the selection function is valid in
        this region. To account for uncertainty in the scattering time
        measurement, the fiducial model is fit using bins that are wider than
        those shown here, with linear bins of width 0.5\,ms up to 2\,ms and
        logarithmic bins thereafter. To events for which only an upper limit on
        scattering is measured, we assign a value of 1/2 the 2$\sigma$ upper
        limit.
        The log-normal fit for the fiducial model is a marginal match to
        the data,
        although the data are affected by observational uncertainties and the
        large portion of events for which we measure only upper limits.
        We find there is severe selection bias against events with
        scattering time larger than 10\,ms and the handful of highly scattered
        events we observed imply there may be a substantial unobserved population of
        highly scattered FRBs.
    }
    \label{fig:scat_dist}
\end{figure}

\begin{figure}
    \centering
    \includegraphics[width=0.9\textwidth]{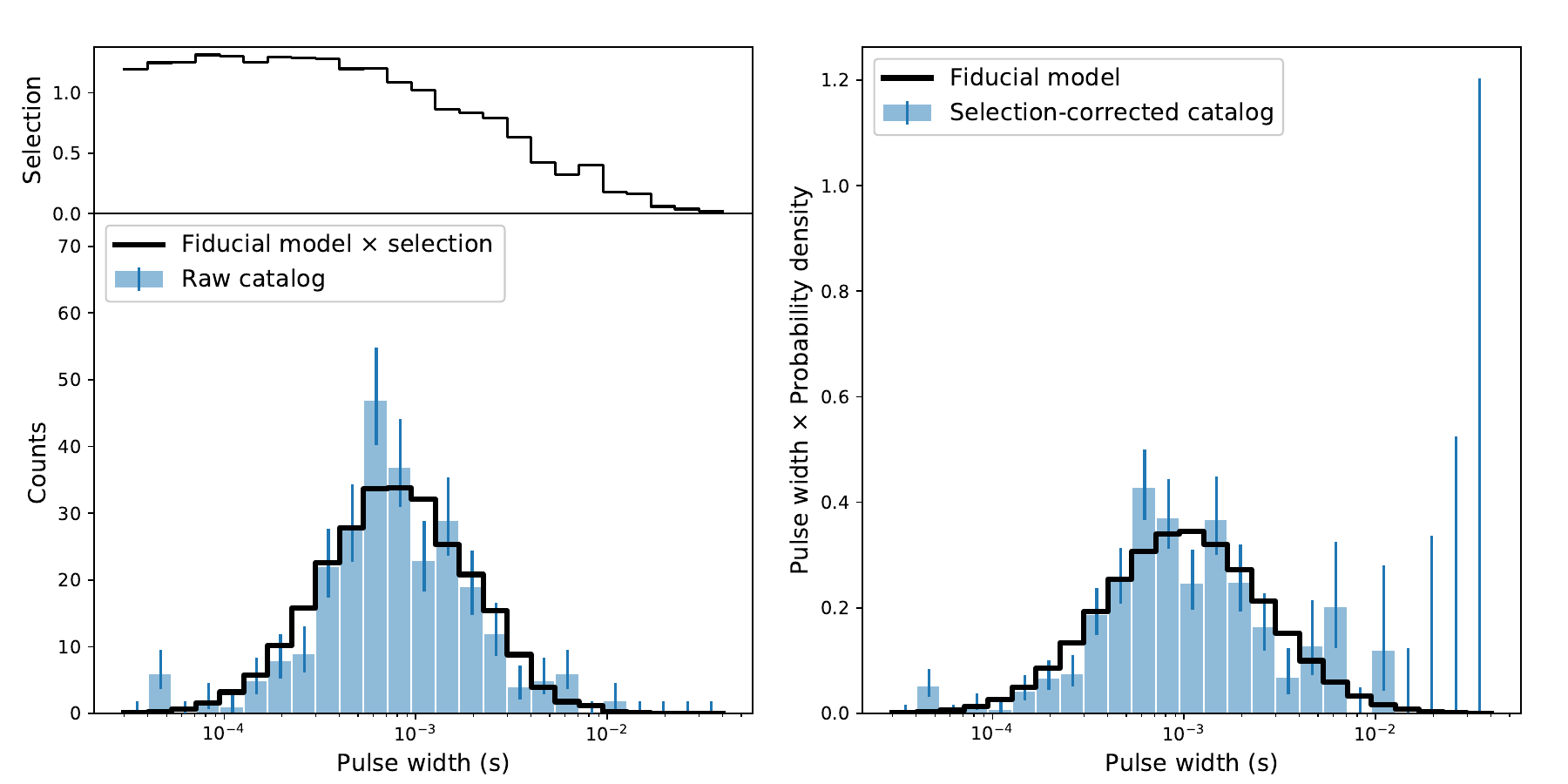}
    \caption{
        {Intrinsic width ($w$, defined as $\sigma$ of a Gaussian profile)
        distribution before (left) and after (right)
        compensating for selection effects.} Histogram, error bar, and line
        meanings are analogous to Figure~\ref{fig:dm_dist}. Complex-morphology
        bursts are assigned an effective width as described in  Appendix~\ref{app:fid}.
        The treatment of
        measurement uncertainty for fitting (linear 0.5-ms-wide bins below
        2\,ms) and upper limits is the same as in the scattering case. While
        there is a strong selection bias against wide events, we find a
        log-normal distribution to be a satisfactory fit for the full range of
        widths, with little evidence for an unobserved, intrinsically very wide
        population.
    }
    \label{fig:width_dist}
\end{figure}

The injections and fiducial model are used to calculate selection
functions for the properties DM, $\tau$, and $w$. These are shown in
Figures~\ref{fig:dm_dist}, \ref{fig:scat_dist}, and \ref{fig:width_dist},
along with both the uncompensated and selection-compensated distributions
for the properties of observed events.

\begin{figure}
    \centering
    \includegraphics[width=0.95\textwidth]{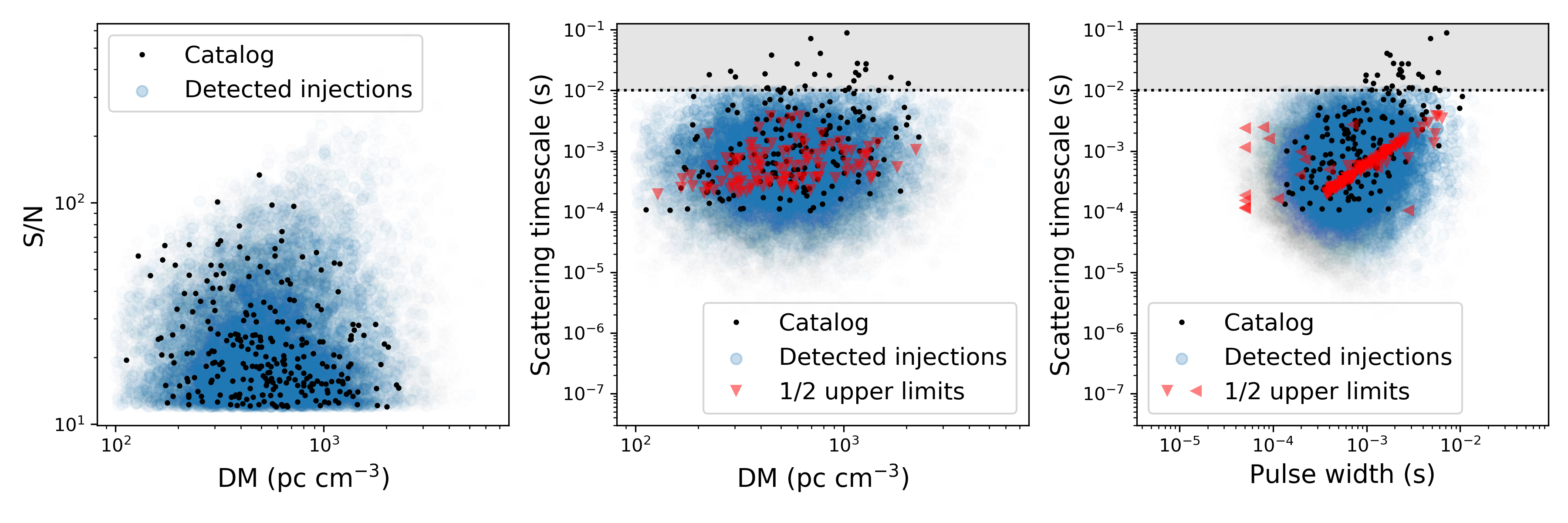}
    \caption{
        {Catalog~1 events in a few property subspaces compared to the fiducial
        model with selection effects included.} For
        injected events, the opacity is proportional to the weight derived from
        the fiducial model $W(F, \dm, w, \tau)$. Cases for which we only
        measure upper limits on a property are plotted at half the upper-limit
        value, and
        are denoted with red triangles. Property measurement effects (which are
        distinct from selection effects) are apparent
        here. Particularly, for cases where no scattering is detected we have
        assigned an upper limit equal to 2 times the width (see
        Section~\ref{sec:properties}).
    }
    \label{fig:property_scatter}
\end{figure}

In cases where intrinsic correlations might be expected, it is instructive
to plot events in property subspaces. Motivated by the jackknife tests in
Appendix~\ref{sec:jackknife}, we show a number of these subspaces in
Figure~\ref{fig:property_scatter}. Note that when viewed in this way, intrinsic 
correlations (as opposed to correlated selection effects) should manifest as
discrepancies between the catalog sample and the intrinsically
uncorrelated fiducial model (which may have selection-induced correlations).
Our ability to measure the properties of catalog bursts complicates this
comparison, although no such correlations are visually obvious. The exception
might be SNR and $\dm$, for which there is a deficit of events at high DM and
high SNR compared to the fiducial model. We study this in more detail in 
Section~\ref{sec:flux_rate}.

\begin{figure}
    \centering
    \includegraphics[width=0.9\textwidth]{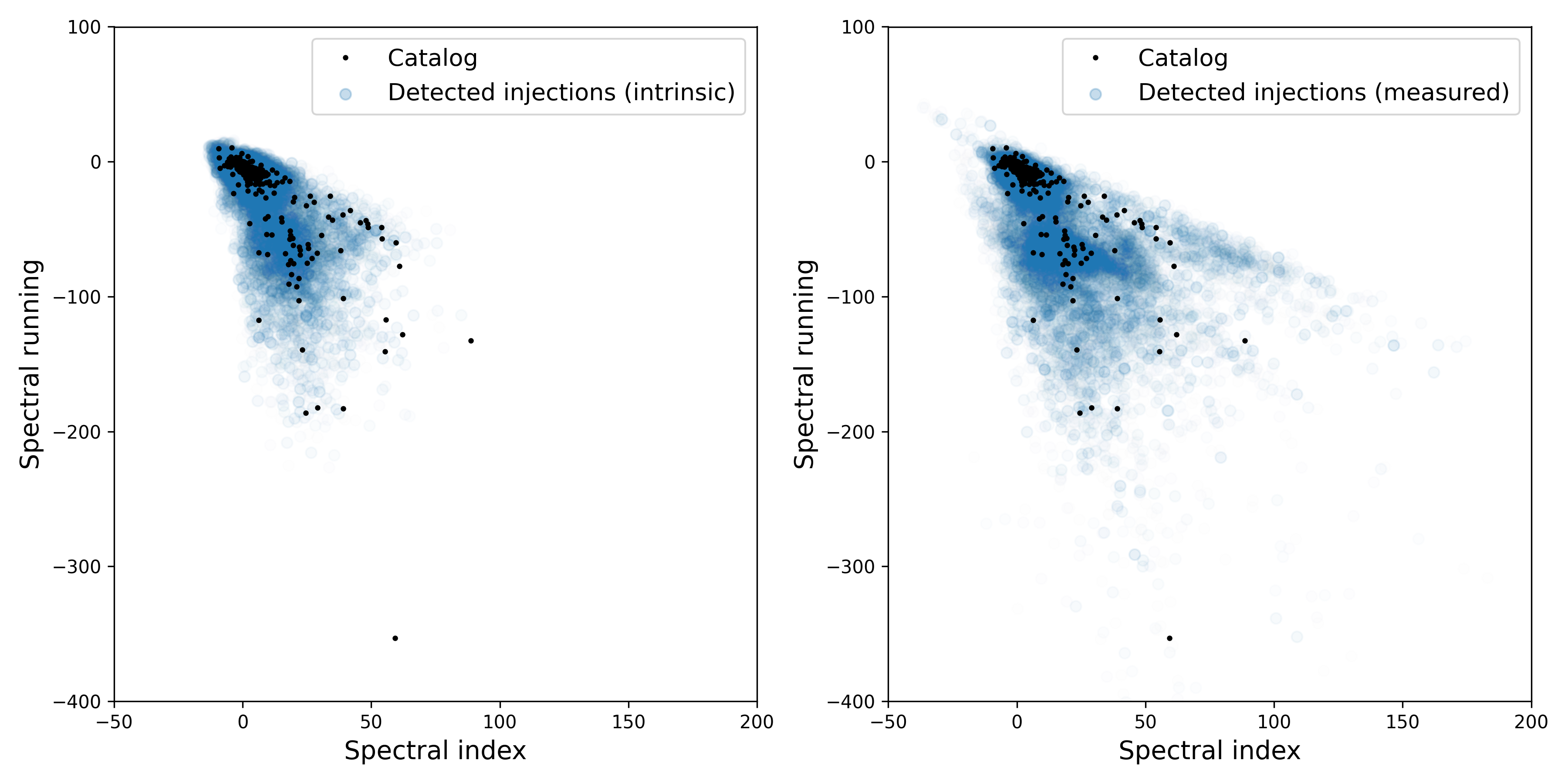}
    \caption{
        {Spectral index ($\gamma$) and spectral running ($r$) for the
        catalog and detected injection events.} Each data point represents a single
        event, with the Catalog~1 data identical in both panels. In the left panel,
        we plot the intrinsic spectral parameters of the injected event, and in
        the right panel we plot recovered spectral parameters after simulating
        the measurement processes using a model for the telescope beam. For
        injected events, the opacity is proportional to the weight derived from
        the fiducial model $W(F, \dm, w, \tau)$. The detected injections sample provides a
        reasonable match to the spectral parameters of the catalog events,
        especially after accounting for the measurement process.
        Bursts with multiple subcomponents from the catalog (which have multiple values of
        $\gamma$ and $r$) are omitted in this comparison.
    }
    \label{fig:spectral_params}
\end{figure}

As mentioned above, the spectral index ($\gamma$) and spectral running ($r$) parameters
are observed to be strongly correlated and due to the observed complexity
in this space, we do not attempt to fit a functional form to $P(\gamma, r)$.
Instead we use kernel density estimation directly on the catalog measurements.
Here, there is potential for a large mismatch between the observations and the
fiducial model once accounting for selection biases using injections.
Indeed, in the left panel of Figure~\ref{fig:spectral_params}, the catalog spans
a larger space of spectral parameters than the detected injections.
Fortuitously, forward modeling the measurement process of the spectral
properties of the detected
injections makes them a better match to the catalog.
Each injected burst has an associated ``intrinsic'' spectral parameter.
After going through the injections system, the fluence spectrum gets modulated by the beam model
and ``measured'' spectral properties $\gamma$ and $r$ are fit. These measured
properties are more directly comparable to the catalog values. The right panel
of Figure~\ref{fig:spectral_params} shows these to be a reasonable match to the
catalog distribution.

\subsection{Fluence distribution and sky rate}
\label{sec:flux_rate}

Here we perform a detailed study of the fluence distribution of FRBs, including
a measurement of the absolute rate on the sky.
We parameterize the fluence distribution
by $\alpha$, the power-law index for the
cumulative distribution such that the number of events occurring above
some fluence threshold is $N(>F)\propto F^\alpha$.
We study our population in the property space of fluence and DM (with SNR as
the observable proxy for fluence).
This is
motivated by the fact that we expect DM to be strongly correlated with distance,
as in the Macquart relation \citep{mpm+20}, which should in turn induce
intrinsic correlations between DM and fluence. Our data are thus the number of
counts $N_{ij}$ in the 2D histogram of the catalog (including 265 events
after cuts) in SNR bins labeled by
index $i$ and DM bins labeled by index $j$.

\begin{figure}
    \centering
    \includegraphics[width=0.48\textwidth]{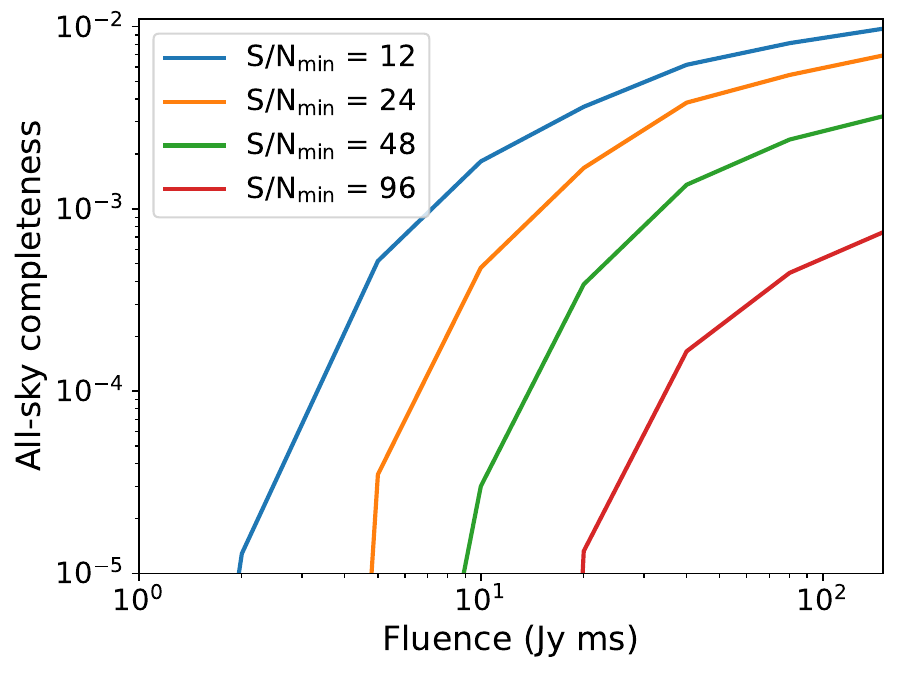}
    \caption{{Injections-calibrated all-sky completeness as a function of fluence and minimum SNR
    included in the sample.}
    Properties other than fluence are drawn from the fiducial property model,
    with the precise equation for the completeness given in
    Equation~\ref{eq:completeness}. Plotted lines are ordered top to bottom in
    the same order as the legend. For high-fluence events, the completeness above
    $\SNR=12$ is of order 0.5\% which can be understood since 
    CHIME/FRB's half-power field of view is roughly 0.3\% of the sky at
    600\,MHz.
    }
    \label{fig:completeness}
\end{figure}

Our first step in modeling the data in this space
is to compute the observation function marginalized over the
other FRB properties using fiducial distributions:
\begin{align}
    P(\SNR|F, \dm) &=
    \int d\tau\,dw\,d\gamma\,dr\,
    P(\SNR | F, \dm, \tau, w, \gamma, r)
    P_{\rm fid}(\tau)
    P_{\rm fid}(w)
    P_{\rm fid}(\gamma, r)
    \\
    &=
    \frac{
        \int d\tau\,dw\,d\gamma\,dr\,
        P(\SNR| F, \dm, \tau, w, \gamma, r) R_{\rm fid}(F, \dm, \tau, w, \gamma, r)
    }{
        \int d\tau\,dw\,d\gamma\,dr\,
        R_{\rm fid}(F, \dm, \tau, w, \gamma, r)
    }.
    \label{eq:obs_func}
\end{align}
In Equation~\ref{eq:obs_func}, both the denominator and the numerator are estimated
from histograms of the
injections and detected injections respectively, using fiducial-model weights.
Note that $P_{\rm fid}(F)$, $P_{\rm fid}(\dm)$, and $R_0$ cancel out in this
expression, although care must be taken to properly account for factors of
$f_{\rm sky}$ and $\epsilon_{\rm inj}$ introduced when generating the injections population (see
Section~\ref{sec:inj_pop}).
As a three-dimensional function, this observation function is difficult to
visualize; however, Figure~\ref{fig:completeness} shows its cumulative version,
integrating out the DM dependence. We term this
the all-sky completeness, given by
\begin{align}
    c(\SNR_{\rm min},F) \equiv \int_{\SNR_{\rm min}}^\infty d\SNR \int
    d\dm\,P(\SNR|F, \dm).
    \label{eq:completeness}
\end{align}

With the required observation function in hand, a prediction can be made
for the DM--SNR distribution of the catalog
\begin{equation}
    \label{eq:p_obs_snr_dm}
    \mathcal{R}_{\rm obs}(\SNR, \dm|\vec{\lambda}) = R_0 \Delta t \int dF P(F,\dm | \vec{\lambda}) P(\SNR|F, \dm),
\end{equation}
for a given model $P(F,\dm|\vec\lambda)$ depending on parameters $\vec\lambda$
(described below). Here, $\Delta t = 214.8$\,days is the survey duration.
Our prediction for the data
$N_{ij}$ is then just this function discretized into finite bins:
\begin{equation}
    \label{e:muij}
    \mu_{ij}(\vec\lambda) = \Delta\SNR_i\, \Delta\dm_j R_0\, \Delta t
    \int dF\,P(F,\dm_j|\vec\lambda)\,P(\SNR_i|F, \dm_j).
\end{equation}
For fitting the model to the data, we use the binned Poisson likelihood
\citep{Zyla:2020zbs}:
\begin{equation}
    \label{e:lik}
    \ln[\mathcal{L}(\vec\lambda)] = -\sum_{ij}\left\{
        \mu_{ij}(\vec\lambda) - N_{ij} + N_{ij}
        \ln\left[\frac{\mu_{ij}(\vec\lambda)}{N_{ij}}\right]
        \right\}
    + \textrm{constant}.
\end{equation}

As a first model, we assume the fluence-DM distribution factorizes into a
power law in $F$ and a free function of DM:
\begin{equation}
     R_0 P(F,\dm_j | \vec{\lambda})\,\Delta t\, \Delta\dm_j =
     -\alpha\left( \frac{F}{F_0} \right)^{\alpha - 1}
     \frac{\eta_j \Delta t}{F_0}.
\label{eq:rate}
\end{equation}
Here, $F_0$ is an arbitrary pivot fluence, and the parameter vector
$\vec\lambda$ contains $\alpha$ and a log rate per DM bin $\ln
\eta_j$. 
In this
parameterization $R_0$ is a derived parameter proportional
to\footnote{Technically, for $\alpha < 0$ the rate in this model 
diverges at low fluence, but we assume a cutoff well below our detection
threshold.}
$\sum_j \eta_j$, the latter of which is the rate of events above the
$F_0$.

We choose logarithmically spaced bins in $\SNR$ and DM with $12$ bins
covering an SNR range of 12 to 200 and $16$ bins covering a DM range of 
100\,pc\,cm$^{-3}$ to 2800\,pc\,cm$^{-3}$, covering the full post-cut Catalog~1
sample. The integral
in Equation~\ref{e:muij} is performed with Riemann sums, with $P(\SNR_i|F,
\dm_j)$ estimated from fiducial-model injections in 100
logarithmic bins
covering 0.2\,Jy\,ms to 50\,Jy\,s. We choose a pivot scale $F_0=5$\,Jy\,ms,
which is substantially higher than the minimum fluence CHIME can detect,
$\sim1$\,Jy\,ms. Choosing this higher pivot scale substantially reduces the
statistical correlations between the inferred rate and $\alpha$.
We employ uniform priors on our parameters
such that the likelihood in Equation~\ref{e:lik} is proportional to the posterior.

To validate the procedure outlined above, we apply it to a suite
of randomly chosen subsamples drawn from detected injected events in place of
Catalog~1 events. In these tests, we use initial distributions instead of fiducial
distributions in Equation~\ref{eq:obs_func} since the scattering, widths, and
spectral parameters of the injections are drawn from these initial
distributions. In all trials we are able to recover the injected fluence
distribution (for which $\alpha=-1$) within statistical uncertainties. This validation
remained true when increasing the sample size to 4000 in order to reduce
statistical uncertainties and search for biases.

In Appendix~\ref{app:rate_alpha_sys} we do an accounting of systematic errors
in this measurement. These final systematic errors are dominated by our rough
estimate of the effect of intrinsic property correlations, as assessed through
jackknife tests in the fiducial model fitting procedure (see Appendix~\ref{app:fid}).
For the rate, uncertainty in the beam model is also
significant. The net systematic uncertainty for the rate is
$^{+27\%}_{-25\%}$ and for $\alpha$ it is
$^{+0.060}_{-0.085}$.

\begin{figure}
    \centering
    \includegraphics[width=0.48\textwidth, trim={5, 0, 10, 0}, clip]{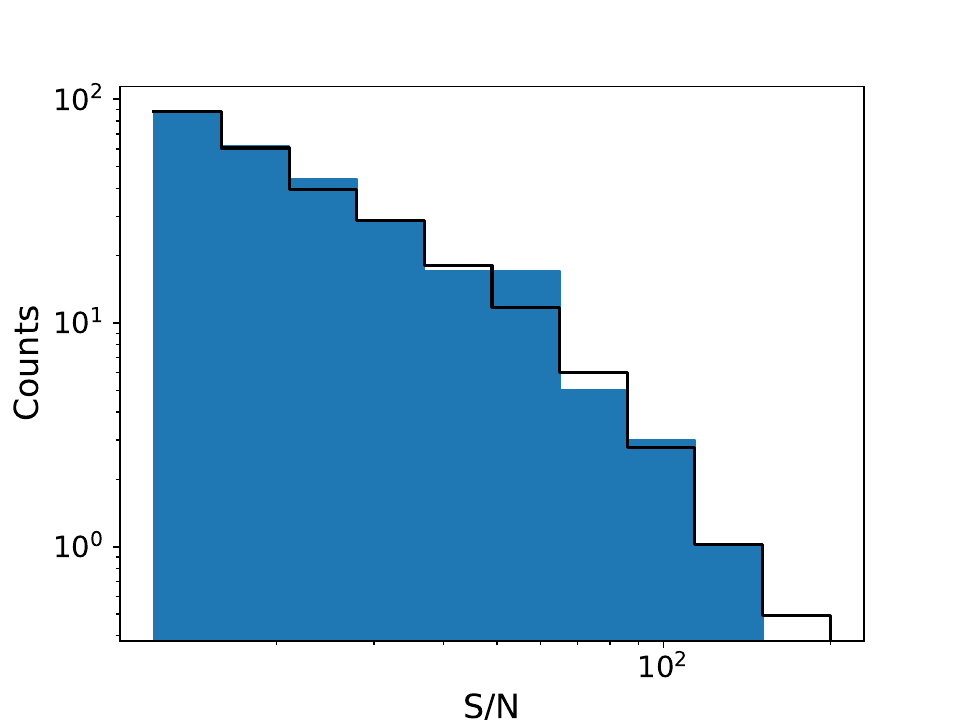}
    \includegraphics[width=0.48\textwidth, trim={5, 0, 10, 0}, clip]{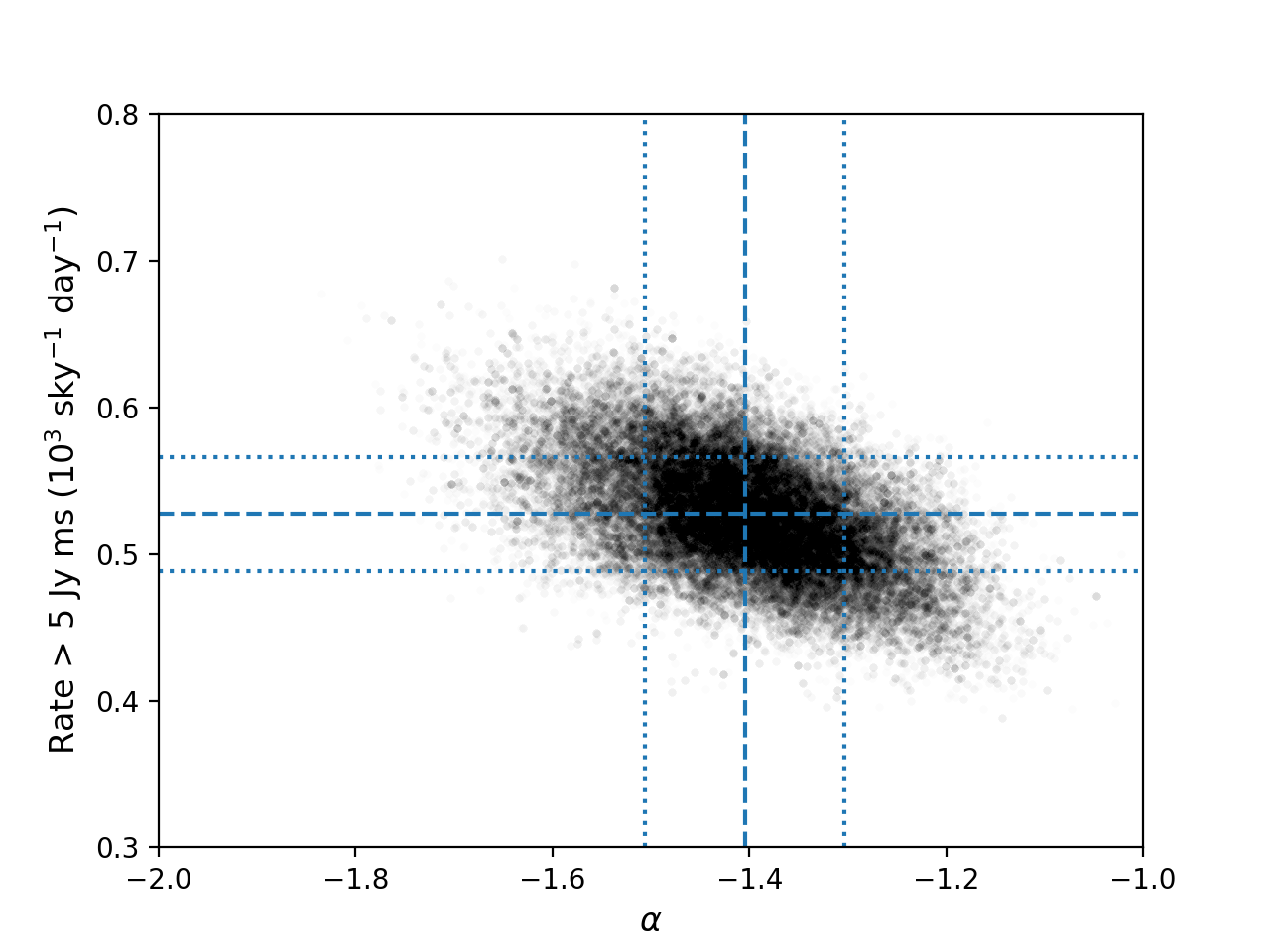}
    \caption{{Fits to the FRB rate and fluence distribution.}
    (Left)
    histogram of the detection SNR and a model fit that includes a DM-dependent
    probabilistic mapping from fluence to SNR calibrated from injections.
    (Right) Markov-Chain Monte Carlo samples for the model-fit posterior
    distribution. Model parameters are the all-sky rate of FRBs with fluence
    above 5\,Jy\,ms and the cumulative fluence distribution index $\alpha$.
    Dashed lines show the mean parameter values and dotted lines span
    $\pm1\sigma$ statistical uncertainty. The distribution includes bursts DM above 
    $100$\,pc~cm$^{-3}$ and with scattering times less than 10\,ms at 600\,MHz.
    Not included in the distributions is a systematic error of
    $^{+27\%}_{-25\%}$
    on the rate, and $^{+0.06}_{-0.09}$ on $\alpha$.
    This constitutes a statistically precise measurement of the FRB rate in the
    CHIME band and at this fluence scale, and indicates the fluence
    distribution is consistent with the Euclidean expectation of $\alpha=-3/2$.}
    \label{fig:alpha_rate_fit}
\end{figure}

Figure~\ref{fig:alpha_rate_fit} shows $N_{ij}$ summed over DM
bins $j$ as well as $\mu_{ij}$ for the best-fit parameters. Markov-Chain Monte
Carlo (MCMC) samples of the posterior, generated using the \texttt{emcee} package
\citep{fhlg13} are also shown. We find $\alpha=-1.40 \pm
0.11({\rm stat.}) ^{+0.06}_{-0.09}({\rm sys.})$, and
the rate $\sum_j \eta_j = [525 \pm 30 ({\rm stat.}) ^{+140} _{-130} ({\rm
sys.})]~{\rm sky}^{-1}~{\rm day}^{-1}$, above a fluence of 5\,Jy\,ms, with a scattering
time at $600$\,MHz less than 10\,ms and above a DM of 100\,pc~cm$^{-3}$.

We note that the resulting value of $\alpha$ is not identical to that fit for
the fiducial model, which contains the same parameter, in Appendix~\ref{app:fid}.
However, the analysis in
this section is more thorough, with notable differences being: 1)~we are using a
different model for the DM distribution; 2)~we are simultaneously fitting DM
and fluence distributions; and 3)~we are sampling the likelihood using MCMC
instead of maximum likelihood. Because of this, differences in best-fit values
of a fraction of a $\sigma$ are expected.

To understand our rate measurement, it is instructive to
determine a rough expectation based on our injections-calibrated completeness
shown in Figure~\ref{fig:completeness}.
This expectation is a strong function of both
the fluence of the burst and the SNR threshold for inclusion in the sample.
We see that even for very
bright bursts with fluence $\sim 100$\,Jy\,ms, our completeness is below 1\%.
This result is not surprising since CHIME/FRB's field of view is about 0.3\% of the
sky.
For our chosen SNR threshold of 12, the completeness is sharply rising at a
fluence of $\sim5$\,Jy\,ms so we are mainly sensitive to bursts above this
level. If we take $c(F = 10\,\textrm{Jy\,ms}) \sim 1\times10^{-3}$ as a
representative value, we obtain a rate of
$N / (\Delta t\,c) \sim 1200\,{\rm sky}^{-1}\,{\rm day}^{-1}$,
above a fluence of 5\,Jy\,ms, close to the value obtained from the full
analysis.

\begin{figure}
    \centering
    \includegraphics[width=0.7\textwidth]{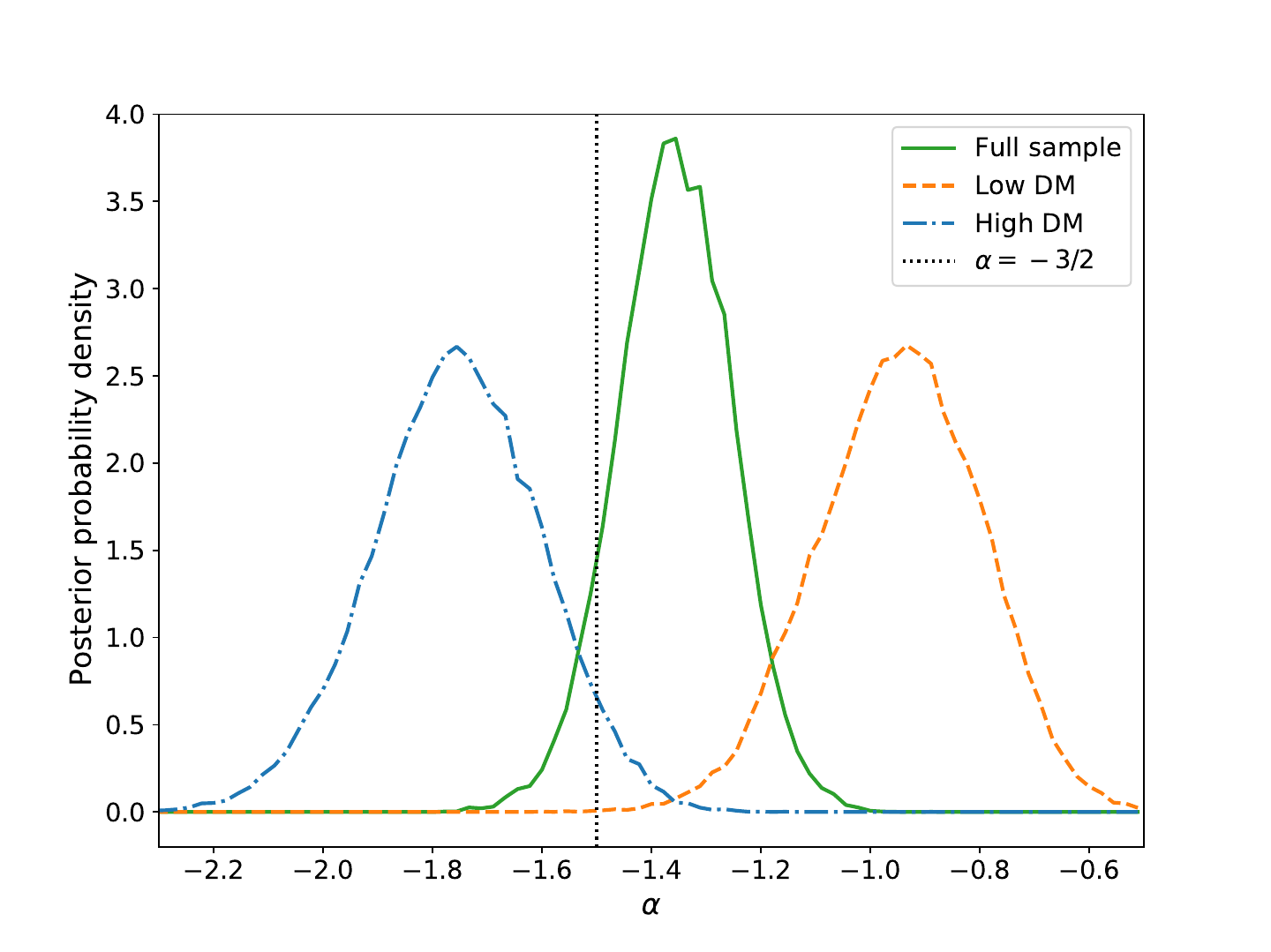}
    \caption{{Marginalized posterior distributions for the fluence
    distribution index $\alpha$ for different ranges of DM.}  The low DM range is $100$ to
    $500$\,pc~cm$^{-3}$ and the high DM range is above $500$\,pc~cm$^{-3}$.
    Not included is a 
    systematic uncertainty of $^{+0.030}_{-0.085}$\,pc~cm$^{-3}$, which is shared amongst DM ranges.
    When considering
    nearly the full range of DM detectable by CHIME/FRB, the fluence distribution is
    consistent with being Euclidean, as expected for a population dominated by source
    redshifts less than $z=1$. However, $\alpha$ is significantly steeper for
    high-DM events than for low-DM events. This is consistent with the
    expectation that high-DM events are more distant and we thus
    preferentially sample the high end of the luminosity function, which, if
    well behaved, must be steeper than the low end (see
    Section~\ref{sec:flu_dist}).
    }
    \label{fig:alpha_posterior}
\end{figure}

To search for distance-scale induced correlations between fluence and DM
(as suggested by our catalog events prior to correction for selection
biases---see Section~\ref{sec:strengthcomp}), we
subdivide our sample by DM, splitting at $\dm=500$\,pc~cm$^{-3}$, which is close to
the median value of our catalog after cuts. As shown in
Figure~\ref{fig:alpha_posterior}, the low-DM sample has 
$\alpha=-0.95 \pm 0.15({\rm stat.}) ^{+0.06}_{-0.09}({\rm sys.})$
whereas the high-DM sample has
$\alpha=-1.76 \pm 0.15({\rm stat.}) ^{+0.06}_{-0.09}({\rm sys.})$.
Noting that the systematic error should be mostly common between the two
samples, the difference between them is significant at the 3.8$\sigma$ level.

\begin{figure}
    \centering
    \includegraphics[width=0.95\textwidth, trim={0, 0, 0, 0}]{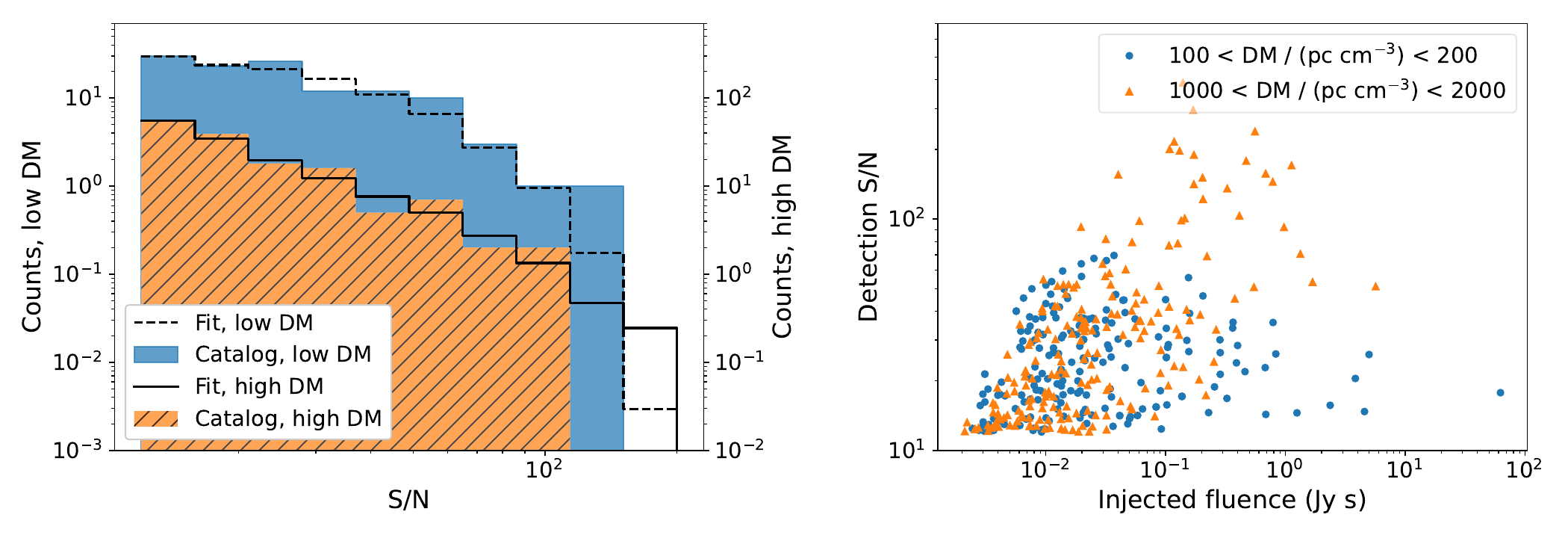}
    \caption{{Data and selection effects leading to the fluence
    distribution measurements in the high and low DM ranges.} (Left)
    Histogram of the detection SNR and a model fit that includes selection
    effects, as in the left panel of Figure~\ref{fig:alpha_rate_fit}, and with
    vertical scales offset for clarity.
    The low DM range is $100$--$500$\,pc~cm$^{-3}$ and the high DM range is $>500$\,pc~cm$^{-3}$. The
    fit model for the low-DM events appears to be steeper than that for the
    high-DM events, despite being derived from a fit value of $\alpha\sim-1$
    compared to $\alpha\sim-2$. The apparent inconsistency is due to a DM-dependent
    probabilistic mapping from fluence to SNR which we have calibrated from injections.
    This is illustrated in the right panel, which shows the SNR and fluence of 200
    injected and
    subsequently detected events in each of two DM ranges, drawn from otherwise
    identical property distributions. It can be seen that our detection system has a shallower
    mapping from fluence to SNR for low-DM events compared to high-DM events,
    with very few low-DM events achieving an SNR above 100.
    This latter effect is likely due to our wideband RFI mitigation strategies, which preferentially
    remove signal from bright, low-DM events.
    }
    \label{fig:alpha_hi_lo_dm}
\end{figure}

Properly accounting for the observation function $P(\SNR|F, \dm)$ is critical
to making this measurement, and shown in Figure~\ref{fig:alpha_hi_lo_dm}. It
can be seen that the SNR distribution of the catalog events is not strikingly
different for the low-DM and high-DM sample. However, 
the 
low-DM events have a shallower mapping from fluence to SNR. This is somewhat
expected since our wideband RFI mitigation clips significant signal from very bright
events, an effect which is substantially stronger at low DM.

\section{Discussion}
\label{sec:discussion}

Our first CHIME/FRB catalog presents 536 FRB events, 
detected over a 371-day period, of which 492 are unique sources, and 18 are repeating sources.
This is
the largest catalog of FRBs detected by a single instrument, allowing their characterization in the context of a single set of carefully studied selection biases. 
We have measured burst properties in a systematic, uniform way.  As such, the catalog represents a unique resource for studying the FRB population, such as
statistical comparisons between repeaters and apparent non-repeaters, and the determination of the fluence distribution and rate of FRBs, as presented in this work.
Additional analyses of the catalog data are ongoing, including a study of repetition statistics, as well as a volumetric rate analysis.  We also have archived complex voltage data on many of the FRBs in this catalog; the analysis of these data is ongoing and beyond the scope of this paper, but will permit polarimetry, high time- and frequency-resolution studies of burst morphology, as well as improved localizations for the relevant bursts.

In what follows, we discuss the main results of this paper, namely the repeater versus apparent non-repeater comparison, the intrinsic DM, width and scattering distributions, as well as the fluence distribution and rate.

\subsection{Are repeaters a different FRB population?}

In Section~\ref{sec:repvsnonrep}, we compared distributions of burst properties for repeaters and apparent non-repeaters to determine whether the two groups represent different astrophysical source populations.

In terms of sky distribution, we found no significant difference in right ascension or declination distributions.
The latter is perhaps more interesting, given the strong CHIME exposure dependence on declination.  Indeed, we find no evidence for a declination-dependence of the first-repeaters to non-repeaters ratio histogram.  This
exposure.  Eventually, we may reach the regime wherein additional exposure no longer results in as many repeating source detections 
because the majority of the detectable repeaters will have been found, with the brightest and most active repeaters having been found first.  We do not yet seem to be in that regime in Catalog~1.
Although we have already reported on 18 new repeater discoveries \citep{abb+19b,abb+19c,fab+20}, we have in fact detected
more and will report on these elsewhere, along with a detailed analysis of the distribution of burst rates.

The DM distributions of repeaters and apparent non-repeaters are consistent
with originating from the same underlying distribution. Roughly, with 18
repeating sources, we would expect to be able to detect $\sim 1/\sqrt{18} \sim
25\%$ differences in the mean DM between the samples (at 1-$\sigma$).

If extragalactic DMs are dominated by plasma in the intergalactic medium \citep[e.g.,][]{mpm+20}, then our results suggest no difference between the distribution or host type of repeaters and non-repeaters. On the other hand, if the extragalactic DMs are dominated by the host's ISM, or its local environment, then the repeaters and non-repeaters must share very similar host properties. In this case, the results from our sample would indicate that any disparities in hosts must coincidentally conspire to yield no significant net different in extragalactic DM distribution between the two types of repeater.

Furthermore, we find no strong evidence for differences in signal strength of repeaters and apparent non-repeaters, nor in scattering properties.
In principal it would be instructive to measure $\alpha$ for the two
populations separately. However, the spectro-temporal differences between the
two populations imply there could be differences in observation biases,
making absolute measurements of $\alpha$ challenging for the sub-populations. Also, the repeater sample is still relatively small for a meaningful analysis. In any case,
the similarity of the SNR distributions implies there is unlikely to be a
statistically significant difference in $\alpha$.

On the other hand, we find strong differences in burst widths and bandwidths,
with repeaters having on average significantly broader widths and narrower
bandwidths, at least in the CHIME band.  The differences are not subtle; they
are apparent by eye (see Figs.~\ref{fig:repcomp_width} and \ref{fig:repcomp_bwscatt}).  They are investigated in more detail by Pleunis et
al. (2021, submitted).
These differences are strongly suggestive of differing
emission mechanisms. This could imply either different source populations, or a single population in which pulse morphology strongly correlates with
repeat rate \citep[e.g.,][]{cmg20}.

Different source populations can have identical spatial and local environment properties. For example, some FRBs may have massive stellar progenitors \citep[as in models requiring isolated neutron stars such as magnetars, e.g.,][]{bel17,mm18}, while others may manifest as FRBs due to interactions with a massive companion star \citep[e.g.,][]{iz20,lbg20}, with both populations found preferentially in regions of star formation within young galaxies. Harder to imagine are examples with one population requiring nearby AGNs \citep[e.g.,][]{2017ApJ...844..162T,vrb+17} and another not (most models); such scenarios seem unlikely, given our results, and also given the absence of FRBs near centers of galaxies \citep{bsp+20}.  Distinct populations in which one is very young \citep[e.g., highly active magnetars;][]{bel17,mm18} and the other very old \citep[e.g., colliding compact objects;][]{ytk20,bel20} also seem unlikely for identical spatial and local environment properties.
However, the DMs and scattering times of both classes could be heavily dominated by the IGM, and the difference in typical host galaxy type may not be not large.

Conversely, a single population can have sources of vastly different observational emission properties.
For example, the Sun itself produces many different types of radio bursts \citep[see, e.g.,][]{kah92}. 
Perhaps more energetically relevant, radio pulsars exhibit a variety of radio pulse phenomenology, ranging from mode changing and nulling to giant pulses \citep[see, e.g.,][]{mt77,lk12}.
 Emission properties in neutron stars can vary with age as well, with young magnetars being highly X-ray and gamma-ray active, but perhaps subsequently evolving to more stable, fainter ``Isolated Neutron Stars'' \citep{kas10}, though in this case, differing local environments would be expected.

Thus, although we have found strong evidence for differing types of emission when comparing repeaters and apparent non-repeaters (see Pleunis et al. 2021, submitted, for further details), it is merely suggestive, but not proof, of different source populations, particularly noting the otherwise similar property distributions.

\subsection{Accounting for selection biases}

Through an extensive program of signal injections (see Section~\ref{sec:injection}), we have characterized the
selection effects in our FRB survey, as described in Section~\ref{sec:cordists}. So far, we have only injected events using a relatively simple signal model. As such, 
events with complex pulse shapes or spectral structure may not be adequately
characterized. In addition, event properties may be correlated, either due to
intrinsic correlations in the population or correlated selection effects. We have mitigated the
latter by matching to a fiducial
model, such that the observed
statistics of the injected sample are a good match to those of the catalog.
However, we have explicitly ignored most intrinsic
correlations.
Another area for concern is that we have
assumed a log-normal functional form for the DM, width, and scattering
distributions, and only roughly modelled the distributions of the spectral
parameters using a kernel density estimator.
All of these assumptions are likely
highly simplistic compared to reality.

Nonetheless, bursts with complex structure represent a small minority of events, and we believe
ignoring correlations in the populations to be a reasonable approximation when
estimating the selection function.
In addition, our best-fit fiducial model is reasonably well matched to the
catalog data as shown in the figures presented in Section~\ref{sec:cordists}.
This indicates our simplistic model provides a decent description at the
level of our current statistical precision.

As such, we are able to
draw conclusions about the FRB population with unprecedented statistical
precision and control of systematics.
Further improvements to the multi-dimensional property-space
modeling of the FRB population should be a focus of future work in the
field as the data continue to become more constraining.

The methods for accounting for selection bias through Monte Carlo-style real-time injections
described here are relatively new.
\citet{gff+21} have implemented an injections system analogous to that used
here, although they do not attempt to model the effects of the telescope beam and
have not yet propagated their measured selection effects to detailed population
inferences. The most complete treatment of FRB selection effects to date was
performed by \citet{jpm+21a, jpm+21b}, who follow an equivalent statistical
formalism to that employed here but construct a model for the observation
function rather than calibrating it through injections.
A failure to account for selection effects will, for any
sizable FRB sample, result in incorrect assertions about population
distributions. As such, our methods are an important outcome of this work,
together with the data presented.

\subsection{Intrinsic DM, width, and scattering distributions}
\label{sec:disc_intrin_dist}

From Figure~\ref{fig:dm_dist}, it is clear that the selection effects in DM are
modest and that, at least naively, we appear to be detecting the full range of DMs
represented in the population detectable at CHIME/FRB's sensitivity. On one hand, a log-normal distribution peaking at
$\sim500$\,pc~cm$^{-3}$ with tails extending to $\sim3000$\,pc~cm$^{-3}$ is a good
fit; on the other hand, this should be interpreted in the context of our assumption of
negligible intrinsic property correlations. The observed DM distribution could
be skewed by a correlation between scattering and DM (which is physically well-motivated) or fluence and DM (which we have demonstrated to be present). The
latter is of particular concern, since at a lower fluence scale we expect to
detect more distant FRBs, having higher DMs.

Interpreting the DM distribution with respect to the Macquart relation \citep{mpm+20} taken at
face value, roughly half of CHIME/FRB-detected FRB sources
have a redshift less than
0.5
with a tail extending to $\sim2$. However, it is possible that the high-DM tail is
dominated by host/source-local plasma and that the maximum redshift probed by our data is
considerably lower.
Regardless, what is clear is that CHIME/FRB is not detecting many $z>3$
sources so may not be helpful for studies
requiring such objects \citep[e.g.,][]{bkmq20}.  Future sensitive, higher-frequency
telescopes like the Square Kilometer Array\footnote{\url{skatelescope.org}}
or the
CHORD telescope \citep{vlg+19} may be useful in this regard.

Unsurprisingly,
selection effects in the distribution of
intrinsic widths are strong: 
temporally broad events have a more diluted signal, and many of our RFI mitigation strategies have the effect of
filtering out signals with long time-scales from our data.
As such, the median width of the population increases by a
factor of two once selection effects are accounted for.
Nonetheless, even accounting for selection effects, the rate of events with
width in the 10 to 20\,ms range is small compared to those below 10\,ms, and
appears to be falling as width increases further
(although statistical errors are large in this
highly selection-attenuated region). As such, it seems unlikely that there is a large number of FRBs that are undetectable due to large intrinsic width.
We urge some caution
when making interpretations here,
as intrinsic width is the parameter that is most likely to be affected by the limitations of an injection campaign that used only simple burst morphologies.
Our inferences about the width distribution are particularly dependent on our
assumption that it is uncorrelated with fluence, rather than uncorrelated with intrinsic peak
flux (which is proportional to $F/w$). Either assumption is astrophysically well-motivated, and the choice of one over the other depends further on whether FRB emission is an energy-limited or
time-limited process. This in itself is an important question that our
data should be able to address; however, we defer such an investigation to future work.

In contrast, correcting for selection effects in scattering indicates that
there is a substantial population of FRBs with very high scattering that are
challenging for CHIME/FRB to observe. In particular, we detected two
events\footnote{The catalog contains a third event that did not make the cuts
for population inference.} with
scattering time $>50$\,ms and our injections indicate that for these to have
been observed, a huge number of highly scattered events must have gone
undetected. Indeed,
in
Figure~\ref{fig:scat_dist}
we do not see much
evidence that the event rate is falling in the $10$ to $100$\,ms range, and
there could be a large population beyond $100$\,ms which is essentially
unconstrained by our data
due to the difficulty of detecting these events.
A population synthesis analysis based on the first 13 CHIME-detected FRBs,
all of which exhibited scattering timescales $<$10 ms, suggested that FRBs
must be located in environments with stronger scattering properties than the
Milky Way ISM \citep{abb+19a}. We are performing a similar analysis
to explore the astrophysical implications of the existence of a large
population of highly scattered events, results of which will be presented in future work.

Highly scattered events are easier to
detect at higher frequencies due to the steep power-law dependence of
scattering time on observing frequency. Scattering timescales have been
measured for 18 of the 71 FRBs observed at gigahertz
frequencies\footnote{As reported as of 2021 Jan 2 at \url{https://www.wis-tns.org}}. 
Eight of these
FRBs have measured scattering times which scale to $>$ 100 ms at 600 MHz,
assuming a power-law index of $-$4 for the frequency dependence. Observations of FRBs at frequencies above 1 GHz are thus consistent with the existence
of a large population of highly scattered events. We
note that the observed number highly-scattered events
could also be the result of intrinsic
correlations in the population. For instance, if there were a strong
correlation between fluence and scattering, we might observe these
highly-scattered events
without their population being particularly large.
However, astrophysically,
such a correlation seems unlikely since there is no particular reason fluence, an intrinsic property,
should be related to scattering, an extrinsic propagation effect.

A correlation between scattering time and extragalactic DM might be
expected, as discussed in Section~\ref{sec:repvsnonrep}, and would contradict
our assumption of independence of these variables (see Section~\ref{sec:abs_pop}).
We note that such a correlation is present in the catalog events prior
to correction for instrumental biases, though it does not appear strong.
As discussed in Appendix~\ref{sec:jackknife}, our jackknife tests also hint
at a correlation between DM and scattering time after compensating for
selection effects.  The
investigation of the degree of correlation after correction for instrumental
biases is beyond the scope of this paper, but will be discussed in future work.

\subsection{Fluence distribution}
\label{sec:flu_dist}

When considering a wide range of DMs, we find the fluence
distribution index to be
$\alpha=-1.40 \pm 0.11({\rm stat.}) ^{+0.06}_{-0.09}({\rm sys.})$,
which is an excellent match to
the expected value of $-3/2$ for a non-evolving, constant-density source population in
Euclidean space \citep{her1785}.
This agreement is expected, because the
peak of our DM distribution is $500$\,pc~cm$^{-3}$, implying a redshift
distribution that peaks at
$z\lesssim0.5$. Cosmic evolution of the population, as well as effects from the
expansion of the universe, are not expected to result in a significant deviation
from the Euclidean expectations 
at these moderate redshifts.

There have been a number of measurements of the FRB fluence distribution
\citep{ocp16, vrhs16, abb+17, lvl+17, smb+18, bkb+18, pab+18},
the most recent of which have been consistent with the $\alpha=-3/2$, with some
exceptions \citep{als+20, jem+19}. \citet{als+20}
analyzed a heterogeneous set of surveys, with most data coming from the
Parkes \citep[e.g.,][]{tsb+13, cpk+16, bkb+18}
and Australian Square Kilometre Array Pathfinder (ASKAP) Telescopes
\citep{smb+18}. They found $\alpha=-0.91 \pm 0.34$, a nearly
2$\sigma$ disagreement with the Euclidean expectation. A central challenge of
meta-analyses of this type is the comparison of samples from different
instruments and surveys. In particular, the effects of beam-model errors tend to cancel in
the measurement of $\alpha$ so long as the sample is detected through a single
beam \citep{con19}.
\citet{jem+19} separately
analyzed samples detected at Parkes (finding $\alpha = -1.18\pm 0.24$) and ASKAP
(finding  $\alpha = -2.20\pm 0.47$) and, while 
each of these measurements is apparently $\sim$
1.5$\sigma$ consistent with Euclidean,
they asserted that the two populations have different $\alpha$ at the 2.6$\sigma$ level
(after accounting for the
non-Gaussianity of the likelihoods).

Recently, \citet{lp19} analysed an ASKAP-detected sample, modeling the FRB energy distribution
function, redshift evolution of the volumetric rate, and assuming a one-to-one
DM-distance relation. In such models, $\alpha$ is a
derived quantity that cannot deviate strongly from the Euclidean value except in
extreme regions of parameter space or for populations much more distant than the
ASKAP sample. Thus, their analysis is not completely
comparable to direct fits for $\alpha$, although does better incorporate the
astrophysical effects.
However, they do find their model to be consistent with the ASKAP fluences once
they account for completeness.
Likewise \citet{jpm+21a, jpm+21b} jointly analysed the Parkes and
ASKAP datasets, modeling selection effects, the FRB energy distribution
function, and a stochastic DM-distance relation simultaneously. Their best-fit model
predicts $\alpha = -1.5$ over most of the observed fluence range,
shifting to $-1.3$ for the dimmest (and thus most distant) bursts detectable by Parkes.

Because CHIME/FRB observes at significantly
lower frequencies than the 
1.4-GHz surveys, it is non-trivial to compare
the fluence-scales of
the populations seen at Parkes and ASKAP to that of CHIME/FRB. 
Nonetheless, given that
the median DM of the Parkes sample is $\sim900$~pc~cm$^{-3}$ compared to
$\sim500$~pc~cm$^{-3}$ for CHIME/FRB, it is plausible that Parkes is seeing a more distant
population than is CHIME/FRB, and is thus seeing cosmological and/or evolutionary
effects that flatten the fluence distribution. Indeed, this is the interpretation
given by \citet{jem+19} and \citet{jpm+21a, jpm+21b}. 
However, ASKAP is significantly less sensitive than
Parkes and sees a sample with median DM $\sim400$~pc~cm$^{-3}$,
and yet the \citet{jem+19} measurement of $\alpha$ is apparently more discrepant from $-3/2$.
Thus, strong
departures from the Euclidean value seem difficult to explain for that sample.
We note that the $\alpha=-3/2$ expectation holds only
after aggregating over FRBs with all DM values
and for samples that are complete in DM. We have shown that the
DM-completeness of our catalog increases by more than a factor of 2 between
$100$ and $1000$\,pc~cm$^{-3}$, an effect for which we have compensated.
In addition, the non-linear and stochastic mapping between
fluence and SNR had to be carefully calibrated for our measurement.
The other analyses listed above have not compensated for either effect and it is unknown
whether these effects are strong for the search pipelines at Parkes and ASKAP, although
\citet{jpm+21b} assert them to be negligible.

Splitting the sample by DM, we find that the CHIME/FRB low-DM sample has a significantly shallower
slope with $\alpha\approx -1$ compared to the high-DM sample with
$\alpha\approx -1.8$. We argue that this too is qualitatively what we would have
expected.
To understand this, consider the model where DM is exactly proportional to distance.
Then, the energy of each FRB is
$E = C\,\dm^2\,F$ where $C$ is the constant of proportionality.
Thus, at
fixed DM the joint fluence--DM distribution function, $P(F, \dm)$,
is directly proportional to the FRB energy distribution
function, $n_E(E)$:
\begin{equation}
    P(F, \dm) \propto \dm^4\,n_E( C\,\dm^2\,F),
    \label{e:dm-F-static}
\end{equation}
with the DM-dependent prefactor $\propto \dm^4$ coming from the
geometry and the change of variables. In order for the total energy
output of FRBs ($\sim \int_0^\infty dE\,E\,n_E(E)$)
to be finite, the energy distribution must fall more steeply than $E^{-2}$ at high
energy and be more shallow than $E^{-2}$ at low energy.
Integrating $P(F, \dm)$ over DM yields the expected $-3/2$ power law;
however, when considering only low-DM events, we preferentially sample the
lower end and
shallower part of the energy distribution, yielding $\alpha > -3/2$.
Likewise for high-DM events, the higher end and steeper part of the energy
distribution is preferentially sampled and we obtain $\alpha > -3/2$.
Thus, that we observe this expected behavior
indicates that we are
seeing the distance evolution of the FRB population, sampling different parts
of the energy distribution function at different DMs. Note that this
interpretation is not necessarily at odds with the interpretation above that
Parkes may observe a shallower brightness distribution because it is seeing a
more distant, higher-DM population. The Parkes sample is at a different DM
\emph{and fluence} scale than our sample.

Our findings are qualitatively consistent with those of \cite{smb+18}, who
found that the Commensal Real-time ASKAP Fast Transients Survey (CRAFT, which is shallower
and wider compared to surveys using the Parkes Telescope) observed a population
of FRBs with comparatively lower DMs. The median DM for the CRAFT sample was
roughly
400\,pc~cm$^{-3}$ compared to 900\,pc~cm$^{-3}$ for the Parkes sample, which is
a factor of 50 more sensitive than CRAFT. 
It is promising that we can now detect the the DM dependence of the fluence distribution
without the complication of comparing heterogeneous
surveys.

In principle, we should be able to measure the full energy distribution
function: Equation~\ref{e:dm-F-static} contains an unknown single variate
function $n_E(E)$ from which we derive a bivariate observable $P(F, \dm)$. In
practice the fact that DM is a noisy proxy for distance---with a degree of
noisiness that has yet to be well established---makes this measurement
non-trivial. What we have presented here represents a first step along this
path, and we are actively pursuing a more complete analysis.

\subsection{FRB rate}

In this section we
compare our measured sky rate to others published in the literature.  
We note, however, that we have determined our rate using methods that are quite
different from how other rates in literature were estimated, specifically in
our forward-modeling of the multi-dimensional selection function (determined using 
injected bursts; see Section~\ref{sec:injection}), including our complex beam response.  Direct comparison with other reported rates is therefore dangerous, 
since other rate measurements did not account for 
instrumental effects with the same methods.
Nevertheless we proceed with such a comparison, first considering the implications of a simple, naive comparison of published rates, but ultimately recognizing that rate disparities may be a result of different measurement methods and can guide future work on the subject.

As our survey is uniquely in the 400--800-MHz band, we first consider what
average spectral index, $\bar\gamma$, is reasonable to assume when comparing
with rates at other frequencies, 
where, after accounting for the fluence distribution, the rate scales
as
\begin{equation}
R_5^2 = R_5^1 \left( \frac{f_2}{f_1} \right)^{-\bar\gamma\alpha},
\end{equation}
where $R_5^n$ is the rate above 5\,Jy\,ms at radio frequency $f_n$.
\citet{ckj+17} constrained the spectral index
to be $\bar\gamma > -0.3$, fairly flat, using the lack of detections in the
300--400-MHz band from the Green Bank North Celestial Cap (GBNCC) survey and the 1.4-GHz rate from \citet{crt+16}.  This constraint was obtained assuming scattering is important from sources other than the Milky Way and IGM, which seems consistent with our 
preliminary simulation analyses, to be described in future work.
Note that the GBNCC spectral index constraint was not altered by the
recent detection of GBNCC's first FRB
\citep{pck+20}.
In contrast, \citet{msb+19} report a much steeper spectral index, $\bar\gamma = -1.5^{+0.2}_{-0.3}$, based on spectral analysis of 23 ASKAP bursts detected at 1.4 GHz.  However,
\citet{ffb+19} argue that either ASKAP-derived spectral index is too steep given
the low inferred rate from the UTMOST telescope, which operates at 843 MHz, or
perhaps there is a spectral turnover below $\sim$1 GHz.  Furthermore, non-detections of bright ASPA FRBs at the 
Murchison Widefield Array (MWA) yielded a constraint $\bar\gamma \gapp -1$
\citep{sbm+18}, also somewhat flatter than the ASKAP value.
Here we will start by assuming a simple flat spectral index ($\bar\gamma = 0$) in the absence of strong evidence otherwise.

Past rate measurements have usually been in the 1.4-GHz band, dominated by the Parkes and ASKAP telescope samples.  Here, we consider the most recent measurements only, as early values were based on low statistics.  Specifically,
\citet{bkb+18} report a sky rate
of $1.7^{+1.5}_{-0.9} \times 10^3$~sky$^{-1}$~day$^{-1}$ above a fluence limit of 2~Jy~ms based on Parkes 1.4-GHz FRB surveys.
\citet{smb+18} report a rate of
$37 \pm 8$~sky$^{-1}$~day$^{-1}$ above
26~Jy~ms from ASKAP FRB surveying at 1.4 GHz.
However, \citet{ffb+19} report a rate
of $98^{+59}_{-39}$~sky$^{-1}$~day$^{-1}$  above a fluence of 8 Jy~ms at 843 MHz from six FRB events detected with Molonglo/UTMOST, a factor of $\sim$7 below the Parkes and ASKAP rates. 
More recently, \citet{pck+20}
report $3.4^{+15.4}_{-3.3} \times 10^3$~sky$^{-1}$~day$^{-1}$ above a flux of 0.42 Jy for a 5-ms pulse,
equivalent to a fluence limit of $\sim$2~Jy~ms in the 300--400-MHz range, from GBNCC.

\begin{figure}
    \centering
    \includegraphics[width=0.90\textwidth, trim={0, 0, 0, 0}]{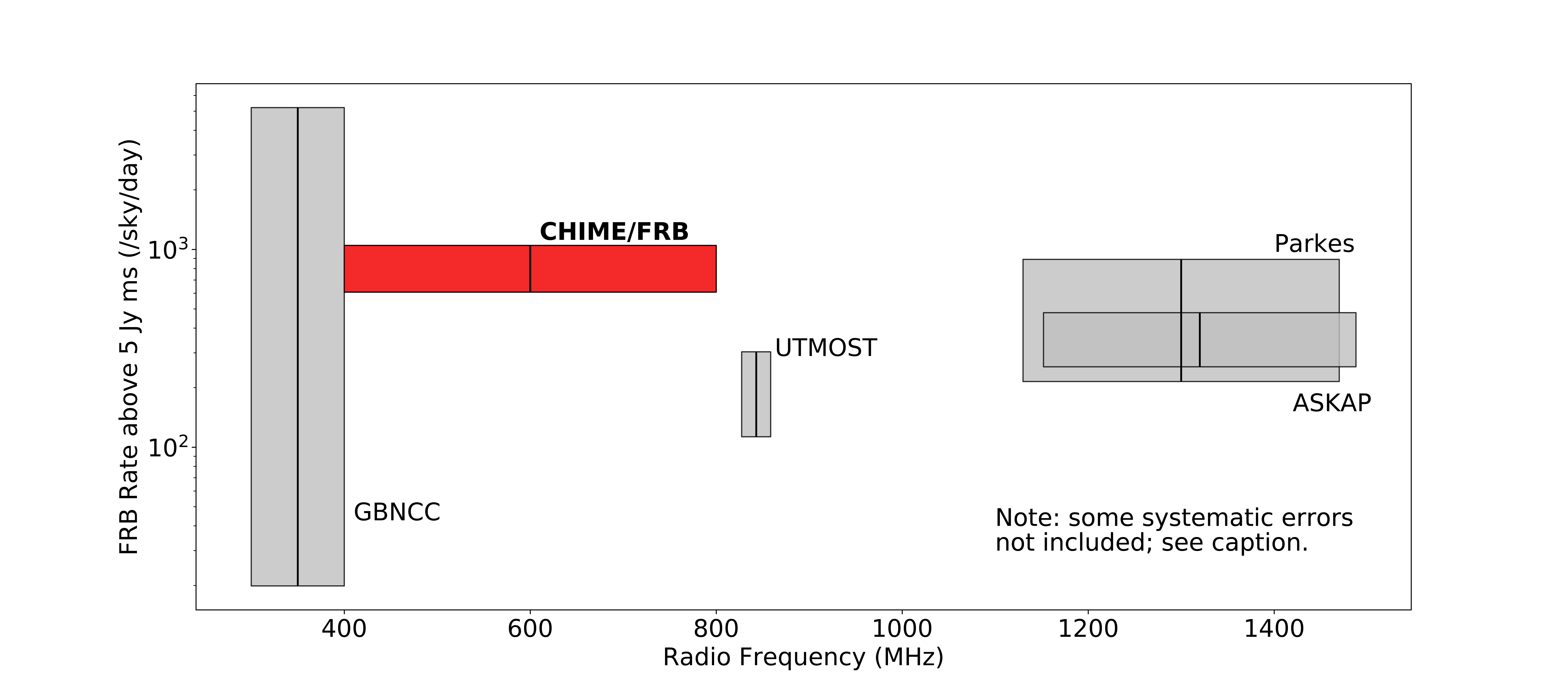}
    \caption{Our new CHIME/FRB rate, along with other recent published rate estimates,
    all scaled to a fluence limit of 5~Jy~ms using our value of 
    $\alpha = -1.41$,
    Equation~\ref{eq:rate}, and assuming a flat spectral index. Uncertainties
    on the scaled rate account for uncertainties on the pre-scaled rate
    and on $\alpha$, including both statistical and systematic values as applicable.
    We caution, however, that these data, originating from different instruments and processed with different techniques, are not directly comparable. Uncertainties in the frequency distribution, rate versus bandwidth and its evolution with frequency, choice of fluence cutoff, fluence distribution index versus DM, and other effects have not been accounted for in this simple comparison. 
    The apparent discrepancies in measured all-sky rates presented in this figure may fall within the systematic differences arising from these effects.
    }
    \label{fig:ratecomp}
\end{figure}

All these rates, along with the CHIME/FRB Catalog~1 rate of 
$525 \pm 30 ^{+140}_{-130}$~sky$^{-1}$~day$^{-1}$ at 600 MHz, are shown in
Figure~\ref{fig:ratecomp}, scaled (if needed) to a fluence limit of 5~Jy~ms using our measured fluence index 
$\alpha=-1.40 \pm 0.11({\rm stat.}) ^{+0.06}_{-0.09}({\rm sys.})$
and Equation~\ref{eq:rate}, and plotted assuming a flat spectral index.
The CHIME/FRB and GBNCC rates are consistent, though the uncertainty on the latter is large.
The CHIME/FRB and UTMOST rates appear to be in mild tension, in spite of the
latter's band being close to the high end of CHIME's. However, this is likely
attributable to differences in the treatments of selection effects.

Most interestingly, the CHIME/FRB rate is naively consistent within
uncertainties with those of Parkes, and ASKAP\footnote{However,
\citet{lp19} argue the ASKAP completeness threshold is 50~Jy~ms rather than
26~Jy~ms, which does not affect the consistency.}.  This supports our assumption of a flat spectral index, and argues
against a spectral turnover below 1 GHz as suggested by \citet{ffb+19}.
However, such a conclusion ignores the possible influence of a large, highly scattered population undetected by CHIME/FRB; for 
$\bar\gamma = 0$, the proximity of the
CHIME/FRB and 1.4-GHz rates suggests such
a population is small, 
as otherwise the 1.4-GHz rate would be higher than that at 600 MHz.
If such a population is large, the spectral index is likely steeper than inferred above.  In this case, 
it would have to be a coincidence that the effects of scattering and spectral index nearly cancel each other out in the relevant frequency range.
As emphasized above, a detailed and conclusive
comparison requires more uniform consideration of selection biases for the various measurements, as well as additional forward modeling to account for yet unmodeled population properties such as bandwidth and frequency distributions.

\section{Conclusions}
\label{sec:conclusions}

We have presented the largest ever sample of FRBs, increasing the
public sample of FRBs by nearly a factor of four. For each burst, we measure
detailed pulse properties, including parameters from a pulse-model fit. This
provides a large sample of bursts, from both repeating and so-far non-repeating
sources, with homogeneous detection pipelines,
selection biases, and property measurements.

The sample has also enabled a direct comparison of bursts from repeating sources to
bursts from sources that have so far not been observed to repeat. We find that 
repeaters and apparent non-repeaters show DM, scattering, sky-location, and
signal-strength 
distributions consistent with coming from the same underlying population. 
However, we confirm
distinct differences in both the intrinsic temporal and spectral morphology of
the two populations. This may suggest that repeaters and one-off FRBs
are distinct populations, or perhaps that repeat rate strongly correlates with
morphology.

In addition, we have developed algorithms and techniques for a synthetic signal injection system to forward model
the selection biases in the CHIME/FRB system as a function of the burst
properties and employed these tools for our analyses. This permits a measurement of FRB property distributions in an
absolute sense, revealing a sizeable population of very highly scattered
events, only a fraction of which are detected.
We have measured the FRB sky rate and fluence distribution, showing that the
latter is consistent with the Euclidean expectation
when including
the full range of observed DMs. However, we find hints of the detailed
shape of the FRB energy distribution in the observed joint distribution of
fluence and DM.

The rich dataset represented by the CHIME/FRB Catalog~1 will be explored in further detailed studies by our team.
The statistics of pulse spectro-temporal morphology, including pulses from both
repeating and non-repeating sources, are presented in
Pleunis et al.\ (2021, submitted).
The sky distribution of FRBs with respect to the Galactic plane is
presented in Josephy el al.\ (2021, submitted).
A cross-correlation analysis of the catalog sources with galaxy catalogs will be presented by Rafiei-Ravandi et al.\ (in preparation) and
a detailed study of the joint distribution of DM and scattering
will be performed in Chawla et al.\ (in preparation).  The flux and fluence calibration techniques we use will be presented in detail
by Andersen et al.\ (in preparation), determination of the CHIME beam model will be presented by Singh et al.\ (in preparation) and Wulf et al.\ (in preparation), and
the injection system we use
to characterize selection effects will be presented in
Merryfield et al.\ (in preparation). The details of our use
of both the catalog and the injections sample to correct for selection biases
and for statistical inference will be presented in Munchmeyer et al.\ (in
preparation). Accompanying these latter two papers will be a public release of the injection sample. Also, we are actively
working on analyzing the joint distribution of fluence and DM and interpreting
it with respect to the FRB energy distribution function.

We look forward to the broader FRB and astrophysical community making use of the first CHIME/FRB catalog
for new interpretations of our results and
for purposes we have yet to envision.
The release of Catalog~1 also marks the start of public,
near real-time alerts of FRB candidates via Virtual Observatory VOEvents\footnote{\url{https://voevent.readthedocs.io/en/latest/}}, which we hope will enable a myriad of
transient FRB follow-up and multi-wavelength science, in a continuing effort to determine the origins of these enigmatic sources.

\bigskip

\acknowledgements

We acknowledge that CHIME is located on the traditional, ancestral, and unceded territory of the Syilx/Okanagan people.

We thank the Dominion Radio Astrophysical Observatory, operated by the National
Research Council Canada, for gracious hospitality and expertise. We thank
Manisha Caleb, Eric Howell, Casey Law, and Aaron Tohuvavohu for providing
external review of the catalog public webpage. CHIME is funded by a grant from the Canada Foundation for Innovation (CFI) 2012 Leading Edge Fund (Project 31170) and by contributions from the provinces of British Columbia, Qu\'ebec and Ontario. The CHIME/FRB Project is funded by a grant from the CFI 2015 Innovation Fund (Project 33213) and by contributions from the provinces of British Columbia and Qu\'ebec, and by the Dunlap Institute for Astronomy and Astrophysics at the University of Toronto. Additional support was provided by the Canadian Institute for Advanced Research (CIFAR), McGill University and the McGill Space Institute via the Trottier Family Foundation, and the University of British Columbia. The Dunlap Institute is funded through an endowment established by the David Dunlap family and the University of Toronto. Research at Perimeter Institute is supported by the Government of Canada through Industry Canada and by the Province of Ontario through the Ministry of Research \& Innovation. The National Radio Astronomy Observatory is a facility of the National Science Foundation (NSF) operated under cooperative agreement by Associated Universities, Inc.  

%
\allacks

\bibliographystyle{aasjournal}

\bibliography{frbrefs,psrrefs}

\appendix

\section{Transient Name Server Names for Fast Radio Bursts}
\label{app:tns}
The Transient Name Server (TNS)\footnote{\url{https://www.wis-tns.org.}} is a public online archive and alert system for new astrophysical transients that was officially adopted by the IAU as of 2016 January. The CHIME/FRB Collaboration, in consultation with the FRB research community, has selected an FRB naming scheme that is now officially maintained through the TNS and by which all Catalog~1 discoveries are named. For the benefit of the wider FRB community, here we provide instructions for observers to catalog their FRB discoveries through the TNS.

New FRB discoveries can be submitted to the TNS by a human through the use of
an online web form submission, or in an automated fashion by a computer program
referred to as a \textit{bot} within the TNS. The first step is to create TNS
user accounts for oneself and team members, and subsequently a TNS group
account to which said members should be added. Each TNS group is given access
to a TNS bot for which a unique API\footnote{An Application Programming
Interface (API) can be described as a standardized path typically specified by
a URL-like token and commonly used to provide controlled, one-off, or
asynchronous access to algorithms and databases that are not maintained by the user.} key will
be generated and provided by the TNS. The TNS exposes a limited set of
archiving functions, allowing the user to submit or retrieve FRB discoveries
automatically with a script, through a selection of URL endpoints. The TNS
provides official templates for such scripts in select programming languages,
including
\texttt{python}\footnote{\url{https://www.wis-tns.org/content/tns-getting-started}},
and it is here that the user is required to plug in their API key. 

A minimal set of measurements and identifiers must be provided with every FRB discovery to be accepted by the TNS. These include Right Ascension, Declination, the discovery date and time (topocentric) of the event, the DM, the instrument or observatory, the reference frequency of the burst, the instrument bandwidth, the number of frequency channels, and the sampling time (among other administrative details). Further information can be provided including a machine-readable list of vertices for an elliptical or polygonal sky localization region. Virtually any supporting file type (e.g., a PNG file for the burst dynamic spectrum in the observing band) can be uploaded as supporting evidence.

FRBs are submitted through the TNS API in bulk discovery reports formatted as JSON\footnote{JavaScript Object Notation (JSON) is a standard way of storing attributes or parameters of data-oriented objects in a key-value structure similar to pythonic dictionaries, especially useful for data transmission protocols over the internet.} documents. The default script for accessing the TNS API returns either the TNS object name for each new FRB object that was added, or a list of error messages (one per offending discovery entry) in the event that the submission data were incomplete or malformed. Under normal conditions, the TNS API updates their database within $\sim 1~\text{s}$ per bulk submission, though it is considered good practice to verify automated submissions after allowing enough time for the updating to finish. New TNS users should exercise caution during bulk submission: if a single FRB within the bulk report is rejected by the TNS, all discoveries in the report are rejected. Therefore, validation of the discovery data before submission, and of the TNS response submission, is an essential component of any system designed to submit in real-time to the TNS.

New TNS users are strongly encouraged to practice their FRB submissions using the TNS sandbox site\footnote{\url{https://sandbox.wis-tns.org}} \textit{before} proceeding to the live site. Just as with the live site, the sandbox site offer TNS users the option to specify the end date and time of a proprietary period during which their new FRB submissions are not visible outside their TNS group. This can be leveraged for users to practice internal bookkeeping with official TNS names, without exposing their discoveries publicly at least for some time.

\section{Quality of least-squares fits to burst morphology models}
\label{app:fitburst}

Here we provide additional details on the least-squares fitting procedure
as implemented in \fitburst{}.
We define the noise-weighted fit residuals to be
\begin{equation}
    h_{t,f}(\vec\lambda) = \frac{d_{t,f}}{\sigma_f} - S_{t,f}(\vec\lambda),
\end{equation}
where $d_{t,f}$ is the total intensity data as a function of discrete time $t$ and
discrete frequency $f$ (the dynamic spectra), $S_{t,f}$ is the model defined as a function of the
parameter vector $\vec\lambda$ as described in Section~\ref{sec:properties},
and $\sigma_f$ is the noise standard deviation measured in each spectral
channel.
Note that $S_{t,f}(\vec\lambda)$ is not divided by ${\sigma_f}$
in the above equation, meaning that for the morphology fits we are effectively
calibrating the data to uniform noise, a choice we have found to yield the most
robust results if not trying to extract absolute flux information (which is
measured elsewhere).
Prior to fitting the data are detrended with a temporal high-pass filter and
cleaned with an automatic narrowband RFI-detection algorithm
that excises spectral channels based on variance and spectral-kurtosis distributions.
Using the {\tt
optimize.least\_squares} solver provided by the open-source {\tt scipy}
software library \citep{scipy}, \fitburst{} minimizes
$\chi^2(\vec\lambda) = \sum_{t,f} [h_{t,f}(\vec\lambda)]^2$ with respect to the
parameters.

\begin{figure}
    \centering
    \includegraphics[width=0.90\textwidth]{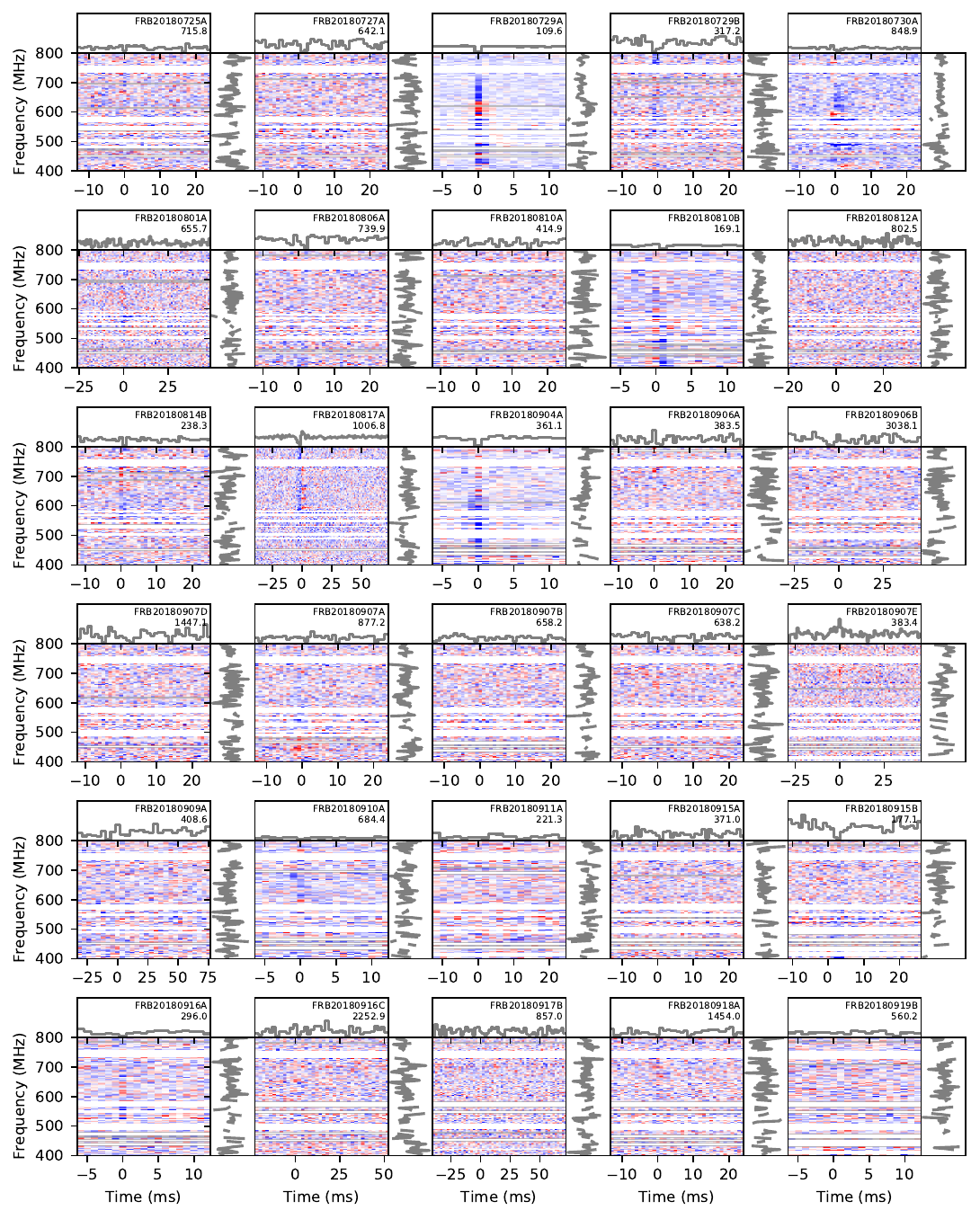}
    \caption{Waterfall plots of {\tt
    fitburst} residuals, down-sampled in frequency, for one-off bursts. Same as
    Figure~\ref{fig:waterfall_oneoffs} except showing the normalized residuals from the
    best-fit model. Color scale saturates at $\pm3\sigma$.
    \arxivonly{Panels for all catalog one-off bursts can be found at
    \url{https://www.canfar.net/storage/list/AstroDataCitationDOI/CISTI.CANFAR/21.0007/data/additional_figures/residuals}.}
    }
    \label{fig:waterfall_oneoffs_residuals}
\end{figure}

\begin{figure}
    \centering
    \includegraphics[width=0.90\textwidth]{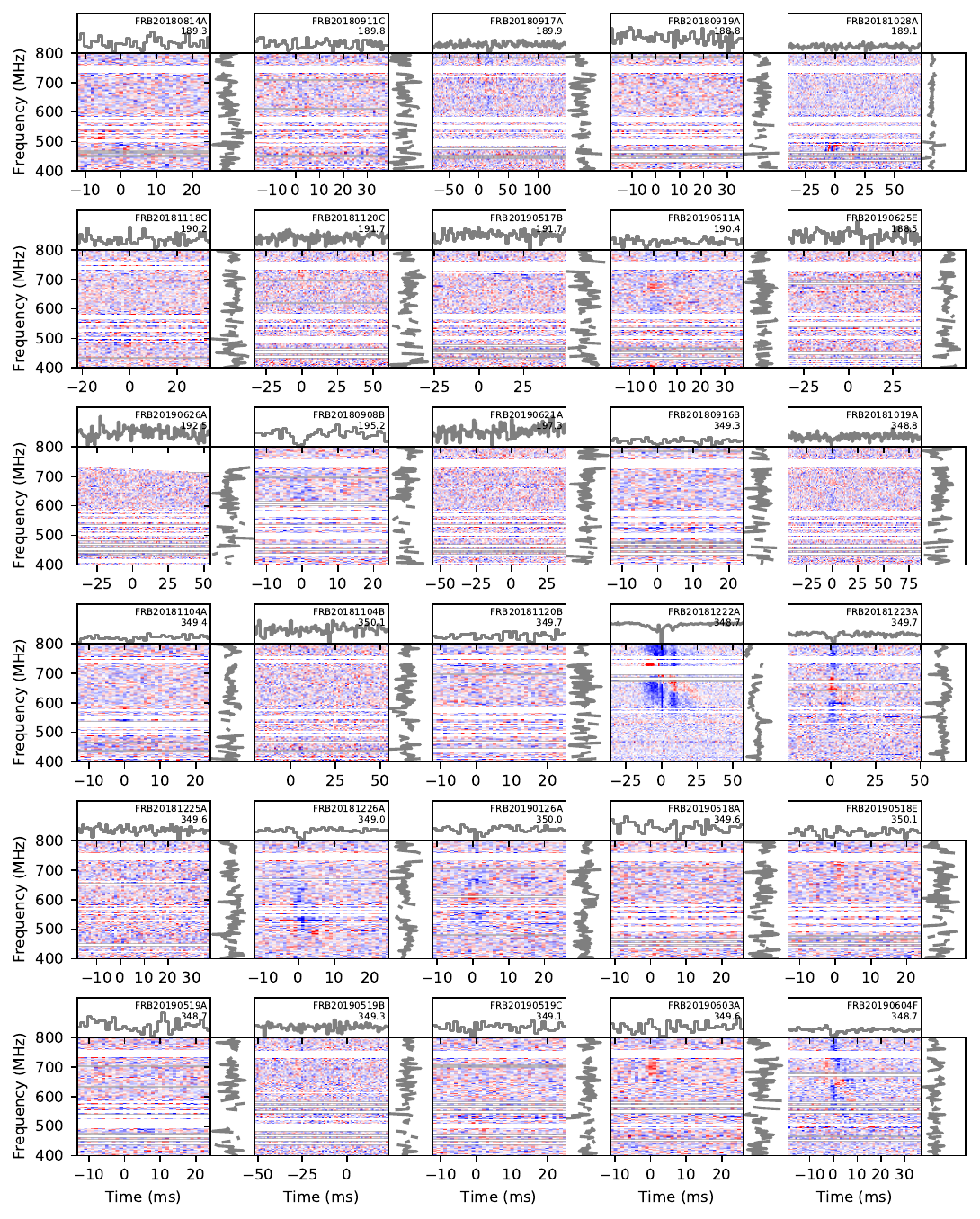}
    \caption{Waterfall plots of {\tt fitburst} residuals for
    repeaters. Same as
    Figure~\ref{fig:waterfall_oneoffs_residuals} except for repeaters (as in
    Figure~\ref{fig:waterfall_repeaters}).
    \arxivonly{Panels for all catalog repeating bursts can be found at
    \url{https://www.canfar.net/storage/list/AstroDataCitationDOI/CISTI.CANFAR/21.0007/data/additional_figures/residuals}.}
    }
    \label{fig:waterfall_repeaters_residuals}
\end{figure}

In addition to best-fit parameters, we tabulate in Catalog~1 several
metrics to help assess the quality of these fits. The first is the fraction of spectral
channels that are missing or flagged as RFI. Second, the \fitburst{} SNR equal to
$\sqrt{\Delta\chi^2}$, where $\Delta\chi^2$ is the difference in $\chi^2$
between the best fit model and the no-burst model ($S_{t,f}=0$).
We also provide the best fit value of $\chi^2$ as well as the number of degrees
of freedom for the fit, allowing the reduced chi-square statistic to be
calculated and chi-squared tests to be performed. Finally, for each burst we
provide waterfall plots of the noise-weighted fit residuals in
Figures~\ref{fig:waterfall_oneoffs_residuals} and~\ref{fig:waterfall_repeaters_residuals} such that the
quality of the fit for each burst may be assessed visually.

As can be seen in the residual plots, the temporal profiles are usually
well modeled with intrinsic or scatter-broadened Gaussian shapes. 
Remaining residual structure tends to be most prominent along the frequency
axis. These features indicate that a smooth, running power-law
spectrum is an imperfect model for many events. Indeed, \citet{msb+19}
analyzed 23 dynamic spectra observed with ASKAP and argued that the diversity in
their measurements arises from effects intrinsic to FRB emission and/or propagation,
such as diffractive scintillation.
Moreover, the telescope beam and non-uniformity in the noise levels as a
function of spectral frequency can introduce spectral structure for which we
have not accounted.
We nonetheless used the running power-law spectral model
in our modeling of all CHIME/FRB events for uniformity in our analysis
and interpretation. Further analysis of fluctuations in total intensity across
the CHIME band will be reserved for future work.

In some cases, structure in the residuals is the result of poorly modeled
temporal profiles or failures of the least-squares solver to converge
to an adequate result, both of which are subject to human interaction and
judgement.
For example, a number of components ($n$ in Section~\ref{sec:properties})
greater than unity were visually
determined; it is likely that
faint bursts with marginally complex morphology were instead modeled as
single-component bursts, which would impact best-fit estimates of several fit
parameters described above. Similarly, even single component bursts may not
be well modelled by an intrinsically Gaussian pulse profile multiplied by a
generalized power law. Moreover, the least-squares algorithm may yield
sub-optimal estimates of parameter estimates and uncertainties that could
instead be sampled
better using the Markov Chain Monte Carlo (MCMC) method. Various efforts for
improving the automated fitting pipeline, including the use of an MCMC sampler for
\fitburst{} and an automatic determination of the number of components,
are ongoing and will be the
subject of future CHIME/FRB catalogs.

Interpretations of the burst properties in Catalog~1 should thus be
taken in the context of these limitations. At the population level, there may
be significant biases in the measured properties (e.g., biases in the
spectral index due to the beam, as shown in Figure~\ref{fig:spectral_params}).
For individual bursts, the quality of a given fit should be accessed through
$\chi^2$ statistics and the residual plots.  Nonetheless, if proper care is
taken to determine the impact of these issues, the data presented here are
adequate for a wide range of analyses including those presented in this article
and in companion works.

\section{Fitting for a fiducial population model}
\label{app:fid}

Here we detail the procedure for finding a fiducial model for the property
distributions $P(F)$, $P(\dm)$, $P(\tau)$, $P(w)$, and $P(\gamma, r)$, as
introduced in Section~\ref{sec:abs_pop}. The purpose of the fiducial model is
not to provide a precise description of the true property distributions, but a
rough one, such that correlated selection effects are accounted for when
performing deeper analyses in property subspaces.

\subsection{Property distribution models and overview of fitting procedure}
\label{app:fid_over}

Empirically, we found that the following models for the intrinsic property
distribution functions $P(\xi)$ provide a reasonable match to the data:
\begin{itemize}
    \item Fluence is described by a power-law distribution
        $P(F) \propto -\alpha(F/F_0)^{\alpha - 1}$ with power-law index $\alpha$.
    \item DM, scattering $\tau$, and pulse width $w$ are described by log-normal distributions with a shape $\sigma$ and scale $m$, given by $P(\xi) = \frac{1}{\sigma (\xi/m) \sqrt{2 \pi}} \exp \left[ - \frac{\ln^2{(\xi/m)}}{2 \sigma^2} \right] $.
    \item For the spectral index $\gamma$ and running $r$, we fit a kernel
        density estimator to their joint observed distribution and equate it to
        the intrinsic distribution without compensating for selection effects.
        However, we do verify that, once accounting for measurement effects from
        the beam, the final population model provides a reasonable description
        of the catalog data (see Figure~\ref{fig:spectral_params}).
\end{itemize}

Rather than performing a joint fit to the parameters of all property
distributions simultaneously, we use an
iterative fitting scheme, where we optimize the parameters of each population
factor $P(\xi)$ independently while keeping the other parameters fixed. This method is
possible because of our assumption that the rate function factorizes, and is
necessary because of our limited injections sample---we do not have enough
injections to fully determine the 7D observation function
$P(\SNR | F, \dm, \tau, w, \gamma, r)$; however, we can determine certain integrals
thereof.  We are currently exploring the use of machine learning to fully
determine the observation function.

Schematically, our fitting procedure is as follows:
\begin{enumerate}
    \item On the $i = 0$ iteration, define an initial model $R_{i=0} = R_{{\rm init}}$ for
        the intrinsic rate composed
        of individual property distribution models $P_{i = 0} = P_{{\rm init}}$.
    \item On each iteration $i$, and for each property $\xi$ in $(F, \dm, \tau, w)$, form the
        selection function $s_i(\xi) = P_{{\rm obs}, i}(\xi) / P_{i}(\xi)$,
        obtaining the observed distribution from the injection system.
    \item Fit a model $P_{\rm obs}(\xi) = s_i(\xi) P_{i + 1}(\xi)$ to the catalog data.
    \item Iterate steps 2 and 3 until convergence. The converged result is the
        fiducial model.
\end{enumerate}

\subsection{Modeling observed population with injections}
\label{app:inj_pop}

The initial probability density functions $P_{\rm init}(\xi)$ are the same as
those used for the injections population described in
Section~\ref{sec:inj_pop} and are designed to both fully sample the range of
observed properties and more densely sample parts of phase space populated by
the catalog.
We take care to apply the same cuts to the injected events as we did to the
catalog, including automated RFI classification, and cuts on galactic DM. Less than half of the 84\,697 detected
injections survive these cuts, with most of the attrition coming from
the $\SNR>12$ requirement for population analysis.
These cuts provide
a sample of
$\mathcal{R}_{\rm obs, init}(\SNR, \dm, \tau, w, \gamma, r)$. The sample is not
dense enough to fully determine this 7D function;
however, integrals thereof such
as $P_{\rm obs, init}(\dm)$ are determined with little sampling noise.

At each iteration $i$ of our fitting procedure described above, a simple
approach would be to generate a new population $R_i$ of
FRBs from the best-fit parameters at the current iteration, inject these into
the pipeline, and thus obtain a new observed distribution
$\mathcal{R}_{\mathrm{obs}, i}$. However, we can avoid re-running the injections system
at each iteration;
using the initial injected population, we can
generate populations for any new distribution as follows.

Our injections sample was drawn from
$R_{\rm init}(F, \dm, \tau, w, \gamma, r)$, providing corresponding samples of
$\mathcal{R}_{\rm obs, init}(\SNR, \dm, \tau, w, \gamma, r)$ after injection. These can be converted to
estimates of $P_{\rm init}(\xi)$ and $P_{\rm obs,init}(\xi)$ by accumulating
these samples from the full and detected injections into histograms
respectively. Further, this can be converted to any new
model $R(F, \dm, \tau, w, \gamma, r)$ by \emph{reweighting} the sample---for
every event, we construct weights
\begin{equation}
    W(F, \dm, \tau, w, \gamma, r)
    = \frac{R(F, \dm, \tau, w, \gamma, r)}
    {R_{\rm init}(F, \dm, \tau, w, \gamma, r)}.
\end{equation}
Accumulating the FRB injection sample into histograms, counting each event with
weight $W$ then
provides estimates of $P(\xi)$ (for the full injection sample), and $P_{\rm obs}(\xi)$
(for the detected events) for the new model
rather than the initial model. $P_{\rm obs}(\xi)$ is thus our prediction for the
catalog, to which we fit the underlying parameters by optimizing a
likelihood analogous to
Equation~\ref{e:lik} (holding the distributions of properties other than $\xi$
fixed on a given iteration).

\subsection{Jackknife tests}
\label{sec:jackknife}

Jackknife tests provide a way to search for inconsistencies in the data
by splitting the sample by some criteria and comparing the changes in the
analysis results to those expected statistically from the cut.
A full study of the uncertainty in the fiducial model (either statistical
or systematic) is beyond the scope of this work, since it is only
used as a starting point for further analysis. However,
we can use our procedure for finding the fiducial model to search
for systematic errors, as well as test our strongest assumption: that the FRB
properties are intrinsically uncorrelated.
We note that we do
perform an accounting of uncertainties for our final $\alpha$ and sky-rate
measurements in Section~\ref{app:rate_alpha_sys}, much of which is informed by the
jackknife tests presented here.

We perform the following jackknife tests:
\begin{itemize}
    \item A set of 50 ``random" jackknives,
        where half the 265 events in the post-cut sample are excluded
        at random to
        give an indication of the expected change in parameters from the
        smaller sample size. For each parameter we quote the mean and standard
        deviation (SD) over the set.
    \item Shifting the SNR cut from 12 to either 10 or 14 to check for
        incompleteness in the human classification step of catalog events.
        This would not be accounted for by the injections, since detected
        injections are
        verified by coincidence with an input injection, rather than human
        classification.
    \item Separate cuts on DM, $w$, and $\tau$ near the median values of the
        sample. These cuts test for \emph{intrinsic} correlations among
        properties in the sample. When cutting on a given property, e.g.,~$\tau$,
        we fix the model $P(\tau)$ at the fiducial model. This better tests for
        correlations in the properties, rather than how the reduced data
        range for a
        given property affects the fit.
\end{itemize}
The results of the iterative model fitting for these jackknives are quoted in
Table~\ref{t:jackknife}.

\begin{table}
\newlength{\pw}
\setlength{\pw}{1.5cm}
\begin{center}
    \caption{Best Fit Parameters for Full Sample and Jackknife Subsamples
    \label{t:jackknife}
    }
\begin{tabular}{l*2{>{\centering\arraybackslash}m{\pw}}|*7{>{\centering\arraybackslash}m{\pw}}}\hline\hline
\centering
Parameter & Fiducial (full) & Random (SD) & Random (mean) & SNR $>10$ & SNR
    $>14$ & DM $>500^{\rm a}$ & $w$\newline $<1$\,ms & $\tau$\newline
    $<1$\,ms \\
\hline
N$_{\rm events}$ & 265 & ... & $-133$ & 94 & $-48$ & $-118$ & $-97$ & $-130$ \\
$\alpha$ & $-1.32$ & 0.11 & $-0.01$ & $-0.09$ & $-0.1$ & $-0.35$ & 0.14 & 0.33 \\
DM scale / $100^{\rm a}$ & 4.95 & 0.22 & 0.02 & 0.17 & $-0.05$ & ... & $-0.37$ & $-0.91$ \\
DM shape & 0.66 & 0.03 & $-0.01$ & 0.02 & 0.01 & ... & 0.01 & $-0.02$ \\
$w$ scale (ms) & 1.0 & 0.08 & $-0.02$ & 0.19 & $-0.13$ & 0.2 & ... & $-0.43$ \\
$w$ shape & 0.97 & 0.08 & $-0.01$ & 0.02 & $-0.06$ & 0.06 & ... & $-0.32$ \\
$\tau$ scale (ms) & 2.02 & 1.31 & 0.31 & 0.02 & 0.24 & 1.11 & 0.14 & ... \\
$\tau$ shape & 1.72 & 0.26 & 0.02 & $-0.15$ & 0.08 & $-0.14$ & 0.75 & ... \\
\hline
\end{tabular}
\end{center}
\textbf{Notes.}
Results of the iterative model fitting procedure for the full
sample (yielding our fiducial model) and jackknife samples.
The fitting procedure and parameter definitions are described in
Appendix~\ref{app:fid_over}.
When cutting on DM, $w$, or $\tau$, we fix the model parameters for that
property's distribution at the fiducial values.
Values to the right of the vertical line are differences from the fiducial values.
\\
$^{\rm a}$ DMs are in units of pc\,cm$^{-3}$
\end{table}

Some caution is required in interpreting the standard
deviation of the random jackknives in terms of the expected shift in parameter
values. For the SNR cuts, much less than half the sample is excluded or added,
and the fit lever arm for $\alpha$ also changes. When cutting on  DM, $w$,
and $\tau$, roughly half the sample is cut. However, the cut events have
weights that are systematically different from a typical event, especially for
$\tau$ where the cut events are upweighted by the fitting procedure
because of the low selection
function.
A precise accounting for the expected statistical fluctuation in each jackknife
is beyond the scope of this paper. The standard deviation of the random jackknives
nonetheless gives a rough sense for how much we expect parameters to
vary, and the tests are valuable
to search for strong biases and strong property correlations of the type that
might invalidate conclusions drawn from our analysis.

While the results of our jackknife tests do allow for
order-unity property
correlations, we do not find evidence for very strong systematic errors or
property correlations (i.e.,\ above 90\%).

Shifting the SNR cut does
produce a shift in $\alpha$ of $\sim0.1$, which is larger than one might expect
given the modest change in event numbers. This may be an indication of residual
bias from human classification, although this explanation is at odds with the fact
that the shift in $\alpha$ is in the same direction when either raising or
lowering the threshold.
There would also seem to be a significant change in the width scale parameter.
This might indicate a negative correlation between fluence (which is
correlated with SNR) and width, although we find this to be
physically implausible. Another explanation is that this is a measurement
effect, with narrow widths difficult to measure to low SNR events,
yielding upper limits that may be significantly higher than their true
values. We note that our analysis fully compensates for completeness as a
function of properties such as $w$, but not in errors in the measurement of the
properties themselves.

Jackknives in  DM, $w$, or $\tau$ all result in changes in the $\alpha$
parameters. For DM, this change in $\alpha$ is studied in detail in Section~\ref{sec:flux_rate}.
The shift for
the $w$ jackknife
is not particularly significant compared to the random jackknives.
Interpreting the shift for the $\tau$ jackknife is complicated by the
highly non-uniform weighting of events as a function of $\tau$, making the
expected statistical fluctuation difficult to calculate.

These jackknives also result in shifts in the  DM, $w$, or $\tau$ scale
parameters consistent with order-unity correlations between these properties.
In particular, we see hints of a positive DM--$\tau$ correlation, both from the change in
DM scale in the $\tau$ jackknife and the change in the $\tau$ scale in the DM
jackknife. Such a correlation is well motivated physically and will be studied
in detail in future work. Likewise, there is evidence for a $w$--$\tau$
correlation, although this is likely a measurement effect since
these properties are particularly hard to disentangle
observationally.

\section{Systematic errors in the rate and $\alpha$ measurement}
\label{app:rate_alpha_sys}

There are several sources of systematic errors in our measurements. Obtaining
the sky rate requires an absolute accounting of the survey duration $\Delta t$.
As described in Section~\ref{sec:expo}, our effective survey duration is
214.8 days.
Note that periods where the system was operating well below nominal sensitivity
have been excluded from this figure and any bursts discovered during these
periods have been excluded from the population analysis.

Nonetheless, our measurement of the overall rate depends on the sensitivity
during the injections campaign (which occurred in August 2020)
being representative of the full Catalog~1 period. Between the beginning of the
survey and the injections campaign, we estimate our noise levels have improved
by 6\%, based on daily observations of the SNRs of pulsar single-pulse
detections. The change since the midpoint of the survey period was 3\%. To
compensate for this, we reduce our effective survey duration by $|\alpha|
\times 3\%$,
using $\alpha=-3/2$ which is the value for a stationary universe and is
consistent with our measurement in 
Section~\ref{sec:flux_rate}.
We allow for a
corresponding
systematic error in our rate measurement of $|\alpha| \times 3\%=4.5\%$.

Another source of systematic error in the rate is incompleteness in
Catalog~1 not accounted for by injections. Injected events are only processed
by our automated pipeline, and lack a final human classification step required
for true events. However, our human classification should be complete
above an L1 SNR of 12. There are 28 cases where an FRB candidate was
successfully classified by the automated pipeline, but due to a system error no
intensity data were saved. Of these 16 events would have passed
our cuts on SNR and occurred during periods where the system was otherwise
operating nominally. For such events, we have no way of knowing
whether they would have survived human classification or our other various data cuts.
This leads to an additional, one-sided systematic error in the rate of
$+16/265=+6.0\%$.

A large source of uncertainty in our analysis is the telescope beam. The CHIME
beam model used for injections is described in Section~\ref{sec:injection}.
This model was verified by comparing to measurements of the Sun over the
declination range $\pm 23.4$ degrees, as well as
holographic measurements of bright persistent sources for a few specific
declinations (see Section~\ref{sec:beam_model}).
The discrepancy between these measurements and our primary beam model, in units
of the peak power response at the declination of Cygnus A, is of
order 0.1 in the main lobe and 0.01 in the near side lobes.
These discrepancies vary in magnitude and sign as a function of both frequency
and declination. As such, we expect a high degree of cancellation in averages
over frequencies and beams. Nonetheless, to conservatively account for
errors in the main-lobe response, we assign a systematic
uncertainty in our rate measurement of $|\alpha| \times 10\%=15\%$.
We are insensitive to beam-model errors in the side lobes since we
are able to identify side-lobe FRB detections from their spectral
characteristics and have cut them from our
analysis. However, some possibility of side-lobe contamination from narrow band
bursts does exist.

In our $\alpha$ measurement, we do not account for systematic error from the
beam, which should be small
because our analysis uses $\SNR$ as our observable rather than fluence
directly. It has been shown that the $\SNR$ distribution power law index is
unaffected by the telescope beam \citep{ocp16, con19, jem+19}, provided
FRB fluences are distributed as a single power law, the instrumental $\SNR$ is
linear in the FRB fluence, and chromatic effects are ignored. All three of
these assumptions are violated, however, we account for these in our
primary analysis, relying on them only at the level of the beam-model
uncertainty. The effect on our $\alpha$ measurement is thus second order (an error on an error)
and can be neglected.

To avoid a high-dimensional fitting problem and to make best use of our finite
injection sample, we have opted to study FRBs in the
property space of fluence and DM, fixing the other properties at their fiducial
distributions. This treatment can induce errors that are not accounted for in
the statistical uncertainty in two ways. First, even if our model without
intrinsic correlations were correct, not fitting simultaneously means we
neglect correlated statistical errors in the distribution parameters. For
instance, a 1$\sigma$ fluctuation in the scattering scale parameter would shift
the scattering distribution to a region with more or less selection bias,
resulting in a different inferred overall rate. Second, our model where the FRB
properties are intrinsically uncorrelated could be incorrect. Indeed,
Table~\ref{t:jackknife} contains evidence for correlations amongst fluence, DM,
width, and scattering.
As an estimate for how large these errors could be, we perturb the scale
parameters for both width and scattering. For the scattering scale, we perturb
by $\pm1$\,ms and for the width scale by $\pm0.3$\,ms, inspired by the
magnitude of deviation we see in Table~\ref{t:jackknife}.  The perturbation in
scattering scale has the larger effect, resulting in changes in $\alpha$ of
$\pm0.03$ and in the overall rate of $^{+18\%}_{-20\%}$. This
treatment conservatively captures the issue of correlations in the parameter
fits; however, we caution that the effects of intrinsic property correlations
could be more subtle and will need to be studied further. In addition, our only
treatment of the spectral properties $\gamma$ and $r$ is to show that the
distribution of the
observed injections population roughly matches that of Catalog 1, a treatment that is
far from complete.

As previously discussed, another concern is incompleteness in Catalog~1
from human classification that is not reflected in the injections. Shifting our
SNR cut from 12 to 16 changes $\alpha$ by $-0.08$ and the rate by $+10\%$.
While these changes are largely within the expectation for statistical
fluctuations from the number of cut events, they are in the direction one would
expect for residual incompleteness at low SNR. We thus conservatively include
these
figures in our systematic error budget.

Prior to the addition of the RFI classification bypass for high-SNR events
described in Section~\ref{sec:obs}, there were three events with $\SNR\gtrsim100$ that
were likely to have been astrophysical but are not included in Catalog~1 as
they were classified as RFI, and thus their data were not retained. Including these
events in the population analysis shifts $\alpha$ by $+0.05$ (which we include
in our error budget) but has a negligible effect on the rate.

During the Catalog~1 period and between then and the injections period, there
have been a number of other changes to our detection pipeline, including changes to
the RFI cleaning and the addition of automated classifiers employing machine learning.
In addition, there have been
changes in the RFI environment at the telescope site. While these changes are difficult
to quantify, they should be an overall small effect above our relatively
high SNR cutoff for statistical analysis of 12. The assessment of the effect of
changing the SNR cutoff in the previous paragraph also partially accounts for
this effect as does our tracking of the overall telescope sensitivity using
pulsar pulses.

All systematic errors are added in quadrature. For the rate, this yields a
net systematic error of $^{+27\%}_{-25\%}$.
For $\alpha$ the systematic error is
$^{+0.060}_{-0.085}$.

\section{Catalog excerpt}
\label{app:cat_excerpt}

In Table~\ref{tab:cat_sample}, we provide an excerpt from Catalog~1.
We show all fields for four Catalog~1 entries. The first entry is a so-far non-repeating source detected during the pre-commissioning period, the second entry is another so-far non-repeating source, and the latter two are sub-bursts from the same event from a repeating FRB. Field
descriptions can be found in Table~\ref{ta:catalog} and the full Catalog~1 data
accompanies the online version of this article and at
the CHIME/FRB Public Webpage\footnote{\url{https://www.chime-frb.ca/catalog}}.

\begin{table}
\footnotesize
\begin{center}
    \caption{Excerpt from Catalog~1
    }
    \label{tab:cat_sample}
\hspace{-1.5cm}
%
%
\begin{tabular}{cccccccc}
\Xhline{4\arrayrulewidth}
\texttt{tns\_name} & \texttt{previous\_name} & \texttt{repeater\_name} & \texttt{ra} & \texttt{ra\_err} & \texttt{ra\_notes} & \texttt{dec} & \texttt{dec\_err} \\
& & & (degrees) & (degrees) & & (degrees) & (degrees) \\
\hline
FRB20180817A & 180817.J1533+42 & ... & 233.200 & 0.052 & ... & 42.20 & $0.16$ \\
FRB20180915B & ... & ... & 225.23 & 0.25 & ... & 25.02 & $0.23$ \\
FRB20180917A & ... & FRB20180814A & 65.54 & 0.19 & ... & 73.63 & $0.27$ \\
FRB20180917A & ... & FRB20180814A & 65.54 & 0.19 & ... & 73.63 & $0.27$ \\
\end{tabular}
\begin{tabular}{cccccccc}
\Xhline{4\arrayrulewidth}
\texttt{dec\_notes} & \texttt{gl} & \texttt{gb} & \texttt{exp\_up} & \texttt{exp\_up\_err} & \texttt{exp\_up\_notes} & \texttt{exp\_low} & \texttt{exp\_low\_err} \\
& (degrees) & (degrees) & (hours) & (hours) & & (hours) & (hours) \\
\hline
... & $68.3$ & $53.98$ & $22.7$ & $2.5$ & ... & ... & ... \\
... & $36.17$ & $60.93$ & $13.1$ & $7.6$ & ... & ... & ... \\
... & $136.46$ & $16.58$ & $41$ & $24$ & ... & $43.5$ & $7.9$ \\
... & $136.46$ & $16.58$ & $41$ & $24$ & ... & $43.5$ & $7.9$ \\
\end{tabular}
\begin{tabular}{cccccccc}
\Xhline{4\arrayrulewidth}
\texttt{exp\_low\_notes} & \texttt{bonsai\_snr} & \texttt{bonsai\_dm} & \texttt{low\_ft\_68} & \texttt{up\_ft\_68} & \texttt{low\_ft\_95} & \texttt{up\_ft\_95} & \texttt{snr\_fitb} \\
& & (pc cm$^{-3}$) & (Jy ms) & (Jy ms) & (Jy ms) & (Jy ms) & \\
\hline
... & $45.6$ & $1002.8$ & $0$ & $6.2$ & $0$ & $11.1$ & $65.1$ \\
... & $12.5$ & $176.3$ & $0$ & $2.4$ & $0$ & $4.2$ & $36.1$ \\
... & $19$ & $194.1$ & $43.8$ & $6.9$ & $155.9$ & $25.6$ & $46.6$ \\
... & $19$ & $194.1$ & $43.8$ & $6.9$ & $155.9$ & $25.6$ & $46.6$ \\
\end{tabular}
\begin{tabular}{cccccccc}
\Xhline{4\arrayrulewidth}
\texttt{dm\_fitb} & \texttt{dm\_fitb\_err} & \texttt{dm\_exc\_ne2001} & \texttt{dm\_exc\_ymw16} & \texttt{bc\_width} & \texttt{scat\_time} & \texttt{scat\_time\_err} & \texttt{flux} \\
(pc cm$^{-3}$) & (pc cm$^{-3}$) & (pc cm$^{-3}$) & (pc cm$^{-3}$) & (s) & (s) & (s) & (Jy) \\
\hline
1006.7714 & $0.0085$ & $979.1$ & $982.0$ & 0.01769 & $0.01103$ & $0.00058$ & $2.4$ \\
177.127 & $0.020$ & $154.7$ & $154.1$ & 0.00492 & $0.0001108$ & $0.0000081$ & $0.99$ \\
189.917 & $0.018$ & $102.1$ & $81.5$ & 0.06095 & $0.0080$ & $0.0025$ & $3.2$ \\
189.917 & $0.018$ & $102.1$ & $81.5$ & 0.06095 & $0.0080$ & $0.0025$ & $3.2$ \\
\end{tabular}
\begin{tabular}{cccccccc}
\Xhline{4\arrayrulewidth}
\texttt{flux\_err} & \texttt{flux\_notes} & \texttt{fluence} & \texttt{fluence\_err} & \texttt{fluence\_notes} & \texttt{sub\_num} & \texttt{mjd\_400} & \texttt{mjd\_400\_err} \\
(Jy) & & (Jy ms) & (Jy ms) & & & (MJD) & (MJD) \\
\hline
$1.5$ & ... & $29$ & $16$ & ... & $0$ & $58347.0760879598$ & $0.0000000031$ \\
$0.42$ & ... & $3.8$ & $1.0$ & ... & $0$ & $58376.9753845892$ & $0.0000000021$ \\
$3.1$ & ... & $46$ & $32$ & ... & $0$ & $58378.0323858329$ & $0.0000000068$ \\
$3.1$ & ... & $46$ & $32$ & ... & $1$ & $58378.0323860277$ & $0.0000000097$ \\
\end{tabular}
\begin{tabular}{ccccccc}
\Xhline{4\arrayrulewidth}
\texttt{mjd\_inf} & \texttt{mjd\_inf\_err} & \texttt{width\_fitb} & \texttt{width\_fitb\_err} & \texttt{sp\_idx} & \texttt{sp\_idx\_err} & \texttt{sp\_run} \\
(MJD) & (MJD) & (s) & (s) & & & \\
\hline
$58347.0757860647$ & $0.0000000040$ & $0.00189$ & 0.00017 & $3.35$ & $0.40$ & $-7.46$ \\
$58376.9753314751$ & $0.0000000063$ & $0.001694$ & 0.000075 & $-9.2$ & $1.0$ & $3.0$ \\
$58378.0323288836$ & $0.0000000088$ & $0.00178$ & 0.00026 & $34$ & $18$ & $-19$ \\
$58378.032329078$ & $0.000000011$ & $0.00345$ & 0.00044 & $52.3$ & $7.7$ & $-46.5$ \\
\end{tabular}
\begin{tabular}{cccccccc}
\Xhline{4\arrayrulewidth}
\texttt{sp\_run\_err} & \texttt{high\_freq} & \texttt{low\_freq} & \texttt{chi\_sq} & \texttt{dof} & \texttt{flag\_frac} & \texttt{peak\_freq} & \texttt{excluded\_flag} \\
& (MHz) & (MHz) & & & & (MHz) & \\
\hline
$0.56$ & $800.2$ & $400.2$ & $1069085.403$ & $1057913$ & $0.434$ & $501.1$ & $1$ \\
$2.8$ & $527.4$ & $400.2$ & $278172.462$ & $276025$ & $0.557$ & $400.2$ & $0$ \\
$14$ & $800.2$ & $658.6$ & $908712.749$ & $904687$ & $0.516$ & $800.2$ & $0$ \\
$6.8$ & $800.2$ & $561.8$ & $908712.749$ & $904687$ & $0.516$ & $701.7$ & $0$ \\
\end{tabular}
\end{center}
\end{table}

\end{document}

%% file: auth.tex
\author[0000-0001-6523-9029]{Mandana Amiri}
  \affiliation{Department of Physics and Astronomy, University of British Columbia, 6224 Agricultural Road, Vancouver, BC V6T 1Z1 Canada}
\author[0000-0001-5908-3152]{Bridget C.~Andersen}
  \affiliation{Department of Physics, McGill University, 3600 rue University, Montr\'eal, QC H3A 2T8, Canada}
  \affiliation{McGill Space Institute, McGill University, 3550 rue University, Montr\'eal, QC H3A 2A7, Canada}
\author[0000-0003-3772-2798]{Kevin Bandura}
  \affiliation{Lane Department of Computer Science and Electrical Engineering, 1220 Evansdale Drive, PO Box 6109, Morgantown, WV 26506, USA}
  \affiliation{Center for Gravitational Waves and Cosmology, West Virginia University, Chestnut Ridge Research Building, Morgantown, WV 26505, USA}
\author[0000-0002-4064-7883]{Sabrina Berger}
  \affiliation{Department of Physics, McGill University, 3600 rue University, Montr\'eal, QC H3A 2T8, Canada}
  \affiliation{McGill Space Institute, McGill University, 3550 rue University, Montr\'eal, QC H3A 2A7, Canada}
\author[0000-0002-3615-3514]{Mohit Bhardwaj}
  \affiliation{Department of Physics, McGill University, 3600 rue University, Montr\'eal, QC H3A 2T8, Canada}
  \affiliation{McGill Space Institute, McGill University, 3550 rue University, Montr\'eal, QC H3A 2A7, Canada}
\author[0000-0001-5470-3084]{Michelle M.~Boyce}
  \affiliation{Department of Physics and Astronomy, University of Manitoba, Winnipeg, MB R3T 2N2, Canada}
\author[0000-0001-8537-9299]{P.~J.~Boyle}
  \affiliation{Department of Physics, McGill University, 3600 rue University, Montr\'eal, QC H3A 2T8, Canada}
  \affiliation{McGill Space Institute, McGill University, 3550 rue University, Montr\'eal, QC H3A 2A7, Canada}
\author[0000-0002-1800-8233]{Charanjot Brar}
  \affiliation{Department of Physics, McGill University, 3600 rue University, Montr\'eal, QC H3A 2T8, Canada}
\author[0000-0002-2349-3341]{Daniela Breitman}
  \affiliation{Department of Physics, University of Toronto, 60 St.~George Street, Toronto, ON M5S 1A7, Canada}
  \affiliation{Dunlap Institute for Astronomy \& Astrophysics, University of Toronto, 50 St.~George Street, Toronto, ON M5S 3H4, Canada}
  \affiliation{David A.~Dunlap Department of Astronomy \& Astrophysics, University of Toronto, 50 St.~George Street, Toronto, ON M5S 3H4, Canada}
\author[0000-0003-2047-5276]{Tomas Cassanelli}
  \affiliation{Dunlap Institute for Astronomy \& Astrophysics, University of Toronto, 50 St.~George Street, Toronto, ON M5S 3H4, Canada}
  \affiliation{David A.~Dunlap Department of Astronomy \& Astrophysics, University of Toronto, 50 St.~George Street, Toronto, ON M5S 3H4, Canada}
\author[0000-0002-3426-7606]{Pragya Chawla}
  \affiliation{Department of Physics, McGill University, 3600 rue University, Montr\'eal, QC H3A 2T8, Canada}
  \affiliation{McGill Space Institute, McGill University, 3550 rue University, Montr\'eal, QC H3A 2A7, Canada}
\author[0000-0003-0173-6274]{Tianyue Chen}
  \affiliation{MIT Kavli Institute for Astrophysics and Space Research, Massachusetts Institute of Technology, 77 Massachusetts Ave, Cambridge, MA 02139, USA}
\author[0000-0001-6509-8430]{J.-F.~Cliche}
  \affiliation{Department of Physics, McGill University, 3600 rue University, Montr\'eal, QC H3A 2T8, Canada}
\author[0000-0001-6422-8125]{Amanda Cook}
  \affiliation{Dunlap Institute for Astronomy \& Astrophysics, University of Toronto, 50 St.~George Street, Toronto, ON M5S 3H4, Canada}
  \affiliation{David A.~Dunlap Department of Astronomy \& Astrophysics, University of Toronto, 50 St.~George Street, Toronto, ON M5S 3H4, Canada}
\author[0000-0003-2319-9676]{Davor Cubranic}
  \affiliation{Department of Physics and Astronomy, University of British Columbia, 6224 Agricultural Road, Vancouver, BC V6T 1Z1 Canada}
\author[0000-0002-8376-1563]{Alice P.~Curtin}
  \affiliation{Department of Physics, McGill University, 3600 rue University, Montr\'eal, QC H3A 2T8, Canada}
  \affiliation{McGill Space Institute, McGill University, 3550 rue University, Montr\'eal, QC H3A 2A7, Canada}
\author[0000-0001-8123-7322]{Meiling Deng}
  \affiliation{Perimeter Institute for Theoretical Physics, 31 Caroline Street N, Waterloo, ON N25 2YL, Canada}
  \affiliation{Dominion Radio Astrophysical Observatory, Herzberg Research Centre for Astronomy and Astrophysics, National Research Council Canada, PO Box 248, Penticton, BC V2A 6J9, Canada}
\author[0000-0001-7166-6422]{Matt Dobbs}
  \affiliation{Department of Physics, McGill University, 3600 rue University, Montr\'eal, QC H3A 2T8, Canada}
  \affiliation{McGill Space Institute, McGill University, 3550 rue University, Montr\'eal, QC H3A 2A7, Canada}
\author[0000-0003-4098-5222]{Fengqiu (Adam) Dong}
  \affiliation{Department of Physics and Astronomy, University of British Columbia, 6224 Agricultural Road, Vancouver, BC V6T 1Z1 Canada}
\author[0000-0003-3734-8177]{Gwendolyn Eadie}
  \affiliation{David A.~Dunlap Department of Astronomy \& Astrophysics, University of Toronto, 50 St.~George Street, Toronto, ON M5S 3H4, Canada}
\author[0000-0002-6899-1176]{Mateus Fandino}
  \affiliation{Department of Physics and Astronomy, University of British Columbia, 6224 Agricultural Road, Vancouver, BC V6T 1Z1 Canada}
\author[0000-0001-8384-5049]{Emmanuel Fonseca}
  \affiliation{Department of Physics, McGill University, 3600 rue University, Montr\'eal, QC H3A 2T8, Canada}
  \affiliation{McGill Space Institute, McGill University, 3550 rue University, Montr\'eal, QC H3A 2A7, Canada}
\author[0000-0002-3382-9558]{B.~M.~Gaensler}
  \affiliation{Dunlap Institute for Astronomy \& Astrophysics, University of Toronto, 50 St.~George Street, Toronto, ON M5S 3H4, Canada}
  \affiliation{David A.~Dunlap Department of Astronomy \& Astrophysics, University of Toronto, 50 St.~George Street, Toronto, ON M5S 3H4, Canada}
\author[0000-0001-5553-9167]{Utkarsh Giri}
  \affiliation{Perimeter Institute for Theoretical Physics, 31 Caroline Street N, Waterloo, ON N25 2YL, Canada}
  \affiliation{Department of Physics and Astronomy, University of Waterloo, Waterloo, ON N2L 3G1, Canada}
\author[0000-0003-1884-348X]{Deborah C.~Good}
  \affiliation{Department of Physics and Astronomy, University of British Columbia, 6224 Agricultural Road, Vancouver, BC V6T 1Z1 Canada}
\author[0000-0002-1760-0868]{Mark Halpern}
  \affiliation{Department of Physics and Astronomy, University of British Columbia, 6224 Agricultural Road, Vancouver, BC V6T 1Z1 Canada}
\author[0000-0001-7301-5666]{Alex S.~Hill}
  \affiliation{Department of Computer Science, Math, Physics, \& Statistics, University of British Columbia, Kelowna, BC V1V 1V7, Canada}
  \affiliation{Dominion Radio Astrophysical Observatory, Herzberg Research Centre for Astronomy and Astrophysics, National Research Council Canada, PO Box 248, Penticton, BC V2A 6J9, Canada}
\author[0000-0002-4241-8320]{Gary Hinshaw}
  \affiliation{Department of Physics and Astronomy, University of British Columbia, 6224 Agricultural Road, Vancouver, BC V6T 1Z1 Canada}
\author[0000-0003-3059-6223]{Alexander Josephy}
  \affiliation{Department of Physics, McGill University, 3600 rue University, Montr\'eal, QC H3A 2T8, Canada}
  \affiliation{McGill Space Institute, McGill University, 3550 rue University, Montr\'eal, QC H3A 2A7, Canada}
\author[0000-0003-4810-7803]{Jane F.~Kaczmarek}
  \affiliation{Dominion Radio Astrophysical Observatory, Herzberg Research Centre for Astronomy and Astrophysics, National Research Council Canada, PO Box 248, Penticton, BC V2A 6J9, Canada}
\author[0000-0003-2739-5869]{Zarif Kader}
  \affiliation{Department of Physics, McGill University, 3600 rue University, Montr\'eal, QC H3A 2T8, Canada}
  \affiliation{McGill Space Institute, McGill University, 3550 rue University, Montr\'eal, QC H3A 2A7, Canada}
\author[0000-0002-3354-3859]{Joseph W.~Kania}
  \affiliation{Center for Gravitational Waves and Cosmology, West Virginia University, Chestnut Ridge Research Building, Morgantown, WV 26505, USA}
  \affiliation{Department of Physics and Astronomy, West Virginia University, PO Box 6315, Morgantown, WV 26506, USA }
\author[0000-0001-9345-0307]{Victoria M.~Kaspi}
  \affiliation{Department of Physics, McGill University, 3600 rue University, Montr\'eal, QC H3A 2T8, Canada}
  \affiliation{McGill Space Institute, McGill University, 3550 rue University, Montr\'eal, QC H3A 2A7, Canada}
\author[0000-0003-1455-2546]{T.~L.~Landecker}
  \affiliation{Dominion Radio Astrophysical Observatory, Herzberg Research Centre for Astronomy and Astrophysics, National Research Council Canada, PO Box 248, Penticton, BC V2A 6J9, Canada}
\author[0000-0002-1172-0754]{Dustin Lang}
  \affiliation{Perimeter Institute for Theoretical Physics, 31 Caroline Street N, Waterloo, ON N25 2YL, Canada}
  \affiliation{Department of Physics and Astronomy, University of Waterloo, Waterloo, ON N2L 3G1, Canada}
\author[0000-0002-4209-7408]{Calvin Leung}
  \affiliation{MIT Kavli Institute for Astrophysics and Space Research, Massachusetts Institute of Technology, 77 Massachusetts Ave, Cambridge, MA 02139, USA}
  \affiliation{Department of Physics, Massachusetts Institute of Technology, 77 Massachusetts Ave, Cambridge, MA 02139, USA}
\author[0000-0001-7931-0607]{Dongzi Li}
  \affiliation{Department of Physics, University of Toronto, 60 St.~George Street, Toronto, ON M5S 1A7, Canada}
  \affiliation{Canadian Institute for Theoretical Astrophysics, 60 St.~George Street, Toronto, ON M5S 3H8, Canada}
\author[0000-0001-7453-4273]{Hsiu-Hsien Lin}
  \affiliation{Canadian Institute for Theoretical Astrophysics, 60 St.~George Street, Toronto, ON M5S 3H8, Canada}
\author[0000-0002-4279-6946]{Kiyoshi W.~Masui}
  \affiliation{MIT Kavli Institute for Astrophysics and Space Research, Massachusetts Institute of Technology, 77 Massachusetts Ave, Cambridge, MA 02139, USA}
  \affiliation{Department of Physics, Massachusetts Institute of Technology, 77 Massachusetts Ave, Cambridge, MA 02139, USA}
\author[0000-0001-7348-6900]{Ryan Mckinven}
  \affiliation{Dunlap Institute for Astronomy \& Astrophysics, University of Toronto, 50 St.~George Street, Toronto, ON M5S 3H4, Canada}
  \affiliation{David A.~Dunlap Department of Astronomy \& Astrophysics, University of Toronto, 50 St.~George Street, Toronto, ON M5S 3H4, Canada}
\author[0000-0002-0772-9326]{Juan Mena-Parra}
  \affiliation{MIT Kavli Institute for Astrophysics and Space Research, Massachusetts Institute of Technology, 77 Massachusetts Ave, Cambridge, MA 02139, USA}
\author[0000-0003-2095-0380]{Marcus Merryfield}
  \affiliation{Department of Physics, McGill University, 3600 rue University, Montr\'eal, QC H3A 2T8, Canada}
  \affiliation{McGill Space Institute, McGill University, 3550 rue University, Montr\'eal, QC H3A 2A7, Canada}
\author[0000-0001-8845-1225]{Bradley W.~Meyers}
  \affiliation{Department of Physics and Astronomy, University of British Columbia, 6224 Agricultural Road, Vancouver, BC V6T 1Z1 Canada}
\author[0000-0002-2551-7554]{Daniele Michilli}
  \affiliation{Department of Physics, McGill University, 3600 rue University, Montr\'eal, QC H3A 2T8, Canada}
  \affiliation{McGill Space Institute, McGill University, 3550 rue University, Montr\'eal, QC H3A 2A7, Canada}
\author[0000-0001-8292-0051]{Nikola Milutinovic}
  \affiliation{Department of Physics and Astronomy, University of British Columbia, 6224 Agricultural Road, Vancouver, BC V6T 1Z1 Canada}
  \affiliation{Dominion Radio Astrophysical Observatory, Herzberg Research Centre for Astronomy and Astrophysics, National Research Council Canada, PO Box 248, Penticton, BC V2A 6J9, Canada}
\author[0000-0002-2626-5985]{Arash Mirhosseini}
  \affiliation{Department of Physics and Astronomy, University of British Columbia, 6224 Agricultural Road, Vancouver, BC V6T 1Z1 Canada}
\author[0000-0002-3777-7791]{Moritz M\"{u}nchmeyer}
  \affiliation{Perimeter Institute for Theoretical Physics, 31 Caroline Street N, Waterloo, ON N25 2YL, Canada}
  \affiliation{Department of Physics, University of Wisconsin-Madison, 1150 University Ave, Madison, WI 53706, USA}
\author[0000-0002-9225-9428]{Arun Naidu}
  \affiliation{Department of Physics, McGill University, 3600 rue University, Montr\'eal, QC H3A 2T8, Canada}
  \affiliation{McGill Space Institute, McGill University, 3550 rue University, Montr\'eal, QC H3A 2A7, Canada}
\author[0000-0002-7333-5552]{Laura Newburgh}
  \affiliation{Department of Physics, Yale University, New Haven, CT 06520, USA}
\author[0000-0002-3616-5160]{Cherry Ng}
  \affiliation{Dunlap Institute for Astronomy \& Astrophysics, University of Toronto, 50 St.~George Street, Toronto, ON M5S 3H4, Canada}
\author[0000-0003-3367-1073]{Chitrang Patel}
  \affiliation{Department of Physics, McGill University, 3600 rue University, Montr\'eal, QC H3A 2T8, Canada}
  \affiliation{Dunlap Institute for Astronomy \& Astrophysics, University of Toronto, 50 St.~George Street, Toronto, ON M5S 3H4, Canada}
\author[0000-0003-2155-9578]{Ue-Li Pen}
  \affiliation{Department of Physics, University of Toronto, 60 St.~George Street, Toronto, ON M5S 1A7, Canada}
  \affiliation{Dunlap Institute for Astronomy \& Astrophysics, University of Toronto, 50 St.~George Street, Toronto, ON M5S 3H4, Canada}
  \affiliation{David A.~Dunlap Department of Astronomy \& Astrophysics, University of Toronto, 50 St.~George Street, Toronto, ON M5S 3H4, Canada}
  \affiliation{Perimeter Institute for Theoretical Physics, 31 Caroline Street N, Waterloo, ON N25 2YL, Canada}
  \affiliation{Canadian Institute for Theoretical Astrophysics, 60 St.~George Street, Toronto, ON M5S 3H8, Canada}
\author[0000-0002-9822-8008]{Emily Petroff}
  \affiliation{Department of Physics, McGill University, 3600 rue University, Montr\'eal, QC H3A 2T8, Canada}
  \affiliation{McGill Space Institute, McGill University, 3550 rue University, Montr\'eal, QC H3A 2A7, Canada}
  \affiliation{Anton Pannekoek Institute for Astronomy, University of Amsterdam, Science Park 904, 1098 XH Amsterdam, The Netherlands}
  \affiliation{Veni Fellow}
\author[0000-0002-9516-3245]{Tristan Pinsonneault-Marotte}
  \affiliation{Department of Physics and Astronomy, University of British Columbia, 6224 Agricultural Road, Vancouver, BC V6T 1Z1 Canada}
\author[0000-0002-4795-697X]{Ziggy Pleunis}
  \affiliation{Department of Physics, McGill University, 3600 rue University, Montr\'eal, QC H3A 2T8, Canada}
  \affiliation{McGill Space Institute, McGill University, 3550 rue University, Montr\'eal, QC H3A 2A7, Canada}
\author[0000-0001-7694-6650]{Masoud Rafiei-Ravandi}
  \affiliation{Perimeter Institute for Theoretical Physics, 31 Caroline Street N, Waterloo, ON N25 2YL, Canada}
  \affiliation{Department of Physics and Astronomy, University of Waterloo, Waterloo, ON N2L 3G1, Canada}
\author[0000-0003-1842-6096]{Mubdi Rahman}
  \affiliation{Dunlap Institute for Astronomy \& Astrophysics, University of Toronto, 50 St.~George Street, Toronto, ON M5S 3H4, Canada}
  \affiliation{Sidrat Research, PO Box 73527 RPO Wychwood, Toronto, ON M6C 4A7, Canada}
\author[0000-0001-5799-9714]{Scott M.~Ransom}
  \affiliation{National Radio Astronomy Observatory, 520 Edgemont Rd, Charlottesville, VA 22903, USA}
\author[0000-0003-3463-7918]{Andre Renard}
  \affiliation{Dunlap Institute for Astronomy \& Astrophysics, University of Toronto, 50 St.~George Street, Toronto, ON M5S 3H4, Canada}
\author[0000-0001-5504-229X]{Pranav Sanghavi}
  \affiliation{Lane Department of Computer Science and Electrical Engineering, 1220 Evansdale Drive, PO Box 6109, Morgantown, WV 26506, USA}
  \affiliation{Center for Gravitational Waves and Cosmology, West Virginia University, Chestnut Ridge Research Building, Morgantown, WV 26505, USA}
\author[0000-0002-7374-7119]{Paul Scholz}
  \affiliation{Dunlap Institute for Astronomy \& Astrophysics, University of Toronto, 50 St.~George Street, Toronto, ON M5S 3H4, Canada}
\author[0000-0002-4543-4588]{J.~Richard Shaw}
  \affiliation{Department of Physics and Astronomy, University of British Columbia, 6224 Agricultural Road, Vancouver, BC V6T 1Z1 Canada}
\author[0000-0002-6823-2073]{Kaitlyn Shin}
  \affiliation{MIT Kavli Institute for Astrophysics and Space Research, Massachusetts Institute of Technology, 77 Massachusetts Ave, Cambridge, MA 02139, USA}
  \affiliation{Department of Physics, Massachusetts Institute of Technology, 77 Massachusetts Ave, Cambridge, MA 02139, USA}
\author[0000-0003-2631-6217]{Seth R.~Siegel}
  \affiliation{Department of Physics, McGill University, 3600 rue University, Montr\'eal, QC H3A 2T8, Canada}
  \affiliation{McGill Space Institute, McGill University, 3550 rue University, Montr\'eal, QC H3A 2A7, Canada}
\author[0000-0002-1235-4485]{Andrew E.~Sikora}
  \affiliation{Department of Physics, McGill University, 3600 rue University, Montr\'eal, QC H3A 2T8, Canada}
  \affiliation{McGill Space Institute, McGill University, 3550 rue University, Montr\'eal, QC H3A 2A7, Canada}
\author[0000-0001-7755-902X]{Saurabh Singh}
  \affiliation{Department of Physics, McGill University, 3600 rue University, Montr\'eal, QC H3A 2T8, Canada}
\author[0000-0002-2088-3125]{Kendrick M.~Smith}
  \affiliation{Perimeter Institute for Theoretical Physics, 31 Caroline Street N, Waterloo, ON N25 2YL, Canada}
\author[0000-0001-9784-8670]{Ingrid Stairs}
  \affiliation{Department of Physics and Astronomy, University of British Columbia, 6224 Agricultural Road, Vancouver, BC V6T 1Z1 Canada}
\author[0000-0001-7509-0117]{Chia Min Tan}
  \affiliation{Department of Physics, McGill University, 3600 rue University, Montr\'eal, QC H3A 2T8, Canada}
  \affiliation{McGill Space Institute, McGill University, 3550 rue University, Montr\'eal, QC H3A 2A7, Canada}
\author[0000-0003-2548-2926]{S.~P.~Tendulkar}
  \affiliation{National Centre for Radio Astrophysics, Post Bag 3, Ganeshkhind, Pune, 411007, India}
  \affiliation{Department of Astronomy and Astrophysics, Tata Institute of Fundamental Research, Mumbai, 400005, India}
\author[0000-0003-4535-9378]{Keith Vanderlinde}
  \affiliation{Dunlap Institute for Astronomy \& Astrophysics, University of Toronto, 50 St.~George Street, Toronto, ON M5S 3H4, Canada}
  \affiliation{David A.~Dunlap Department of Astronomy \& Astrophysics, University of Toronto, 50 St.~George Street, Toronto, ON M5S 3H4, Canada}
\author[0000-0002-1491-3738]{Haochen Wang}
  \affiliation{MIT Kavli Institute for Astrophysics and Space Research, Massachusetts Institute of Technology, 77 Massachusetts Ave, Cambridge, MA 02139, USA}
  \affiliation{Department of Physics, Massachusetts Institute of Technology, 77 Massachusetts Ave, Cambridge, MA 02139, USA}
\author[0000-0001-7314-9496]{Dallas Wulf}
  \affiliation{Department of Physics, McGill University, 3600 rue University, Montr\'eal, QC H3A 2T8, Canada}
  \affiliation{McGill Space Institute, McGill University, 3550 rue University, Montr\'eal, QC H3A 2A7, Canada}
\author[0000-0001-8278-1936]{A.~V.~Zwaniga}
  \affiliation{Department of Physics, McGill University, 3600 rue University, Montr\'eal, QC H3A 2T8, Canada}
\newcommand{\allacks}{
CHIME research at WVU is supported by the National Science Foundation under Grant No.~2006548.
A.E.S. is supported by an Natural Sciences and Engineering Research Council (NSERC) CGS-M award.
A.S.H. is supported by an NSERC Discovery Grant.
B.M.G. is supported by an NSERC Discovery Grant (RGPIN-2015-05948), and by the Canada Research Chairs (CRC) program.
C.L. is supported by the U.S. Department of Defense (DoD) through the National Defense Science \& Engineering Graduate Fellowship (NDSEG) Program.
D.C.G. is supported by the John I.~Watters Research Fellowship.
D.M. is a Banting Fellow.
E.P. acknowledges funding from an NWO Veni Fellowship.
G.H. is supported by an NSERC Discovery Grant, CIFAR, and the CRC Program.
G.M.E. is supported by an NSERC Discovery Grant (RGPIN-2020-04554) and by a Canadian Statistical Sciences Institute Collaborative Research Team Grant.
I.H.S. is supported by an NSERC Discovery Grant, CIFAR and a CFI John R.~Evans Leaders Fund Award. 
J.W.K. is supported by NSF Grants (1458952 and AAG-1616042).
K.S. is supported by the NSF Graduate Research Fellowship Program.
K.W.M. is supported by an NSF Grant (2008031).
M. M. is supported by an NSERC PGS-D award.
M.D. is supported by a Killam Fellowship, NSERC Discovery Grant, CIFAR, and by the FRQNT Centre de Recherche en Astrophysique du Qu\'ebec (CRAQ)
P.C. is supported by an FRQNT Doctoral Research Award.
P.S. is a Dunlap Fellow and an NSERC Postdoctoral Fellow. 
S.M.R. is a CIFAR Fellow and is supported by an NSF Physics Frontiers Center Award (1430284).
U.P. is supported by NSERC Grants (RGPIN-2019-067, CRD 523638-201), and by the Ontario Research Fund-Research Excellence Program (ORF-RE), CIFAR, Simons Foundation, Thoth Technology Inc, and Alexander von Humboldt Foundation. 
V.M.K. holds the Lorne Trottier Chair in Astrophysics \& Cosmology, a Distinguished James McGill Professorship and receives support from an NSERC Discovery Grant (RGPIN 228738-13) and Gerhard Herzberg Award, from an R.~Howard Webster Foundation Fellowship from CIFAR, and from the FRQNT CRAQ.
}